\documentclass[twocolumn]{aastex63}

\usepackage{epstopdf}
\usepackage{mathrsfs}
\usepackage{calrsfs}
\usepackage[hang,flushmargin]{footmisc}
\usepackage[autostyle, english = american]{csquotes}
\MakeOuterQuote{"}

\received{--}
\revised{--}
\accepted{--}

\submitjournal{ApJ}

\shorttitle{GPCAL: a generalized calibration pipeline for instrumental polarization in VLBI data}
\shortauthors{Park et al.}

\begin{document}

\title{GPCAL: a generalized calibration pipeline for instrumental polarization in VLBI data}

\correspondingauthor{Jongho Park}
\email{jpark@asiaa.sinica.edu.tw}

\author[0000-0001-6558-9053]{Jongho Park}
\affiliation{Institute of Astronomy and Astrophysics, Academia Sinica, P.O. Box 23-141, Taipei 10617, Taiwan}
\affiliation{Department of Physics and Astronomy, Seoul National University, Gwanak-gu, Seoul 08826, Republic of Korea}

\author[0000-0003-1157-4109]{Do-Young Byun}
\affiliation{Korea Astronomy and Space Science Institute, Daedeok-daero 776, Yuseong-gu, Daejeon 34055, Republic of Korea}
\affiliation{University of Science and Technology, Gajeong-ro 217, Yuseong-gu, Daejeon 34113, Republic of Korea}

\author{Keiichi Asada}
\affiliation{Institute of Astronomy and Astrophysics, Academia Sinica, P.O. Box 23-141, Taipei 10617, Taiwan}

\author[0000-0002-0822-2973]{Youngjoo Yun}
\affiliation{Korea Astronomy and Space Science Institute, Daedeok-daero 776, Yuseong-gu, Daejeon 34055, Republic of Korea}

\begin{abstract}

We present the Generalized Polarization CALibration pipeline (GPCAL), an automated pipeline for instrumental polarization calibration of very long baseline interferometry (VLBI) data. The pipeline is designed to achieve a high calibration accuracy by means of fitting the instrumental polarization model, including the second-order terms, to multiple calibrators data simultaneously. It also allows using more accurate linear polarization models of calibrators for D-term estimation compared to the conventional way that assumes similar linear polarization and total intensity structures. This assumption has widely been used in the existing packages for instrumental polarization calibration but could be a source of significant uncertainties when there is no suitable calibrator satisfying the assumption. We demonstrate the capabilities of GPCAL by using simulated data, archival Very Long Baseline Array (VLBA) data of many active galactic nuclei (AGN) jets at 15 and 43 GHz, and our Korean VLBI Network (KVN) observations of many AGN jets at 86, 95, 130, and 142 GHz. The pipeline could reproduce the complex linear polarization structures of several sources shown in the previous studies using the same VLBA data. GPCAL also reveals a complex linear polarization structure in the flat-spectrum radio quasar 3C 273 from the KVN data at all four frequencies. These results demonstrate that GPCAL can achieve a high calibration accuracy for various VLBI arrays.

\end{abstract}

\keywords{high angular resolution --- techniques: interferometric --- techniques: polarimetric --- methods: data analysis}

\section{Introduction} \label{sec:intro}

Very Long Baseline Interferometry (VLBI) is the technique that enables us to achieve very high angular resolution by using widely separated antennas as elements of an interferometer array. Polarization observations with VLBI have revealed the ordering and orientation of magnetic fields in and around nonthermal radio sources, which are believed to play a critical role in their formation and evolution. For example, observations of orientations of the magnetic fields in the jets of various active galactic nuclei (AGNs) suggest the presence of shocks in the jets \citep[e.g.,][]{LH2005, Jorstad2007}. In some AGN jets, Faraday rotation measure (RM) gradients transverse to the jets were revealed, which indicates the presence of helical magnetic fields wrapping around those jets \citep[e.g.,][]{Asada2002, Gabuzda2004, ZT2005}. Also, the magnitude of RM was found to systematically increase with decreasing distance from the black hole in the jet of the nearby radio galaxy M87 \citep{Park2019} and increase with frequency in the cores of blazars \citep[e.g.,][]{Jorstad2007, OG2009, Hovatta2012, Hovatta2019, Kravchenko2017, Park2018}. These results suggest that AGN jets tend to be in the environments of higher electron densities and/or stronger magnetic fields at shorter distances from the central engines.

An ideal circular polarization feed would respond to only one polarization. However, in reality, any feed will also respond to the other polarization signal, introducing "instrumental" polarization signals to the data. These signals are of the same order as the intrinsic source polarization signals in many cases and must be properly estimated and removed from the data. LPCAL, a task implemented in Astronomical Image Processing System \citep[AIPS,][]{Greisen2003} based on the linearized leakage model \citep[][see also e.g., \citealt{Cotton1993, Roberts1994} for more details of instrumental polarization calibration]{Leppanen1995}, has been widely used for instrumental polarization calibration of VLBI data. It has been very successful for a great deal of studies using various VLBI arrays such as the Very Long Baseline Array \citep[VLBA, e.g.,][]{Jorstad2017, Lister2018}, the Global mm-VLBI Array \citep[GMVA, e.g.,][]{Casadio2017}, the High Sensitivity Array \citep[HSA, e.g.,][]{Hada2016}, the \emph{RadioAstron} space VLBI mission \citep[e.g.,][]{Gomez2016}, and the Korean VLBI Network \citep[KVN, e.g.,][]{Park2018}.

Nevertheless, there are some circumstances that one needs different calibration strategies and improved calibration accuracy. First of all, for global VLBI arrays such as the Event Horizon Telescope \citep[EHT,][]{EHT2019a,EHT2019b,EHT2019c,EHT2019d,EHT2019e,EHT2019f} and the GMVA, a common sky area for a single calibrator for some antennas can be quite limited. In this case, the parallactic angle coverages of the calibrator for those antennas would also be limited, resulting in a relatively inaccurate calibration. Since both the common sky area and parallactic angle coverage are often sensitive to the source's declination (see, e.g., Figure 2 in \citealt{Trippe2010}), using multiple calibrators at different declinations to model the instrumental polarization signals can help to improve the calibration accuracy. 

Secondly, LPCAL relies on the similarity assumption, which assumes that linear polarization structures of calibrators are proportional to their total intensity structures\footnote{LPCAL allows us to split the total intensity models of calibrators into several "sub-models" and apply the similarity assumption for each sub-model (see \citealt{Leppanen1995} and Section~\ref{sec:simil} for more details). This is a good strategy if the calibrators consist of several distinct emission regions and the linear polarization and total intensity structures are similar in each region.} \citep{Cotton1993, Leppanen1995}. However, this assumption may not always hold, especially at high frequencies. Most calibrators for VLBI observations are resolved and show significant variability in their source structures. Thus, it is often challenging to have suitable calibrators satisfying the assumption in the data, even though observers select the calibrators based on the information of the source's linear polarization structures from previous observations.

Thirdly, some VLBI arrays, such as the KVN and the EHT, do not have many antennas \citep{Park2018, EHT2019b}. It is challenging for those arrays to have instrumental polarization removed from the data adequately due to the small number of measurements\footnote{It would be important for those arrays to keep searching for good calibrators that are either very compact or have very low degrees of polarization, although that appears to be challenging at mm wavelengths \citep[e.g.,][]{Casadio2017}.} (baselines). Combining the results from multiple calibrators could mitigate the difficulty. Still, it is generally not straightforward to take into account different signal-to-noise ratios (SNRs) and parallactic angle coverages of various sources for combining the results. 

Lastly, some heterogeneous VLBI arrays such as the EHT and GMVA in conjunction with the Atacama Large Millimeter/submillimeter Array \citep[ALMA, e.g.,][]{EHT2019a, Issaoun2019} and the HSA including the phased-up Very Large Array \citep[VLA, e.g.,][]{Hada2017} have very different sensitivities among different stations. In this case, fitting would be dominated by the baselines to the most sensitive stations. If the antenna gains of those sensitive stations are not well corrected and there are remaining systematic errors in the data, the fitting solutions of all other stations can be distorted. One can scale the visibility weights of the sensitive stations down for fitting to avoid this problem, as done for the imaging of the shadow of the supermassive black hole in M87 with CLEAN\footnote{Finding the "sweet spot" of the relative weights between stations may require a dedicated investigation through e.g., a parameter survey using simulated data \citep{EHT2019d}.} \citep{EHT2019d}.

These motivated us to develop a new pipeline for instrumental polarization calibration of VLBI data, named the Generalized Polarization CALibration pipeline (GPCAL). It allows us to (i) fit the instrumental polarization model to multiple calibrators data simultaneously, (ii) use more accurate linear polarization models of calibrators for fitting, (iii) flexibly change the visibility weights of each station, and (iv) easily check the fitting results and statistics. It is based on AIPS and Difmap \citep{Shepherd1997}, which have been widely used for calibration and imaging of VLBI data for a long time. We implemented external scripts written in Python in the pipeline only for the impossible or difficult parts to deal with AIPS and Difmap. This makes the pipeline more reliable and friendly to the users who are already familiar with those softwares. 

In Section~\ref{sec:model}, we describe the model of instrumental polarization employed in GPCAL. The general calibration scheme of GPCAL is explained in detail in Section~\ref{sec:pipeline}. We verify the pipeline and demonstrate the capabilities of GPCAL by using simulated VLBI data and real data observed with different VLBI arrays at different frequencies in Section~\ref{sec:results}. We present a concluding summary in Section~\ref{sec:summary}.

\section{D-term Model} \label{sec:model}

We follow \cite{Leppanen1995} for description of a model for the interferometer response, which relates the measured cross correlations ($r^{RR}$, $r^{LL}$, $r^{RL}$, and $r^{LR}$) and the true visibilities ($\mathscr{RR}$, $\mathscr{LL}$, $\mathscr{RL}$, and $\mathscr{LR}$) on baseline $mn$.
\begin{eqnarray}
\label{eq:model}
r^{RR}_{mn} &=& G^{R}_{m}G^{R*}_{n}[e^{-j(\phi_m-\phi_n)}\mathscr{RR} + D^R_me^{j(\phi_m+\phi_n)}\mathscr{LR} +\nonumber\\
&& D^{R*}_ne^{-j(\phi_m+\phi_n)}\mathscr{RL} + D^R_{m}D^{R*}_{n}e^{j(\phi_m-\phi_n)}\mathscr{LL}] \nonumber \\
r^{LL}_{mn} &=& G^{L}_{m}G^{L*}_{n}[e^{j(\phi_m-\phi_n)}\mathscr{LL} + D^L_me^{-j(\phi_m+\phi_n)}\mathscr{RL} +\nonumber\\
&& D^{L*}_ne^{j(\phi_m+\phi_n)}\mathscr{LR} + D^L_{m}D^{L*}_{n}e^{-j(\phi_m-\phi_n)}\mathscr{RR}] \nonumber \\
r^{RL}_{mn} &=& G^{R}_{m}G^{L*}_{n}[e^{-j(\phi_m+\phi_n)}\mathscr{RL} + D^R_me^{j(\phi_m-\phi_n)}\mathscr{LL} +\nonumber\\
&& D^{L*}_ne^{-j(\phi_m-\phi_n)}\mathscr{RR} + D^R_{m}D^{L*}_{n}e^{j(\phi_m+\phi_n)}\mathscr{LR}] \nonumber \\
r^{LR}_{mn} &=& G^{L}_{m}G^{R*}_{n}[e^{j(\phi_m+\phi_n)}\mathscr{LR} + D^L_me^{-j(\phi_m-\phi_n)}\mathscr{RR} +\nonumber\\
&& D^{R*}_ne^{j(\phi_m-\phi_n)}\mathscr{LL} + D^L_{m}D^{R*}_{n}e^{-j(\phi_m+\phi_n)}\mathscr{RL}],\nonumber\\
\end{eqnarray}
\noindent where the star denotes a complex conjugate, $G$ the complex antenna gains, $D$ the leakage factors (so-called "D-terms"), and $\phi$ the antenna field rotation angles. Subscripts denote antenna numbers, and superscripts denote polarization. The field rotation angle is a function of the elevation angle ($\theta_{\rm el}$) and the parallactic angle ($\psi_{\rm par}$), depending on antenna mounts:
\begin{equation}
    \phi = f_{\rm el}\theta_{\rm el} + f_{\rm par}\psi_{\rm par} + \phi_{\rm off},
\end{equation}
where $\phi_{\rm off}$ is a constant offset, which is expected when the antenna feed is rotated with respect to the azimuth axis, which is the case for e.g., the ALMA (see Section 4.2 of ALMA Cycle 7 Technical Handbook). Cassegrain mounts have $f_{\rm par} = 1$ and $f_{\rm el} = 0$ and thus the field rotation angle is equivalent to the parallactic angle, except for the constant offset. Nasmyth mounts have $f_{\rm par} = 1$ and $f_{\rm el} = +1$ for Nasmyth-Right type and $f_{\rm el} = -1$ for Nasmyth-Left type. The true cross-hand visibilities are related to the Stokes parameters as 
\begin{eqnarray}
\mathscr{RL}&=&\mathscr{Q}+j\mathscr{U}=\mathscr{P} \nonumber \\
\mathscr{LR}&=&\mathscr{Q}-j\mathscr{U}=\mathscr{P}^*,
\end{eqnarray}
where $\mathscr{Q}$ and $\mathscr{U}$ are the Fourier transforms of the source's Stokes $Q$ and $U$ on the sky, respectively, $\mathscr{P}$ the Fourier transform of the complex polarization $P\equiv pIe^{2j\chi}$, $p$ the fractional polarization, $I$ the total intensity emission on the sky, $\chi$ the electric vector position angle \citep[EVPA, e.g.,][]{Roberts1994}. GPCAL assumes that the field rotation angles were already corrected and the antenna gains were corrected except for the phase offset between RCP and LCP at the reference antenna\footnote{The phase offsets between polarizations in other stations are expected to be removed during the global fringe fitting and self-calibration \citep[see e.g.,][]{CS1983, Roberts1994, Cotton1995a, Cotton1995b, Leppanen1995}. A single phase offset ($e^{j\phi_{RL, {\rm ref}}}$), which is believed to originate from the instrumental phase offset between polarizations at the reference antenna, will remain in all baselines. This offset is usually assumed to be constant during the observations and can be corrected after D-term correction (so-called the EVPA calibration). This is absorbed in different terms in Equation~\ref{model:model} such that $\mathscr{P}\rightarrow\mathscr{P}e^{j\phi_{RL, {\rm ref}}}$, $D^R \rightarrow D^R e^{j\phi_{RL, {\rm ref}}}$, and $D^L \rightarrow D^L e^{-j\phi_{RL, {\rm ref}}}$.}. Then, one can write the model cross-hand visibilities ($\tilde{r}^{RL}_{mn}$, $\tilde{r}^{LR}_{mn}$) for each measurement at $(u,v)$ coordinates as:
\begin{eqnarray}
\label{model:model}
&&\tilde{r}^{RL}_{mn}(u,v) = \mathscr{P}(u,v) + D^R_me^{2j\phi_m} r^{LL}_{mn}(u,v) + \nonumber\\
&&\quad D^{L*}_ne^{2j\phi_n} r^{RR}_{mn}(u,v) + D^R_{m}D^{L*}_{n}e^{2j(\phi_m+\phi_n)} \mathscr{P}^*(u,v) \nonumber \\
&&\tilde{r}^{LR}_{mn}(u,v) = \mathscr{P}^*(u,v) + D^L_me^{-2j\phi_m} r^{RR}_{mn}(u,v) + \nonumber\\
&&\quad D^{R*}_ne^{-2j\phi_n} r^{LL}_{mn}(u,v) + D^L_{m}D^{R*}_{n}e^{-2j(\phi_m+\phi_n)} \mathscr{P}(u,v).\nonumber\\
\end{eqnarray}
The antenna gains are assumed to be perfectly corrected and the true parallel-hand visibilities ($\mathscr{RR}$, $\mathscr{LL}$) are replaced with the measured parallel-hand visibilities ($r^{RR}$, $r^{LL}$) in these equations. GPCAL fits these model equations to the measured cross-hand visibilities to derive the best-fit D-terms.

\section{GPCAL calibration procedures} \label{sec:pipeline}

\subsection{Modeling of source polarization structure}

Equation~\ref{model:model} requires the antenna field rotation angles and a source polarization model ($\mathscr{P}$) for each visibility data point. The former is a purely geometrical quantity depending on antenna positions and mounts and can be easily computed \citep[e.g.,][]{Cotton1993}. However, the latter is difficult to be constrained directly and requires some assumptions and strategies, which will be briefly discussed below.

\subsubsection{Similarity assumption}
\label{sec:simil}
The standard method is to assume that linearly polarized structures of calibrators are proportional to their total intensity structures, so-called the "similarity" approximation \citep{Cotton1993}. However, this might be an oversimplification for most calibrators, especially at high frequencies, which usually show variations in both fractional polarization and EVPA from regions to regions \citep[e.g.,][]{Jorstad2007, Lister2018}. One can roughly take into account these variations by splitting the source's total intensity CLEAN models into several sub-models ($I_s$) and apply the similarity approximation to each sub-model. In other words, each total intensity sub-model has a constant fractional polarization and EVPA across the sub-model region \citep{Leppanen1995}. This can be expressed as
\begin{equation}
    \mathscr{P}(u,v) = \sum_s p_s\mathscr{F}_s(u,v),
    \label{pipeline:simil}
\end{equation}
where $\mathscr{F}_s(u,v)$ is the Fourier transform of the sub-model $I_s$ and $p_s=(Q_s+jU_s)/I_s$ is the complex coefficient. With this approximation, the number of free parameters in the fitting would be $4N_{\rm ant}+\sum_{\rm cal} 2N_{s, {\rm cal}}$, where $N_{\rm ant}$ denotes the number of antennas and $N_{s, {\rm cal}}$ the number of sub-models for each calibrator . The coefficient 4 is from the real and imaginary parts of the D-terms of RCP and LCP for each antenna. $\sum_{\rm cal} N_{s, {\rm cal}}$ is the total number of sub-models for all calibrators used in the fitting. The coefficient 2 comes from the real and imaginary parts of $p_s$.

\subsubsection{Instrumental polarization self-calibration}
\label{sec:selfpol}
The similarity assumption may not hold in some cases. This is difficult to predict before observations because many calibrators used for VLBI observations show significant variability in their source structures. A possible solution to be nearly free from the similarity assumption and achieve a better calibration accuracy is as follows.
\begin{enumerate}
    \item Obtain the best-fit D-terms by using the similarity assumption and remove the D-terms from the data.
    \item Produce model visibilities $\mathscr{P}(u,v) = \mathscr{Q}(u,v) + j\mathscr{U}(u,v)$ and $\mathscr{P}^*(u,v) = \mathscr{Q}(u,v) - j\mathscr{U}(u,v)$ from imaging of source's Stokes $Q$ and $U$ with CLEAN \citep{Hogbom1974} using the D-term corrected data.
    \item Fit Equation~\ref{model:model} to the D-term un-corrected data and solve for the D-terms only by using the model visibilities $\mathscr{P}(u,v)$ constrained in 2. Remove the D-terms from the data using the new best-fit D-term estimates.
    \item Iterate 2 and 3 until the solutions and the fitting statistics are converged.
\end{enumerate}
This scheme is very similar to self-calibration of parallel-hand data, which iterates (i) imaging of source's total intensity structures and (ii) solving for antenna gains using the model visibilities. This approach was therefore named "instrumental polarization self-calibration" \citep[][see Section 15.4.3]{Cotton1995b}. In this case, the number of free parameters in the fitting is $4N_{\rm ant}$ because the source-polarization information is separately obtained by CLEAN prior to the fitting (in step 2).

\begin{figure*}[t!]
\centering
\includegraphics[width = \textwidth]{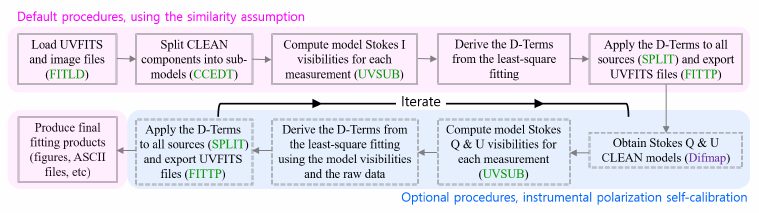}
\caption{Overview of the GPCAL procedures. The pipeline largely consists of two stages: (i) the default procedures using the similarity assumption (magenta color) and (ii) the optional procedures using instrumental polarization self-calibration (blue color). The second stage is iterated as many times as requested by users. GPCAL can use multiple calibrators simultaneously for both stages and different sets of calibrators can be used for different stages. GPCAL uses ParselTongue \citep{Kettenis2006} for running necessary AIPS tasks (green color in the parentheses) and Difmap \citep{Shepherd1997} for obtaining Stokes $Q$ \& $U$ CLEAN models for instrumental polarization self-calibration. \label{fig:schematic}}
\end{figure*}

\subsection{Calibration procedures}
This subsection describes the calibration procedures employed in GPCAL. GPCAL uses ParselTongue, a Python interface to AIPS \citep{Kettenis2006}, for using relevant AIPS tasks. GPCAL reads the input parameters provided by users and runs all the calibration procedures accordingly. First, self-calibrated UVFITS files and image files containing the CLEAN components of calibrators are loaded into AIPS. If self-calibration was performed in Difmap, which assumes that antenna gains for RCP and LCP are the same, then GPCAL can perform an additional self-calibration using CALIB in AIPS to correct the possible remaining antenna gain offsets between polarizations\footnote{GPCAL does not assume $\mathcal{RR} = \mathcal{LL}$, which indicates that it takes into account possible non-zero Stokes $\mathcal{V} \equiv (\mathcal{RR} - \mathcal{LL}) / 2$ in the data (Equation~\ref{eq:model}). However, the circular polarization of AGN jets is expected to be small ($\lesssim1\%$, e.g., \citealt{Wardle1998, HW1999, HL2006}) and one must correct the antenna gains, especially the gain offsets between RCP and LCP, with a good accuracy to obtain the weak source-intrinsic circular polarization signals. The gain offset correction usually requires a careful investigation using many calibrators observed during the same run \citep[e.g.,][]{HW1999, HL2006}. Thus, if there are not many calibrators in the data or obtaining source's circular polarization is not a primary goal of the study, it is a reasonable approach to perform self-calibration assuming $\mathcal{RR}=\mathcal{LL}$ \citep[e.g.,][]{Jorstad2005}. GPCAL allows users to follow the same strategy.}. If requested, GPCAL splits the total intensity CLEAN models into several sub-models using CCEDT in AIPS. The sub-model split can be done manually by providing ASCII files which contain each sub-model's locations on the map or automatically by CCEDT. Then, the model visibility corresponding to each measurement at $(u,v)$ for each sub-model, $\mathscr{F}_s(u,v)$ in Equation~\ref{pipeline:simil}, is computed by the AIPS task UVSUB. For each visibility measurement, the antenna field rotation angles are computed by using the source coordinates, the antenna positions, and the antenna mounts in the headers of the UVFITS files.

Then, GPCAL fits Equation~\ref{model:model} to the observed cross-hand visibility data for each baseband channel (often called an intermediate frequency; IF) using the non-linear least-square fitting algorithms implemented in \texttt{Scipy}\footnote{\url{https://docs.scipy.org/doc/scipy/reference/generated/scipy.optimize.curve\_fit.html}}. The similarity assumption is used at this stage, using the CLEAN sub-models produced by CCEDT. If multiple calibrators are requested to be used, then it assumes the same D-terms for the calibrators and different source polarization terms for different sources. The fitting algorithm uses all the visibilities of the requested calibrators simultaneously, using the visibility weights stored in the UVFITS files. Thus, calibrators having higher SNRs would affect the fitting more, which is a good strategy if systematic uncertainties in the data such as antenna gains have been corrected with a good accuracy. GPCAL loads the UVFITS files of all the sources specified in the input parameters into AIPS for applying the best-fit D-terms. Additional self-calibration with CALIB can also be performed for these sources if requested. The AIPS antenna tables of the UVFITS files are updated with the best-fit D-terms. The D-term corrected UVFITS files are produced and exported to the working directory by SPLIT and FITTP in AIPS, respectively. The D-terms in both the parallel and cross-hand visibilities including the second order terms (Equation~\ref{eq:model}) are corrected.

If users request to perform instrumental polarization self-calibration, then GPCAL executes further calibration procedures. It employs a simple Difmap script, which reads the D-term corrected UVFITS files and the CLEAN windows used for total intensity imaging provided by users. The script performs imaging of calibrators' Stokes $Q$ and $U$ with CLEAN until the peak intensity within the windows in the dirty maps reaches the map root-mean-square noise times a certain factor specified by users. After the imaging is completed for all the calibrators for instrumental polarization self-calibration, the images are loaded into AIPS. The Stokes $Q$ and $U$ models for each visibility measurement are extracted by UVSUB, from which $\mathscr{P}(u,v)$ in Equation~\ref{model:model} is computed. GPCAL fits the model (Equation~\ref{model:model}) again to the visibilities of all the specified calibrators simultaneously but solves for the D-terms only this time by using the model source-polarization visibilities. The best-fit D-terms are applied to the data with SPLIT and the D-term corrected UVFITS files are exported to the working directory with FITTP. 

This procedure, obtaining the model polarization visibilities with CLEAN, solving for the D-terms using the model visibilities and the D-term un-corrected data, and producing new D-term corrected UVFITS files, is repeated as many times as specified in the input parameters. The calibrators for this procedure do not have to be the same as those used for the initial D-term estimation using the similarity assumption. Calibrators with high fractional linear polarization with complex polarization structures can be usable. Nevertheless, selecting good calibrators, having either very low degrees of linear polarization or compact linear polarization structures, is important, especially for the initial D-term estimation. This is because instrumental polarization self-calibration, very similar to total intensity imaging and self-calibration, would work well only when the initial D-term estimates are reasonably close to the real D-terms. Similarly, one should avoid using calibrators having poor field rotation angle coverages for many stations and low SNRs. Those calibrators would easily degrade the D-term estimates. We present the results of a simple test which demonstrates the importance of selecting good calibrators in Appendix~\ref{appendix:goodcal}.

GPCAL produces several kinds of figures: 1. plots showing the antenna field rotation angles of the calibrators, 2. plots showing the Stokes $Q$ and $U$ visibilities (amplitudes and phases) of the calibrators and their best-fit models, and 3. plots showing the fitting residuals in units of visibility errors for each station, averaged over each scan and over all baselines. The first plots allow users to check if the field rotation angles of the calibrators have wide enough coverages for all antennas, which is essential for robust D-term estimation. The second plots are useful for examining the overall goodness of fit for each baseline. The last plots help identify some problematic scans for some stations, showing large fitting residuals. Large residuals could be caused by imperfect antenna gain correction or the elevation dependence of D-terms of some stations, which violates the assumption of constant D-terms during observations used in the model equation.

GPCAL computes the reduced chi-square of the fit ($\chi^2_{\rm red}$) for each step (using the similarity assumption and the $n$-th iteration of instrumental polarization self-calibration) and produces a plot of $\chi^2_{\rm red}$ as a function of the steps. It also produces plots of the fitted D-terms on the real and imaginary plane for each IF. The fitted D-terms are saved in ASCII files for each step. A log file containing all the procedures in AIPS, Difmap, and GPCAL is also produced. Users can investigate the cause of potential errors during the pipeline running with the log file. An overview of the pipeline procedures is summarized in Figure~\ref{fig:schematic}. The pipeline is publicly available at: \url{https://github.com/jhparkastro/gpcal}.

\subsection{Optional functions}

GPCAL provides functions that could be useful in some specific circumstances, which are briefly summarized as follows.

\begin{figure*}[t!]
\centering
\includegraphics[width = 0.49\textwidth]{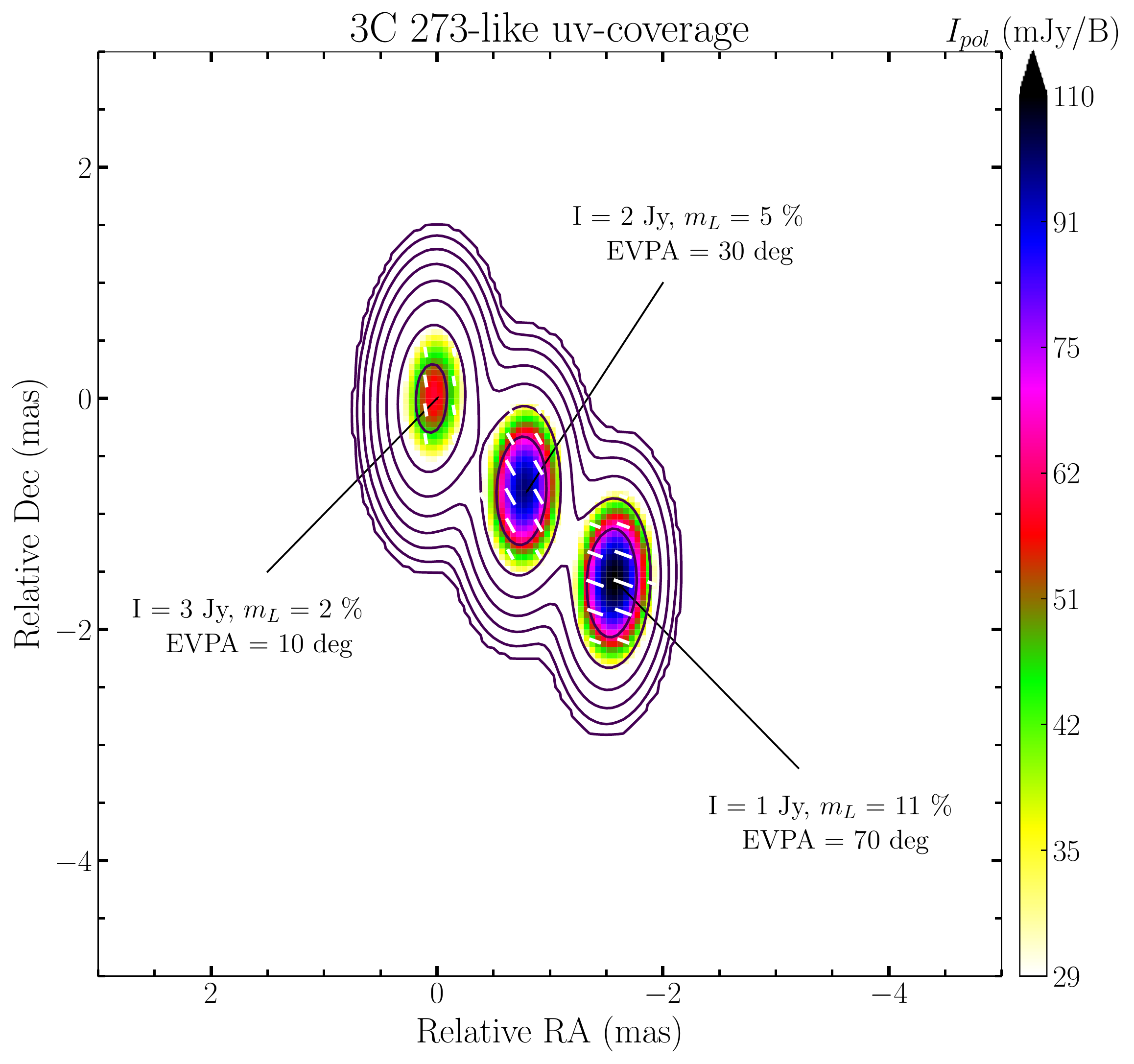}
\includegraphics[width = 0.49\textwidth]{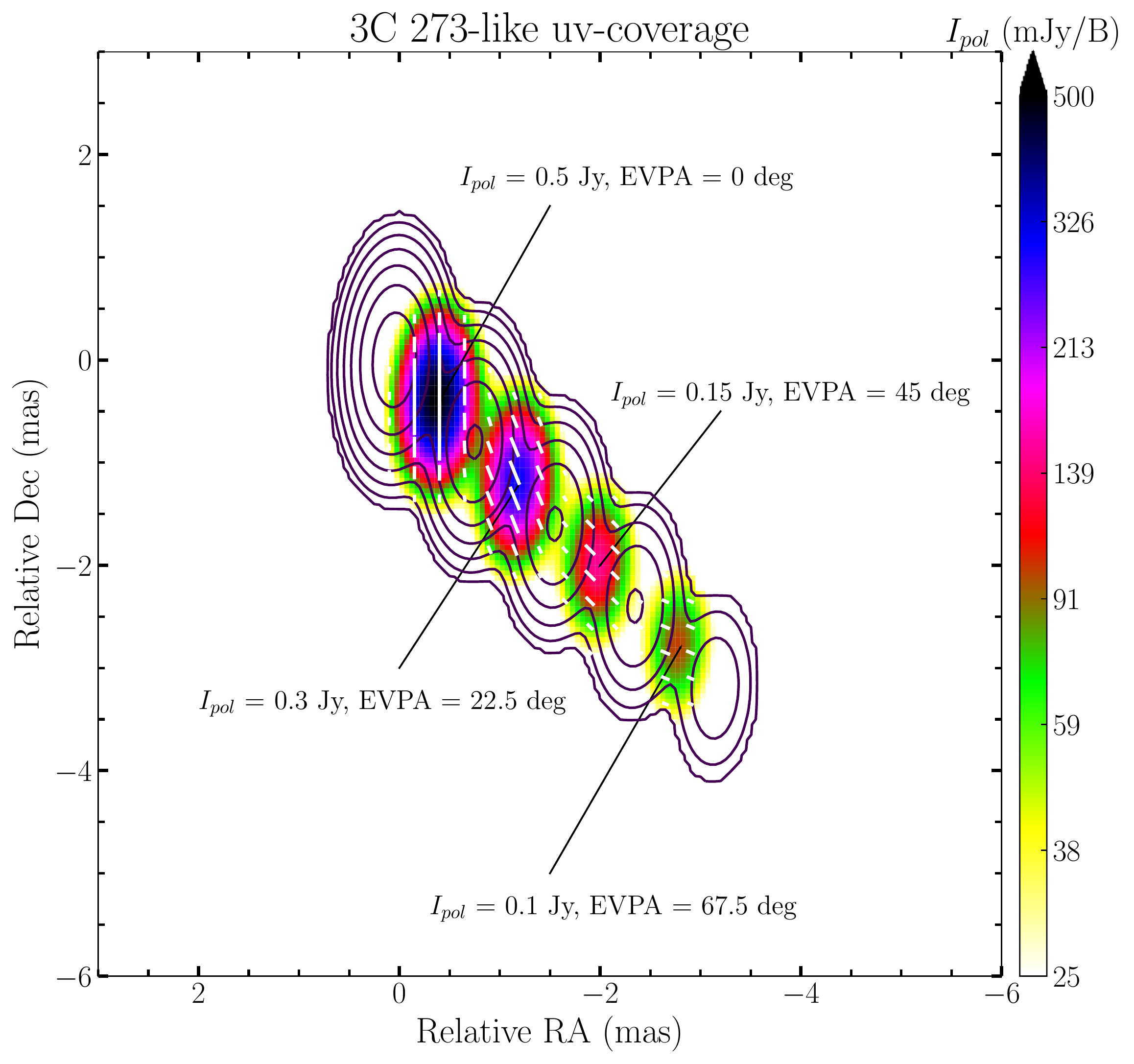}
\caption{Linear polarization maps of the simulated data which assume simple source geometries, a 3C 273-like uv-coverage, and a VLBA-like array. Color shows the distributions of linearly polarized intensity and the white sticks show EVPAs. The length of the white sticks is proportional to the linearly polarized intensity. Stokes $I$, $Q$, and $U$ maps consist of a few point source models. Two kinds of source geometries are assumed: (i) the locations of Stokes $Q$ and $U$ models being coincident with those of Stokes $I$ models (left, $P \propto I$) and (ii) the Stokes $Q$ and $U$ models being significantly shifted from the Stokes $I$ models (right, $P \not\propto I$). The properties of each "component" such as Stokes $I$ flux ($I$), fractional polarization ($m_L$), linearly polarized flux ($I_{\rm pol}$), and EVPA are indicated. \label{fig:synthetic_example}}
\end{figure*}

\begin{itemize}

\item{GPCAL allows to scale the visibility weights of specific antennas up or down by a constant factor. This is particularly useful for arrays having very different sensitivities among antennas such as the EHT+ALMA \citep{EHT2019a}, the GMVA+ALMA \citep{Issaoun2019}, and the HSA including the phased-up VLA \citep{Hada2017}. In this case, fitting can be dominated by the most sensitive stations, and the possible residual systematic errors in those stations can distort the fitting solutions for other stations. One can mitigate this effect by down weighting those stations visibilities for fitting \citep[see e.g.,][]{EHT2019d} and this option is implemented in GPCAL.}

\item{In some cases, the D-terms of some stations can be constrained externally. Users can use that prior knowledge and fix the D-terms of those stations with the known values for fitting.}

\item{Some VLBI arrays have very short baselines, e.g., the phased-up ALMA and the Atacama Pathfinder Experiment (APEX) telescope in Chile in the EHT, which provides a baseline length of 2.6 km \citep{EHT2019c}. Calibrators with compact source geometries will be seen as point-like sources on these short baselines. Therefore, one can estimate the D-terms of the stations comprising the short baselines by assuming point-like sources for the source-polarization terms in the D-term model, i.e., constant $\mathscr{P}(u,v)$ in Equation~\ref{model:model}. Since the model is much simpler and has a much smaller number of free-parameters compared to using the whole arrays at one time, the D-terms of those stations can be robustly constrained. GPCAL allows first obtaining the D-term solutions of those stations using the short baselines and then fixing them in the fitting for the rest of the arrays using the whole baselines. One can use multiple baselines and multiple sources simultaneously for the fitting with short baselines to obtain more accurate D-term solutions.}

\end{itemize}

\section{Verification and Results}
\label{sec:results}

In this section, we evaluate the performance of the pipeline using simulated and real data sets. 

\begin{figure*}[t!]
\centering
\includegraphics[width = 0.495\textwidth]{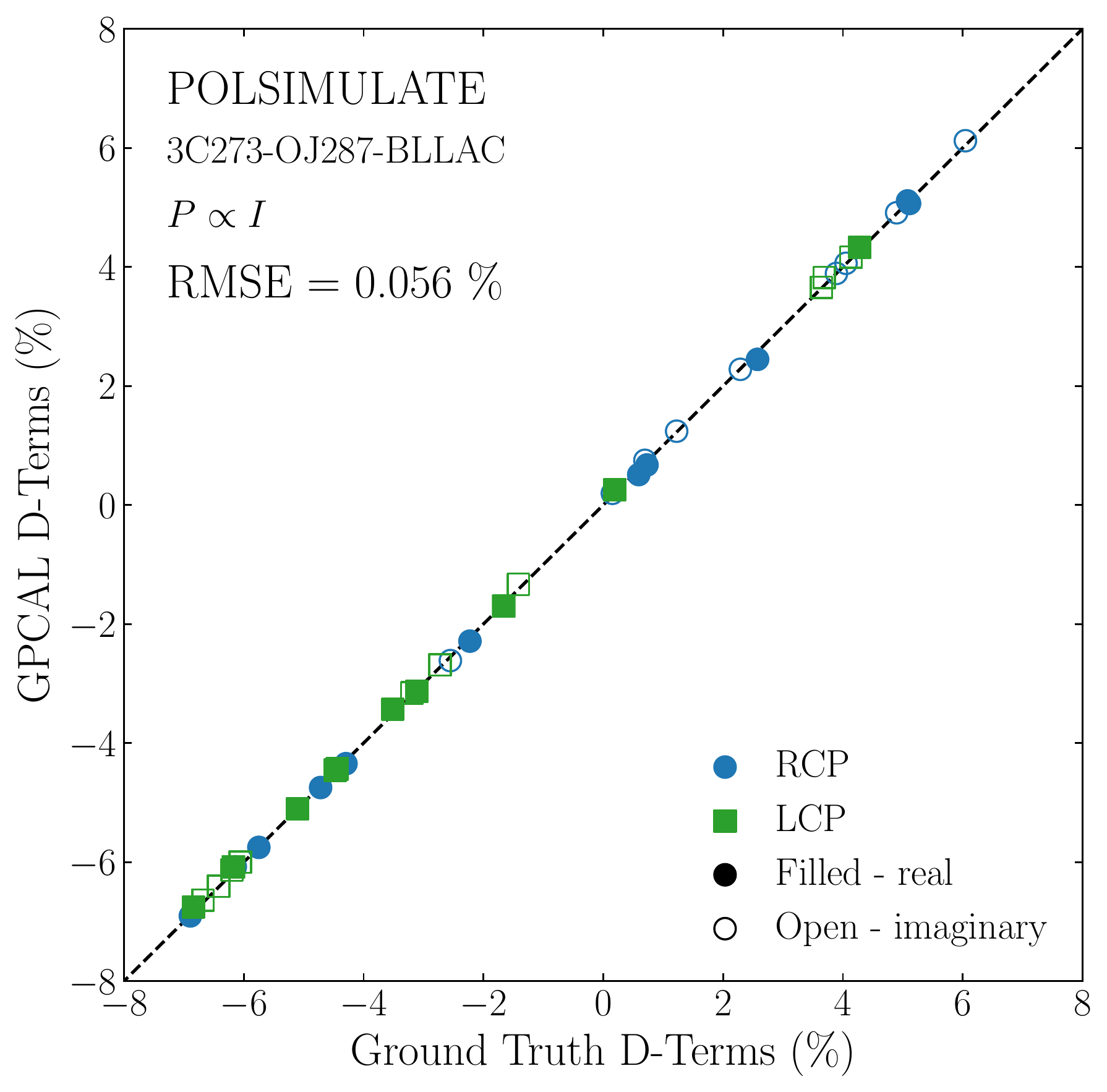}
\includegraphics[width = 0.485\textwidth]{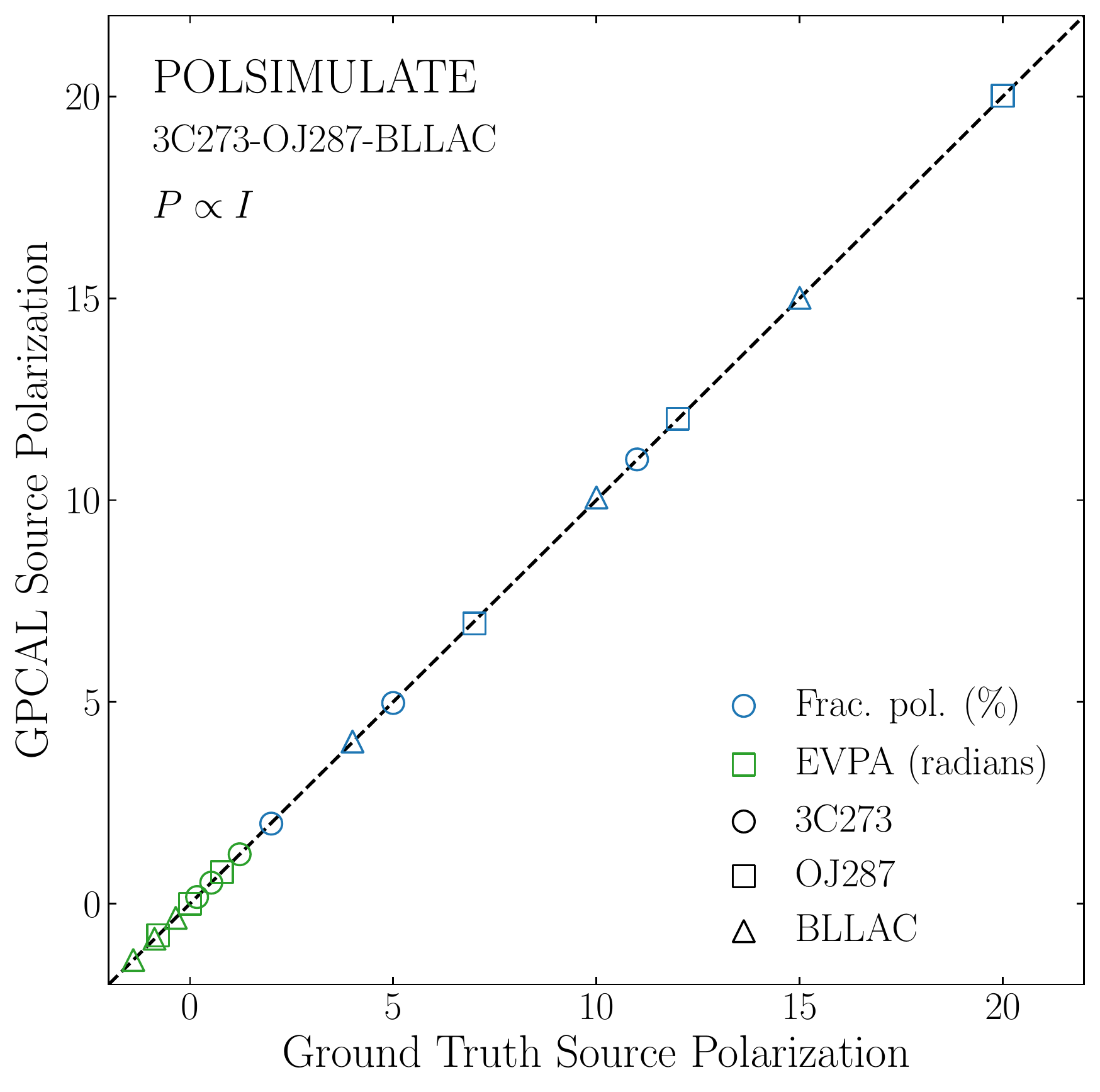}
\caption{Comparison between the ground-truth values assumed in the simulated data and the values reconstructed by GPCAL for D-terms (left) and source polarization properties (right). The results for the $P \propto I$ case are shown. GPCAL used the simulated data sets for the uv-coverages of three sources, 3C 273, OJ 287, and BL Lac, simultaneously to derive the best-fit D-terms and source-polarization terms. The blue and green data points in the left figure denote the RCP and LCP D-terms, respectively. The filled and open data points are for the real and imaginary parts of the D-terms, respectively. The root-mean-square error of the reconstructed D-terms is $\approx0.056\%$. In the right figure, the blue data points are the source fractional polarizations in units of \% and the green data points are the source EVPAs in units of radians. The circles, squares, and triangles are the source polarization properties of the simulated data for the uv-coverages of 3C 273, OJ 287, and BL Lac, respectively. \label{fig:synthetic_simil}}
\end{figure*}

\begin{figure*}[t!]
\centering
\includegraphics[width = 0.32\textwidth]{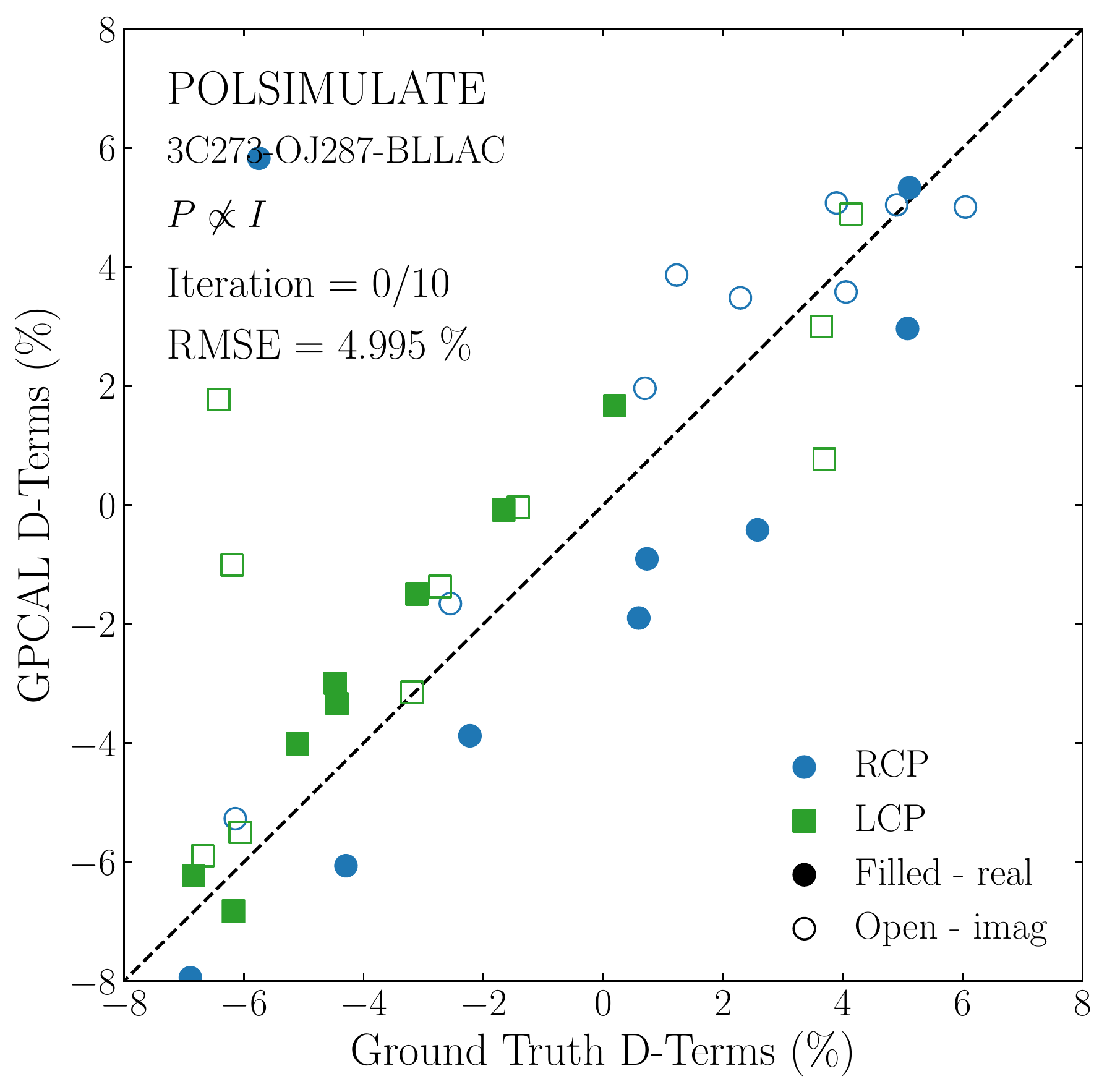}
\includegraphics[width = 0.32\textwidth]{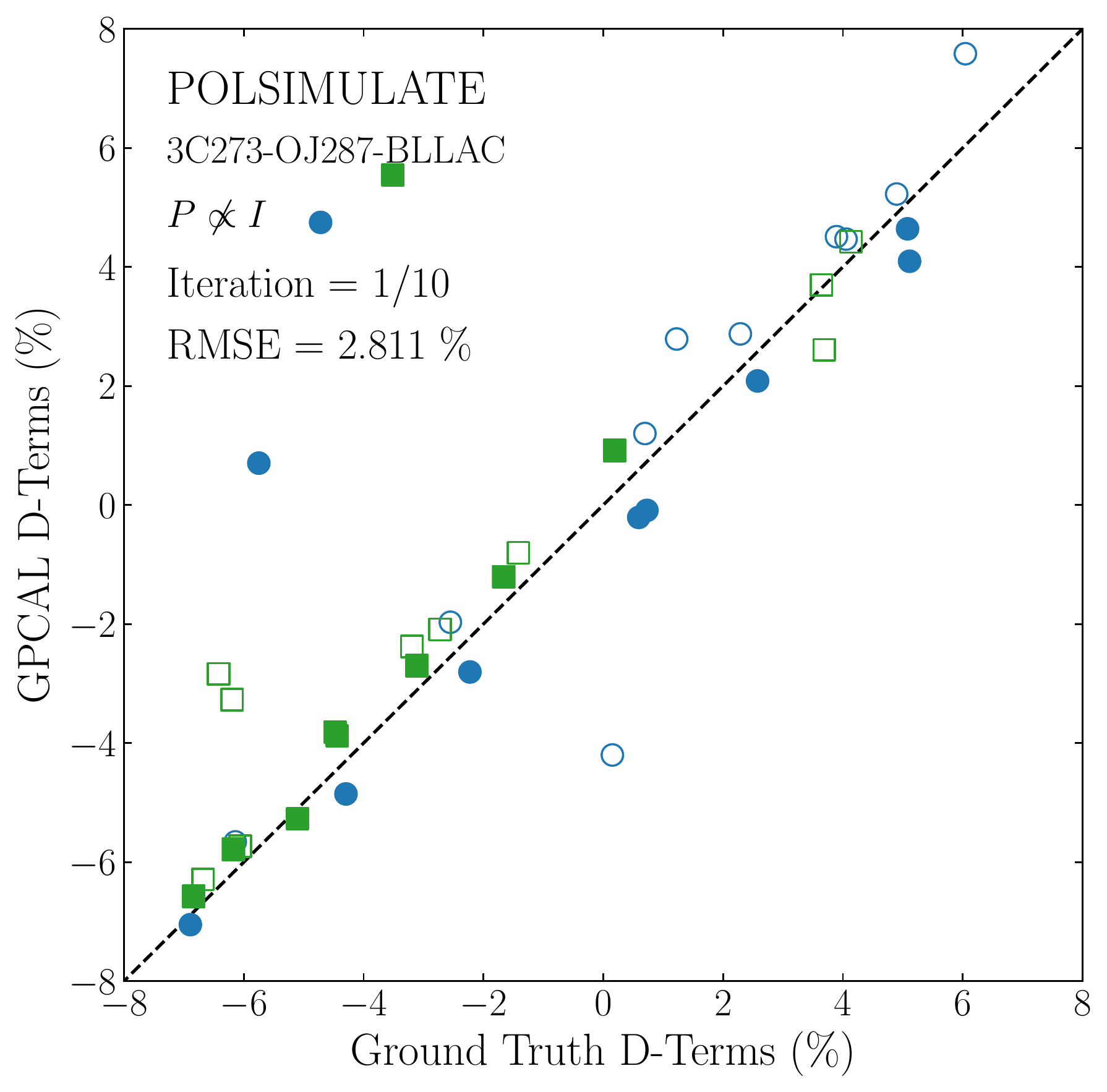}
\includegraphics[width = 0.32\textwidth]{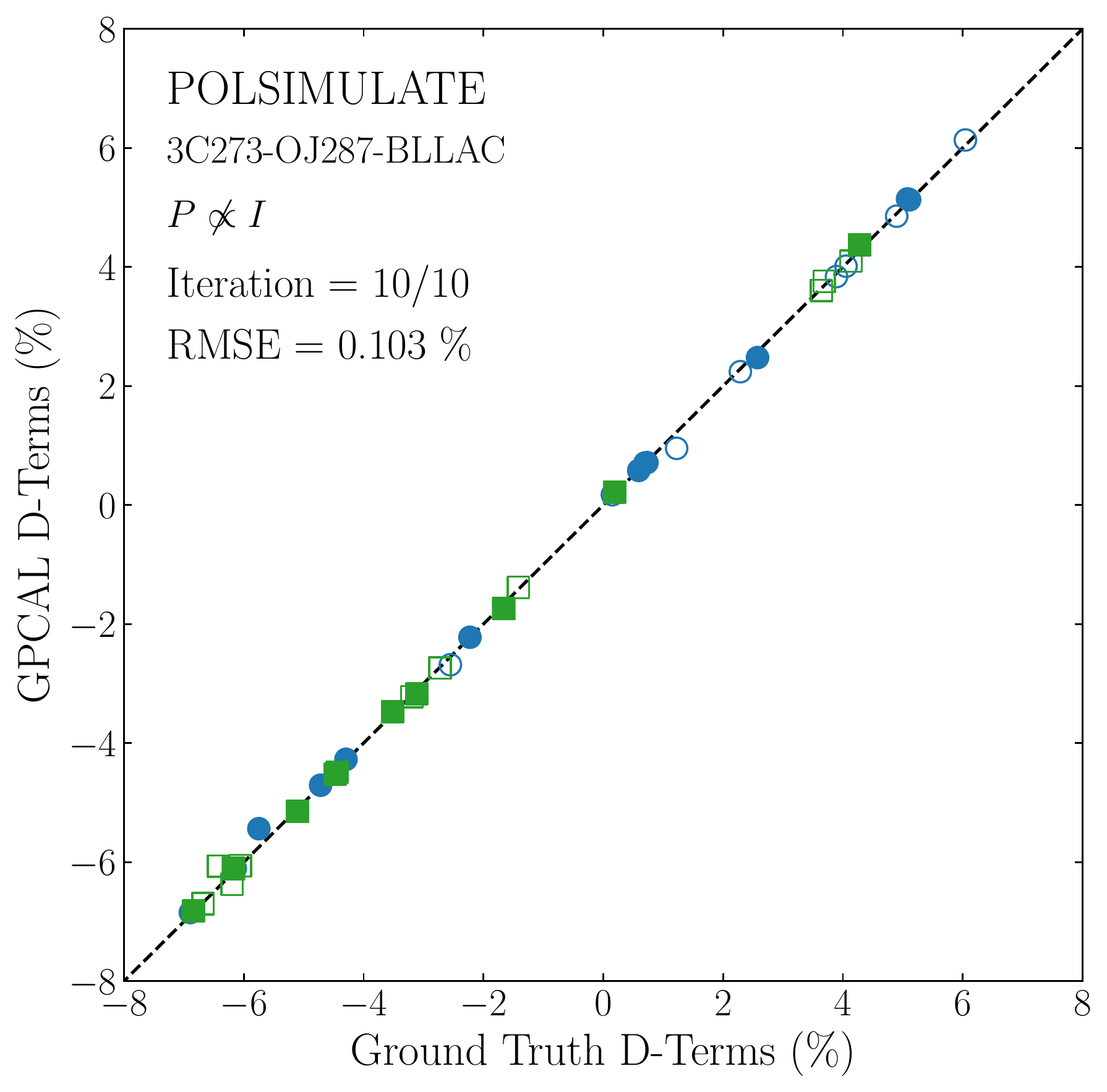}
\caption{Similar to Figure~\ref{fig:synthetic_simil} but for the $P \not\propto I$ case. GPCAL first estimates the D-terms using the similarity assumption (Iteration zero, left) and then improves the estimation with ten iterations of instrumental polarization self-calibration; the D-terms after the first (middle) and tenth iterations are shown (right). The initial D-terms are far from the ground-truth ones because the similarity assumption does not hold in this case. However, the reconstruction is progressively improved with more iterations of instrumental polarization self-calibration, reaching an RMSE of $\approx0.1\%$ after ten iterations. \label{fig:synthetic_pol}}
\end{figure*}

\subsection{Simulated data}
\label{sec:simul}
We used \texttt{PolSimulate}\footnote{\url{https://launchpad.net/casa-poltools}} in the Common Astronomy Software Application \citep[CASA,][]{CASA2007} package to produce simulated data. We assumed (i) an array configuration of the VLBA, (ii) an observing frequency of 15.256 GHz with a bandwidth of 64 MHz, (iii) on-source time of an hour spread over ten scans over 10 hours, (iv) a minimum elevation cutoff of 10 degrees, (v) an atmospheric opacity at the zenith of 0.05, (vi) and the sky, ground, and receiver temperatures of 70, 90, and 50 Kelvins, respectively. The D-terms from a few to about 7\% were assumed, which are the typical amplitudes of the D-terms seen in the VLBA data \citep[e.g.,][]{Attridge2005, Hada2016, Jorstad2017, Lister2018, Park2019}. No antenna gain error was introduced for simulation. We simulated three data sets for the source coordinates of OJ 287, 3C 273, and BL Lac, which have been observed by many VLBI arrays and cover a wide range of right ascensions. The source geometries are assumed to consist of a few point sources for Stokes $I$, $Q$, and $U$. We considered two different cases for the source structures: (i) the locations of Stokes $Q$ and $U$ models being coincident with those of Stokes $I$ models ($P \propto I$) and (ii) the Stokes $Q$ and $U$ models being shifted from the Stokes $I$ models by $\approx60-120\%$ of the full widths at half maximum of the synthesized beams depending on sources ($P \not\propto I$). The former is an ideal case for using the similarity assumption, while the assumption does not hold at all for the latter case. In Figure~\ref{fig:synthetic_example}, we present the example linear polarization maps of the simulated data for a 3C 273-like uv-coverage.

We performed imaging of the simulated data sets with CLEAN in Difmap and ran GPCAL using the similarity assumption for the former case ($P \propto I$). We divided the total intensity source models into several sub-models in such a way that each knot-like structure is regarded as a sub-model. We fitted the D-term model to the data of all three sources simultaneously to verify the multi-source fitting capability of GPCAL. Since no antenna gain error was introduced in the simulation, one can expect to reconstruct the assumed ground-truth D-terms nearly perfectly. Some expected sources of minor deviation from the truth values are (i) deconvolution errors in CLEAN, (ii) parallel-hand visibilities distorted by the D-terms, and (iii) thermal noise in the data. The second source comes from the fact that our model (Equation~\ref{model:model}) assumes that the measured parallel-hand visibilities ($r^{RR}$, $r^{LL}$) are the same as the true visibilities ($\mathscr{RR}$, $\mathscr{LL}$). However, the measured visibilities in the simulated data are slightly distorted by the D-terms and not identical to the true visibilities in reality (Equation~\ref{eq:model}).

\begin{deluxetable}{cccc}
\tablecaption{RMSE of D-term reconstruction using the simulated data for the $P \propto I$ case. \label{tab:synthetic_rmse}}
\tablewidth{0pt}
\tablehead{ & \colhead{one source} & \colhead{two sources} & \colhead{three sources}}
\startdata
\hline
RMSE (\%) & 0.088 & 0.069 & 0.056 \\
\enddata
\tablecomments{Three RMSE values obtained by running GPCAL using one source (3C 273, OJ 287, and BL Lac individually) and two sources (three possible combinations of the three sources) are averaged. The RMSE value in the "three sources" column is obtained by using all three sources together for fitting.}
\end{deluxetable}

Figure~\ref{fig:synthetic_simil} shows the results for the $P \propto I$ case (Figure~\ref{fig:synthetic_example}, left). The reconstructed D-terms are consistent with the ground-truth D-terms with a root-mean-square error (RMSE) of $\approx0.06\%$. The reconstructed source-polarization terms (the fractional polarizations and EVPAs) are also in good agreement with the assumed source polarizations in the simulation. We present the RMSEs obtained by using the simulated data for only one source and two sources in Table~\ref{tab:synthetic_rmse}. The RMSE becomes smaller when we use more sources, as expected. These results verify that GPCAL can derive the D-terms from multiple calibrators data simultaneously when the linear polarization structures are similar to their total intensity structures.

For the $P \not\propto I$ case (Figure~\ref{fig:synthetic_example}, right), we repeated the above procedures on top of which we performed instrumental polarization self-calibration with ten iterations. Figure~\ref{fig:synthetic_pol} shows that the reconstructed D-terms using the similarity assumption (iteration zero) significantly deviate from the ground-truth values with an RMSE of $\approx5\%$. It is because the assumed source geometries are far from the similarity assumption. However, the reconstruction is progressively improved as we iterate instrumental polarization self-calibration and becomes nearly converged to the ground-truth values with an RMSE of $\approx0.1\%$ after ten iterations. This result demonstrates that GPCAL can reconstruct the D-terms even from calibrators having complex linear polarization structures, which has been challenging for existing packages like LPCAL.

\begin{figure*}[t!]
\centering
\includegraphics[width=0.325\linewidth]{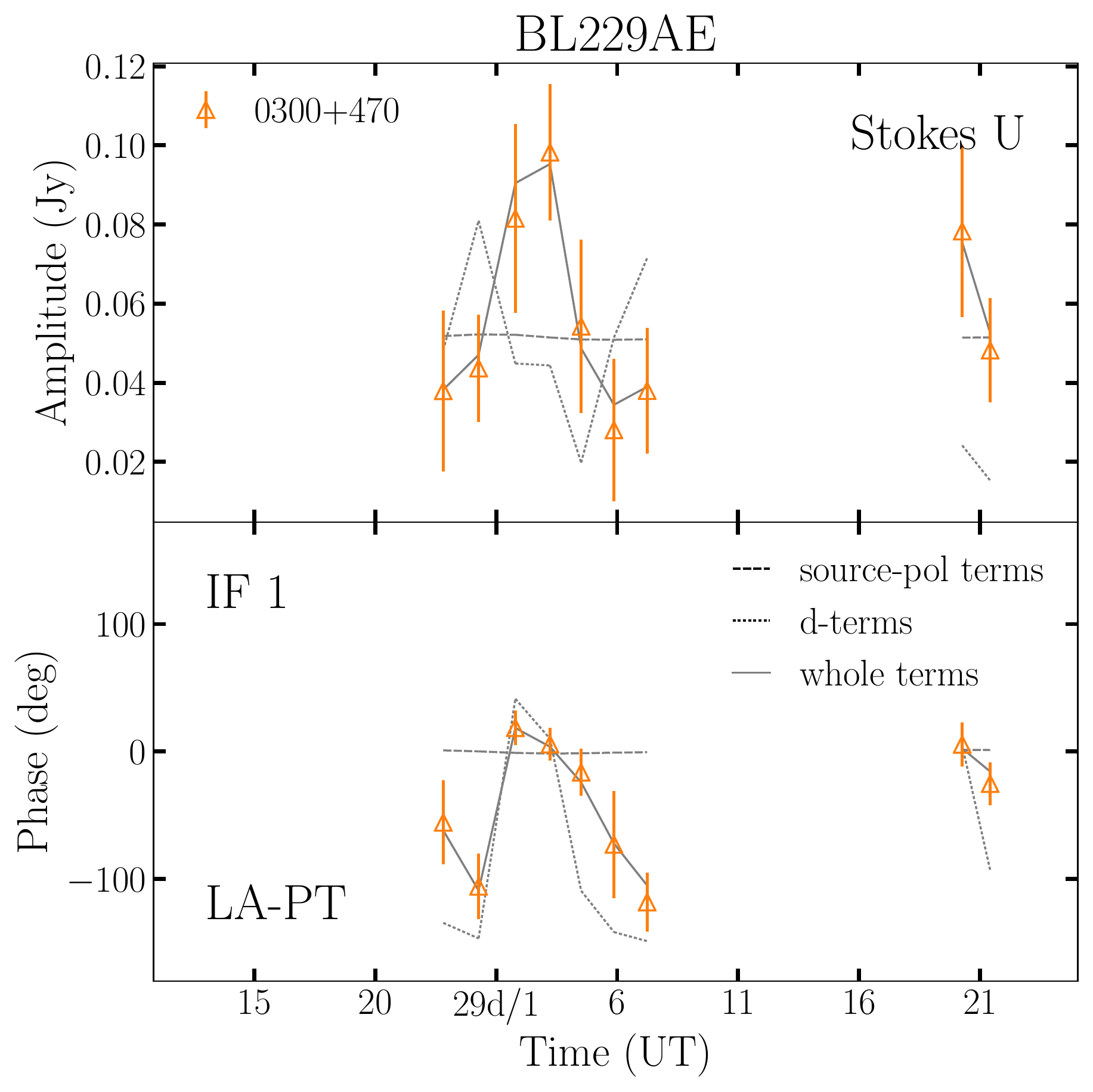}
\includegraphics[width=0.325\linewidth]{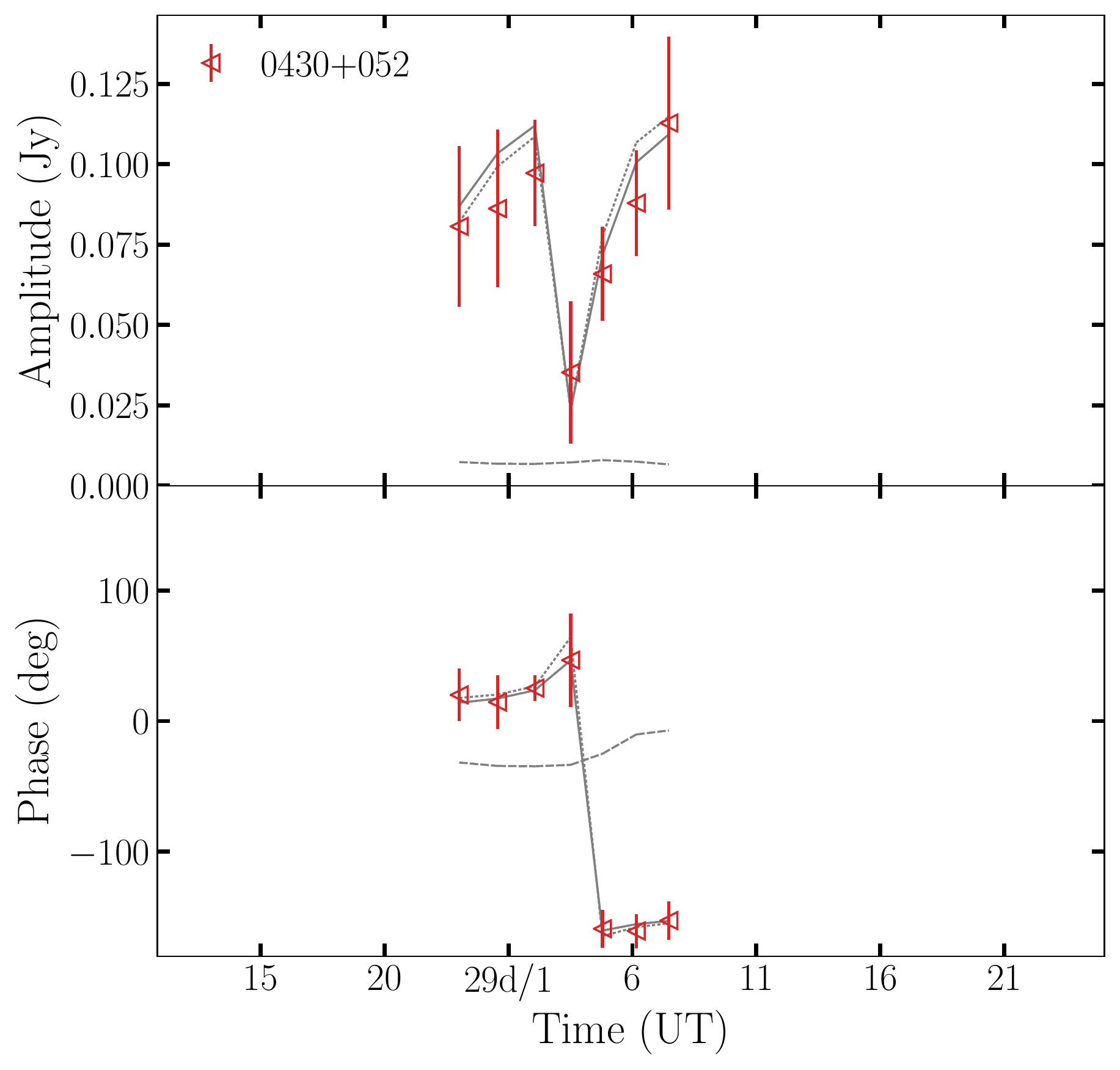}
\includegraphics[width=0.325\linewidth]{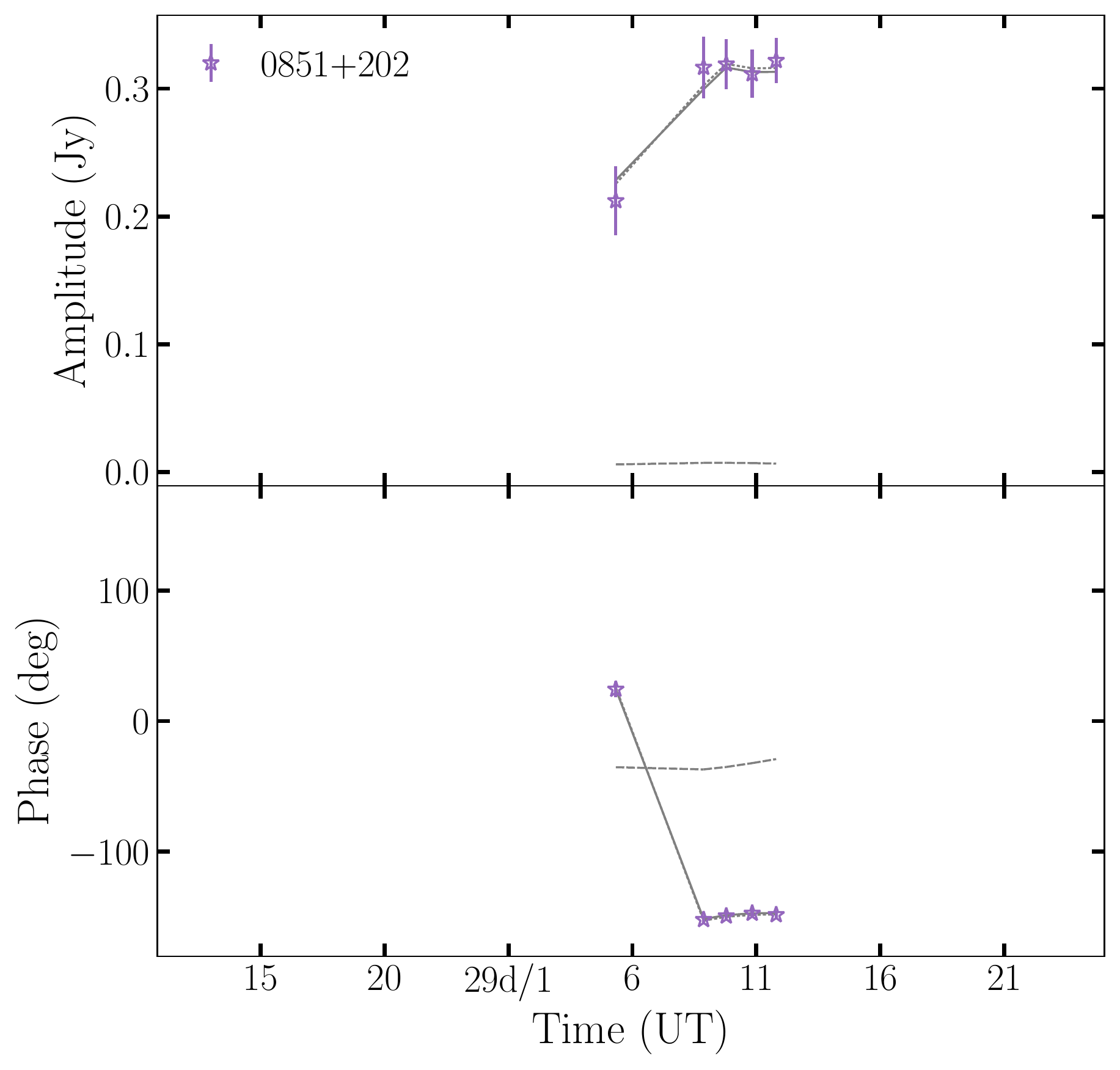}
\includegraphics[width=0.325\linewidth]{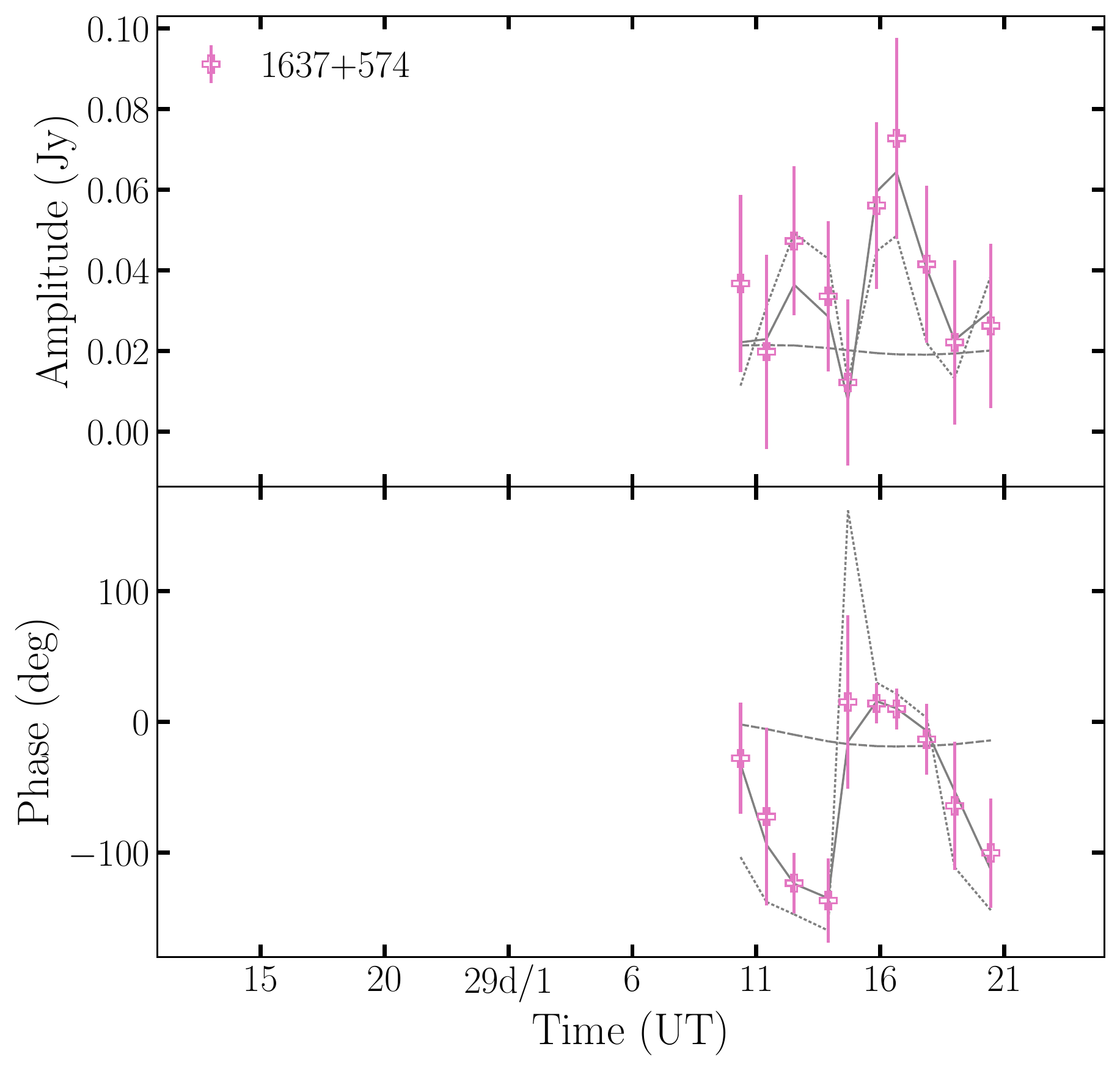}
\includegraphics[width=0.325\linewidth]{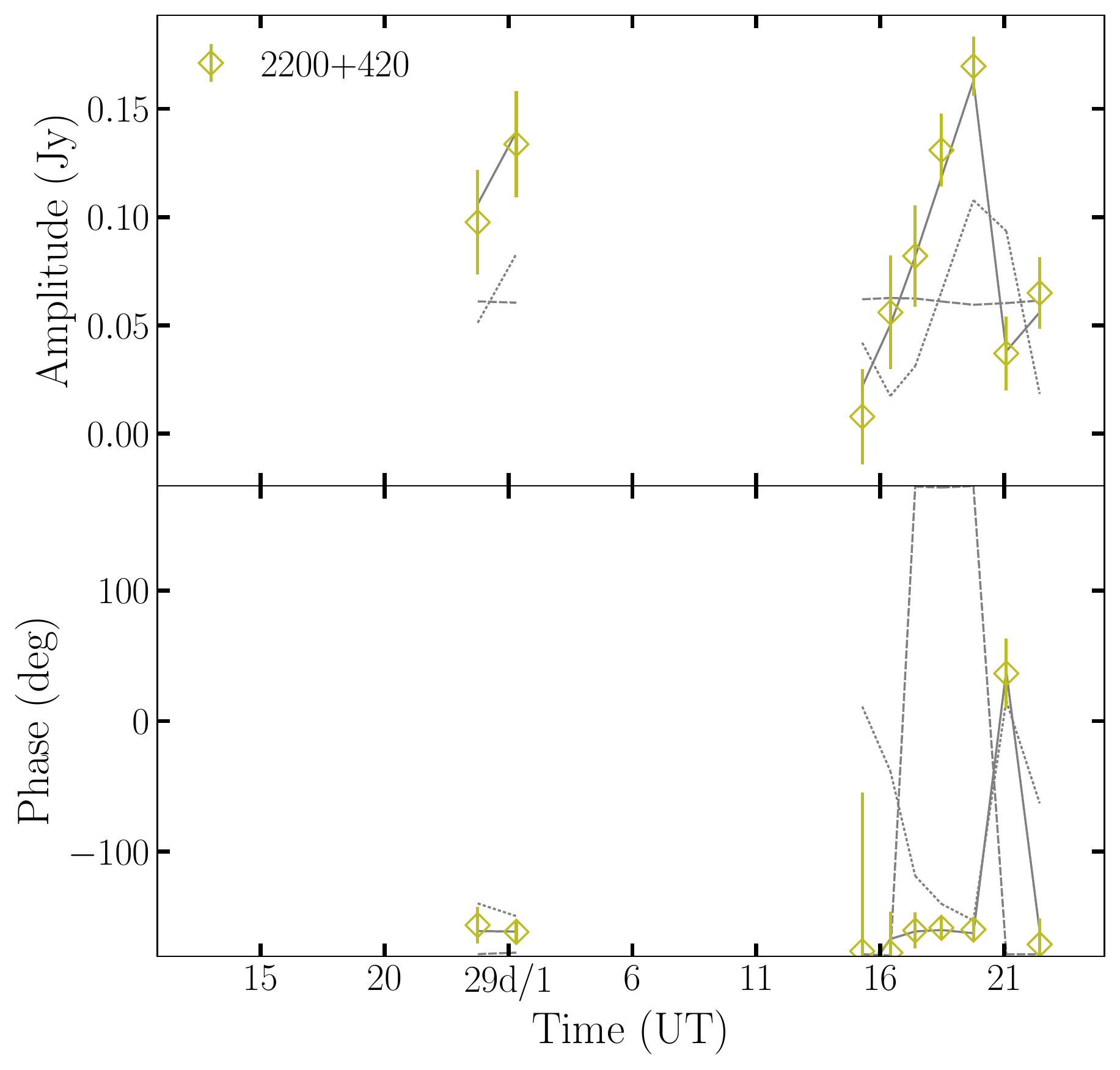}
\includegraphics[width=0.325\linewidth]{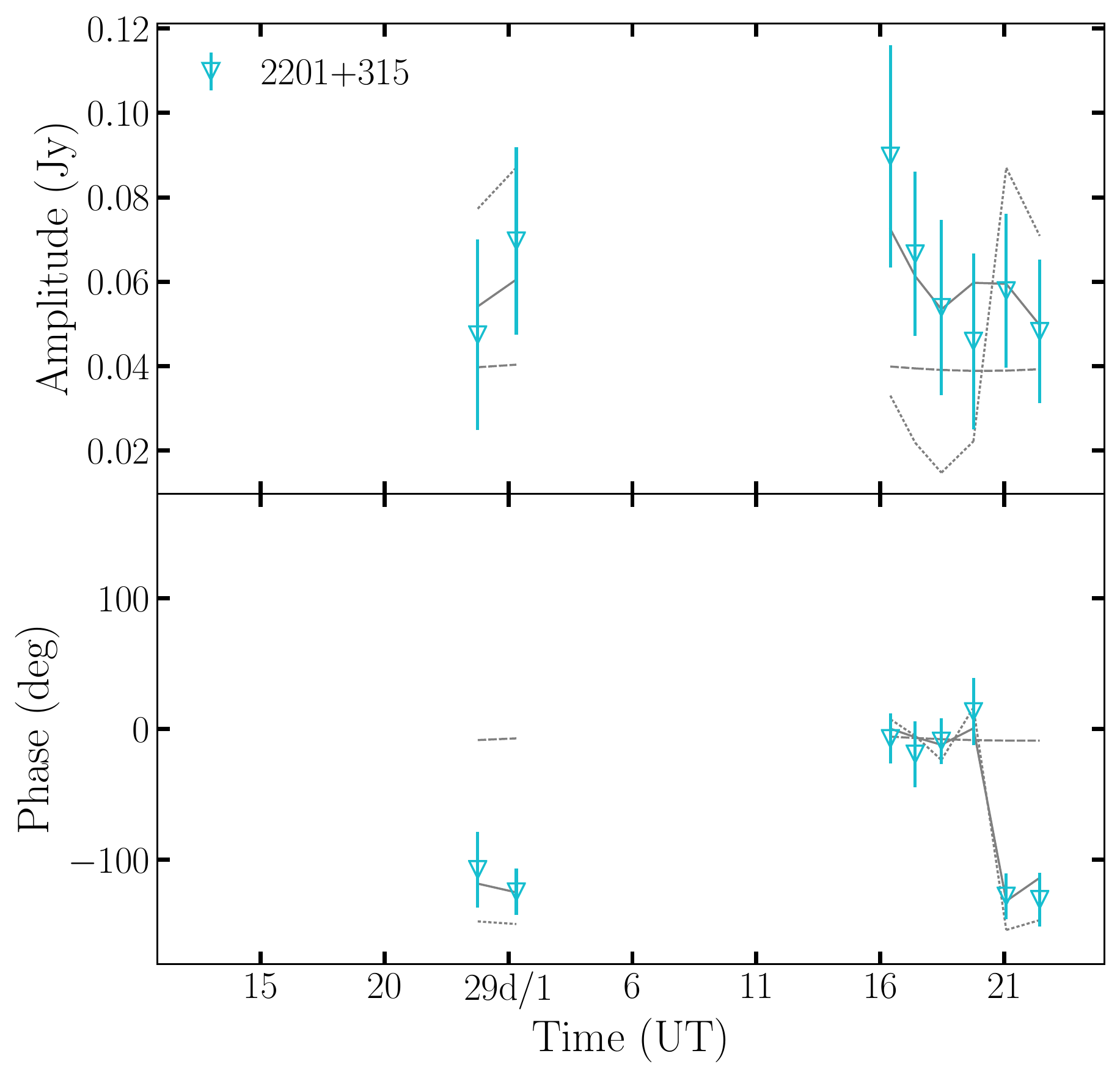}
\caption{GPCAL fitting results on the MOJAVE data observed on 2017 Jan 28--29 for an example baseline of LA--PT for Stokes $U$ and IF 1 data. The Stokes $U$ data's amplitudes and phases for six out of 11 calibrators are shown in different panels. The data are averaged over each scan for better visibility, and the error bars represent the scatter of the data within each scan. The best-fit model (grey solid lines), which consists of the source-polarization terms (grey dashed lines) and the instrumental polarization terms (grey dotted lines), is shown. The source-polarization terms are obtained from CLEAN of Stokes $Q$ and $U$ data for each source during instrumental polarization self-calibration, while the instrumental polarization terms are derived by fitting the model which assumes the same D-terms for different sources (see text for more details). Although the results are shown for each source separately, the model was fitted to the data of all sources simultaneously. The source-polarization terms show smooth variations with time, while the instrumental polarization terms vary rapidly with time due to the change of antenna field-rotation angles. \label{fig:mojave_vplot}}
\end{figure*}

\begin{figure}[t!]
\centering
\includegraphics[width = 0.49\textwidth]{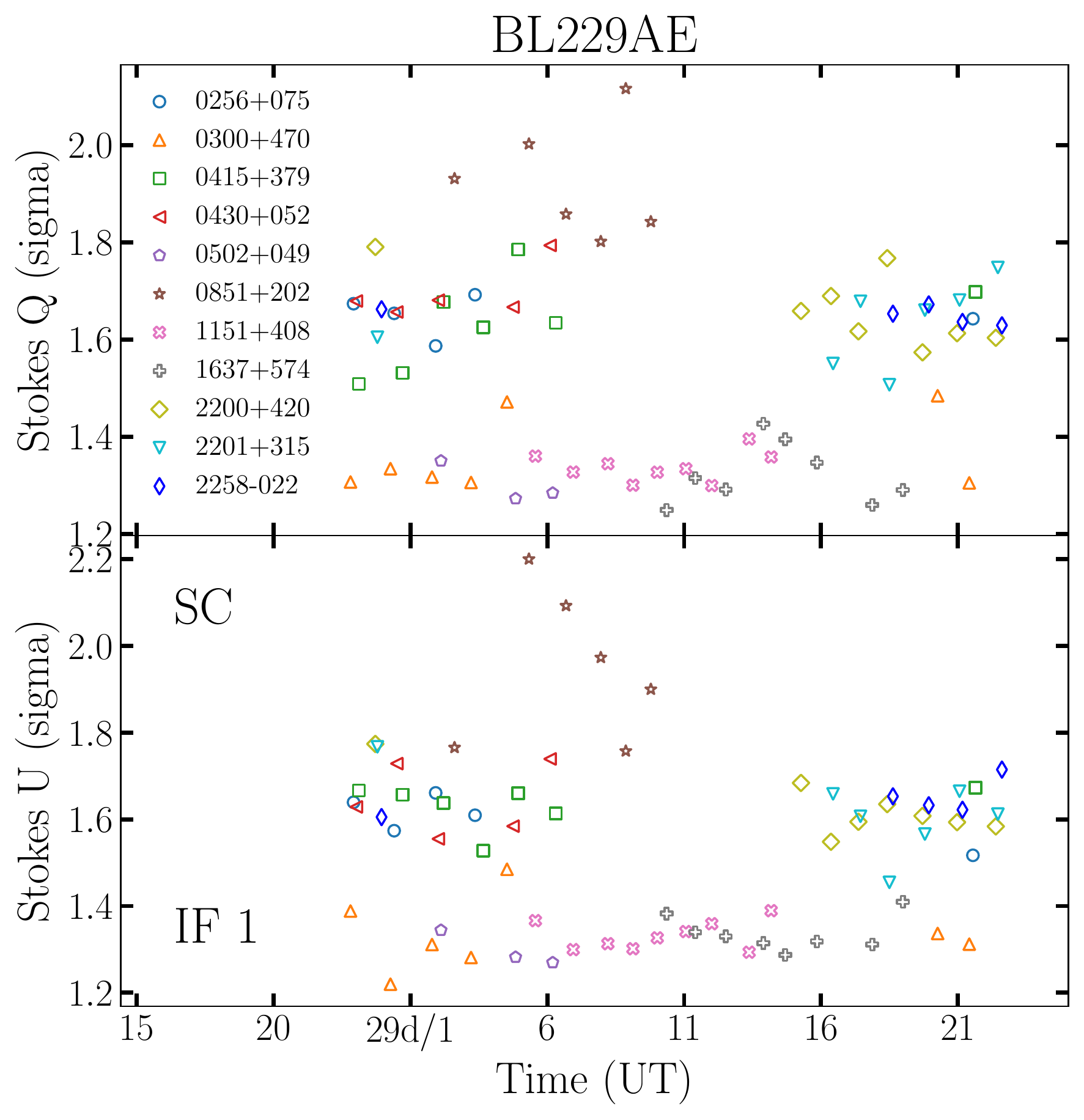}
\caption{GPCAL fitting residuals of the MOJAVE data for an example station. The fitting residuals in units of visibility errors, $|V-\hat{V}|/\sigma$, of the Stokes $Q$ (upper) and $U$ (lower) data are averaged over each scan and over all baselines to SC station. If the data are represented well by the model, then residuals of $\approx1.4\sigma$ are expected, which corresponds to the reduced chi-square of 1. \label{fig:mojave_resplot}}
\end{figure}

\begin{figure}[t!]
\centering
\includegraphics[width = 0.45\textwidth]{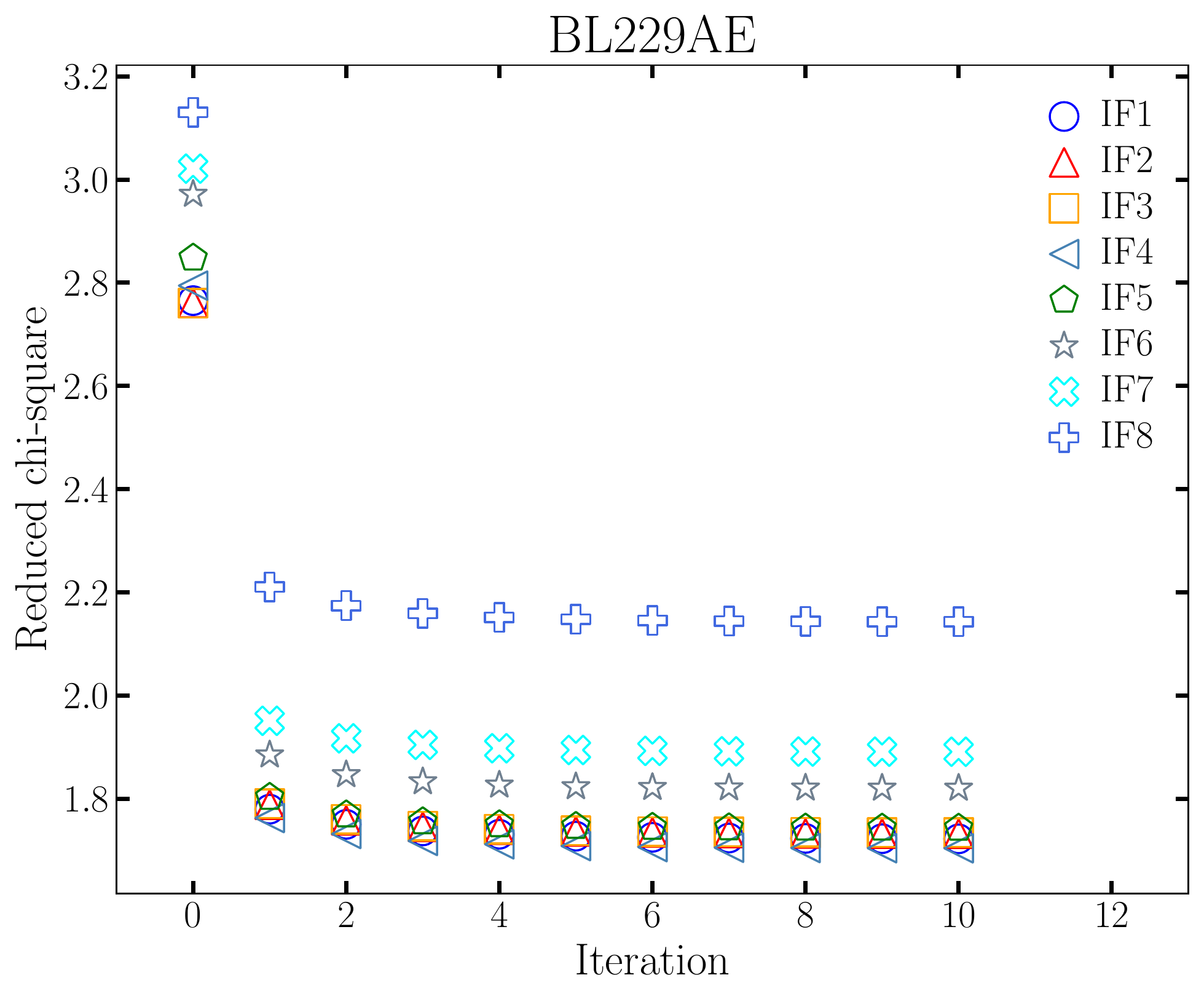}
\caption{Reduced chi-squares of the fitting to the MOJAVE data with GPCAL. Iteration on the x-axis shows different stages of calibration in the pipeline; iteration zero means an initial D-term estimation using the similarity assumption. Positive iterations mean different iterations of instrumental polarization self-calibration. The results for different IFs are shown in different colors and symbols. The goodness of fit is significantly improved between iteration zero and one, which indicates that the similarity assumption does not perfectly hold for the calibrators used for the initial D-term estimation, even though they have simple and core-dominated linear polarization structures. It is gradually improved with more iterations of instrumental polarization self-calibration and becomes saturated after four or five iterations. \label{fig:mojave_chisq}}
\end{figure}

\subsection{Real data}
In this subsection, we apply GPCAL to several real data observed with different VLBI arrays and at different frequencies to evaluate its performance.

\subsubsection{MOJAVE data at 15 GHz}
\label{sec:mojave}

We analyzed one of the Monitoring Of Jets in Active galactic nuclei with VLBA Experiments \citep[MOJAVE,][]{Lister2018} data sets, which have observed many AGN jets with the VLBA at 15 GHz for decades. We selected the observation of 30 sources on 2017 Jan 28 at a recording rate of 2 Gbps (Project code: BL229AE). Since the publicly available data on the MOJAVE database is already fully calibrated, including D-term correction, we analyzed the raw data in the VLBA archive. We performed a standard data post-correlation process with AIPS following \citet{Park2019} and hybrid imaging with CLEAN and self-calibration in Difmap.

\begin{figure*}[t!]
\centering
\includegraphics[width = 0.6\textwidth]{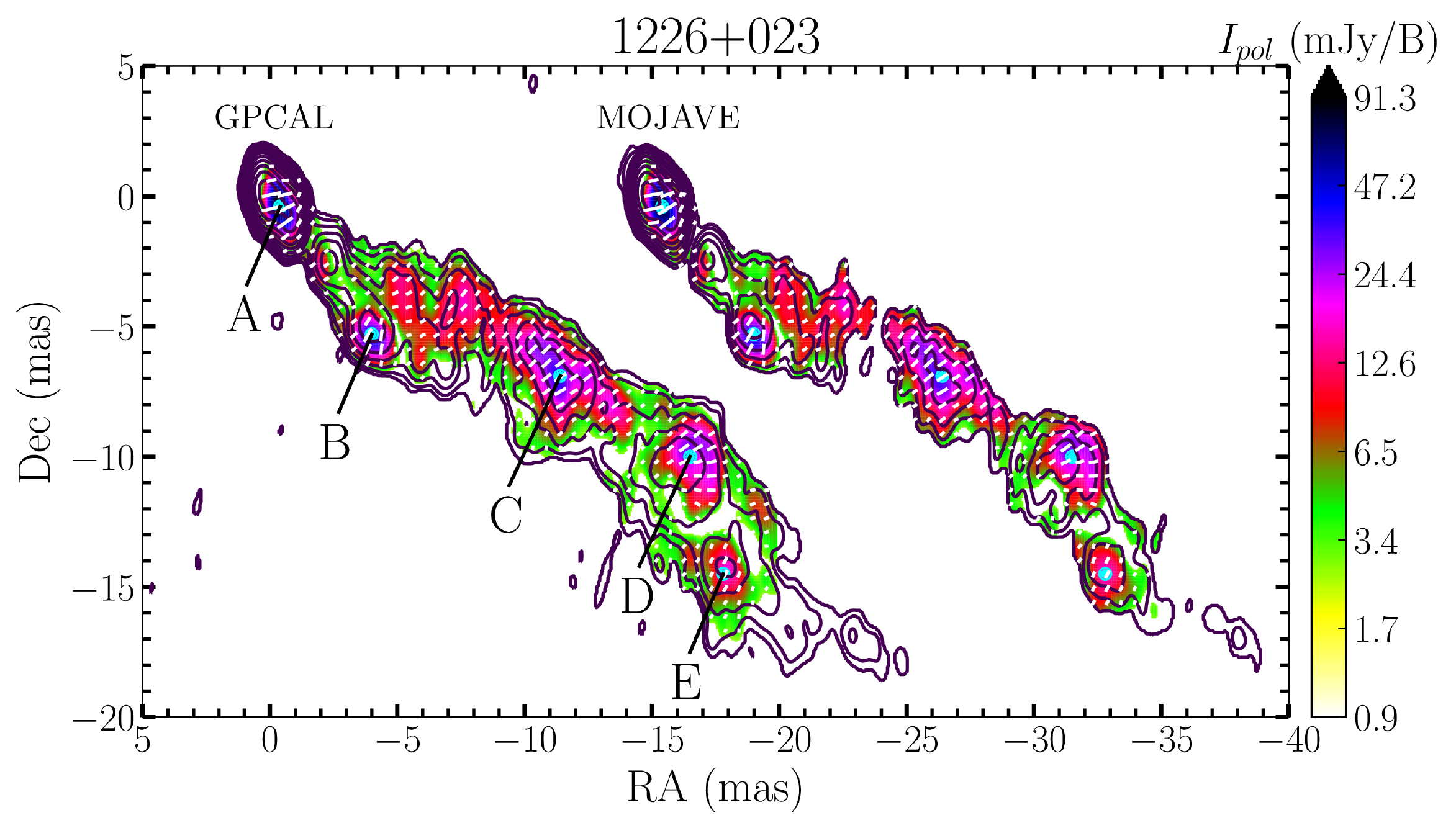}
\includegraphics[width = 0.38\textwidth]{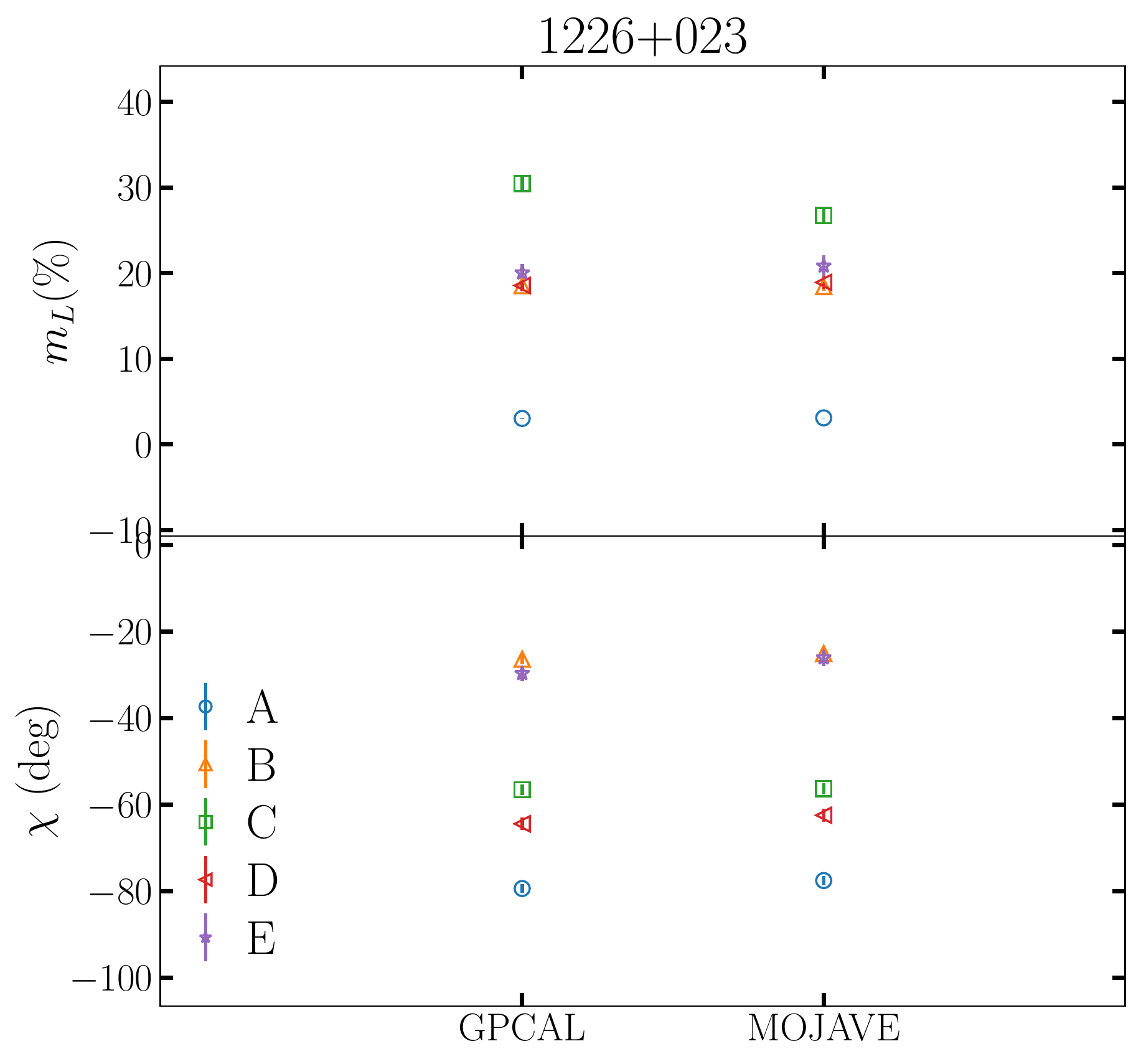}
\includegraphics[width = 0.6\textwidth]{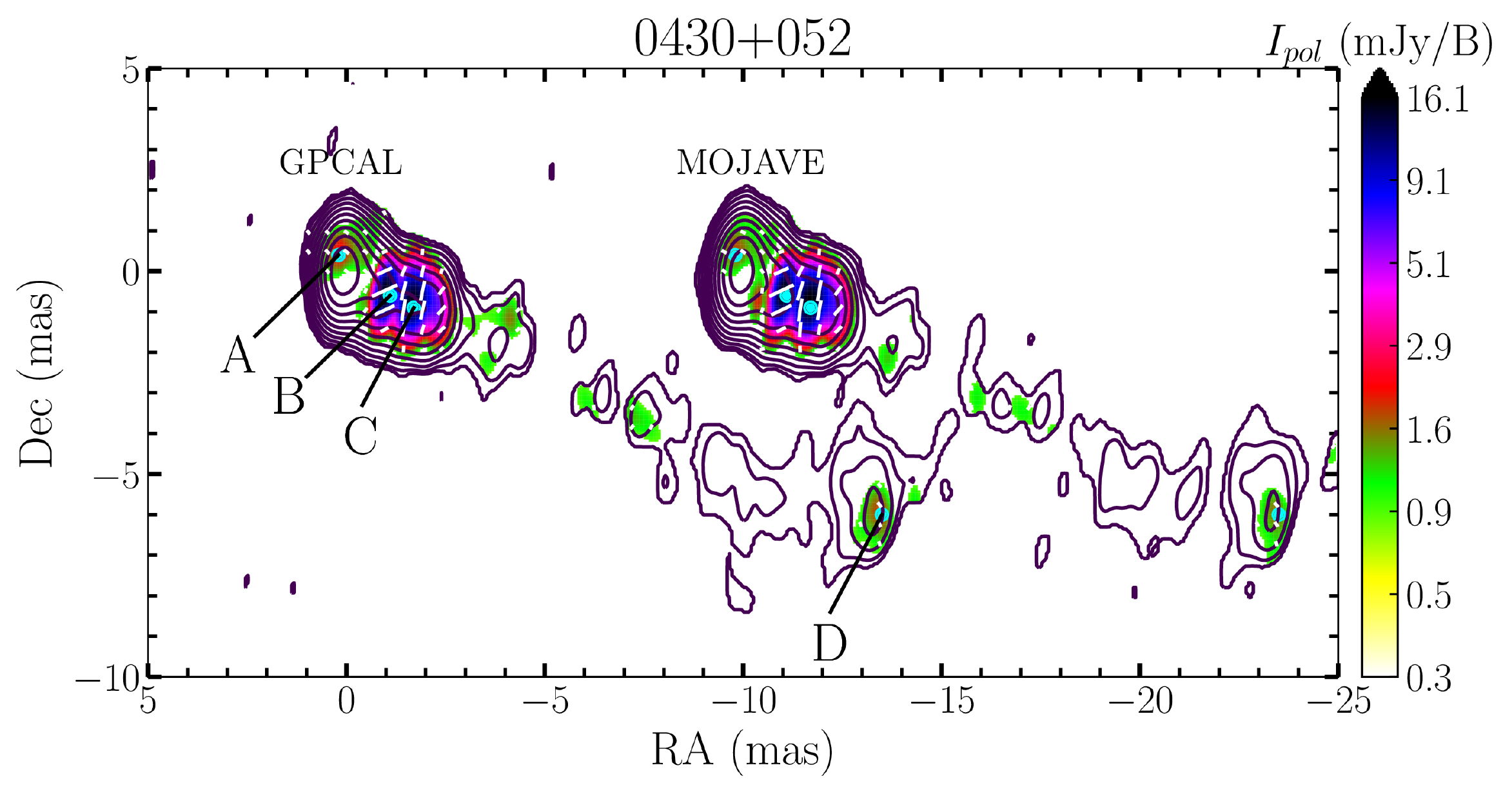}
\includegraphics[width = 0.38\textwidth]{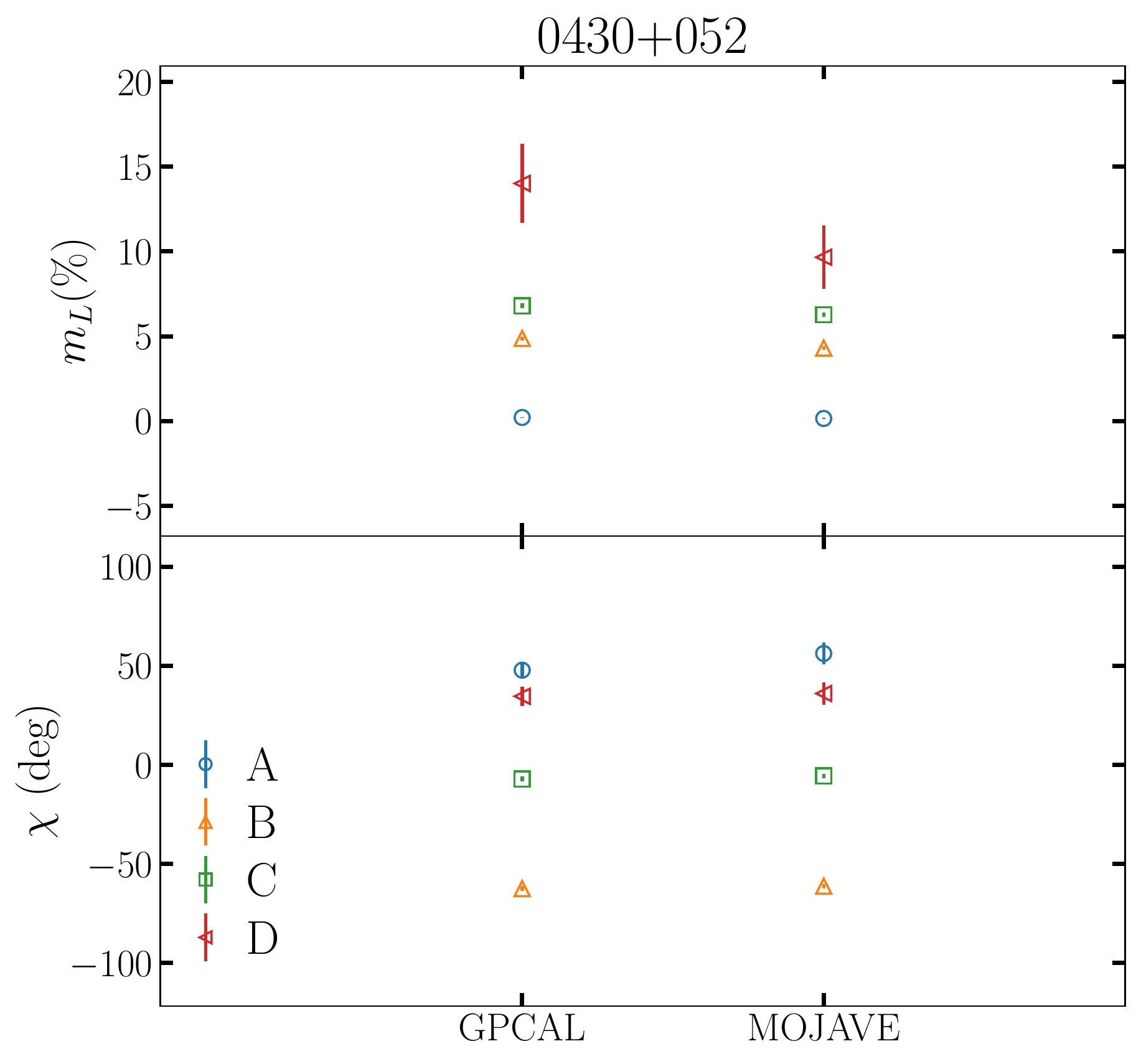}
\caption{Linear polarization maps (left) and polarization properties at several locations marked in the maps (right) of 1226+023 (3C 273, upper) and 0430+052 (3C 120, lower). The results obtained by GPCAL and taken from the MOJAVE database are shown for comparison. Contours start at 1.7 (0.52) and 3.9 (0.69) mJy per beam for the left and right maps of 3C 273 (3C 120), respectively, which are three times the off-source image rms-noise of each map, and increase by factors of two. Both the distributions of linearly polarized intensity and EVPA for both sources are consistent between the two results. The fractional polarizations ($m_L$) and the EVPAs ($\chi$) at several locations of the jets are also in good agreement. \label{fig:mojave_map}}
\end{figure*}

We ran the pipeline on the self-calibrated data. We let GPCAL to perform additional self-calibration with CALIB in AIPS to correct the potentially remaining gain offsets between the polarizations. We selected five calibrators, 0256+075, 0851+202, 2200+420, 2201+315, and 2258-022, which are bright (total flux greater than 0.5 Jy) and have core-dominated linear polarization structures in this particular epoch, for the initial D-term estimation using the similarity assumption. These sources consist of several knot-like structures, which were taken into account for splitting their total intensity CLEAN components into several sub-models. The pipeline performed additional instrumental polarization self-calibration with ten iterations by including six more calibrators, 0300+470, 0415+379, 0430+052, 0502+049, 1151+408, and 1637+574, which are bright but have relatively complex linear polarization structures.

We present an example of the fitting results with the Los Alamos (LA) and Pie Town (PT) baseline for Stokes $U$ data in Figure~\ref{fig:mojave_vplot}. The model derived by GPCAL, obtained by using 11 sources simultaneously, fits the visibilities for all different calibrators quite well. This is the case for many calibrators having complex polarization structures as well, thanks to the instrumental polarization self-calibration. We also present the contributions from the source-polarization terms, i.e., the first terms in Equation~\ref{model:model}, and from the instrumental polarization terms, i.e., the rest terms in Equation~\ref{model:model}. The former shows smooth variations with time, as expected from the smooth changes of $(u,v)$ over time, while the latter varies rapidly with time due to the changes of antenna parallactic angles. In Figure~\ref{fig:mojave_resplot}, the fitting residuals in units of visibility errors, i.e., $|V-\hat{V}|/\sigma$, where $V$ is the visibility data, $\hat{V}$ the model visibility, and $\sigma$ the visibility error, are shown for Saint Croix (SC) station\footnote{The LA-PT baseline is the shortest baseline of the VLBA data and SC station comprises the longest baselines, and they are selected as examples.}. The residuals are averaged over each scan and over all baselines to the station. If the data are represented well by the model, one could expect residuals of $\approx1.4\sigma$, which corresponds to the reduced chi-square of 1. Therefore, this plot helps to identify some problematic scans for specific stations. The most likely causes of a bad fit for some scans are imperfect antenna gain correction or variable D-terms during the observations. The former effect would be more easily seen in very bright calibrators because their systematic errors usually dominate the thermal noise (the error bars). In other words, the fitting would look good for faint calibrators even if there are moderate residual antenna gains in the data because of the large error bars. The latter would usually appear on scans at very low or high elevations.

\begin{figure*}[t!]
\centering
\includegraphics[width = 0.55\textwidth]{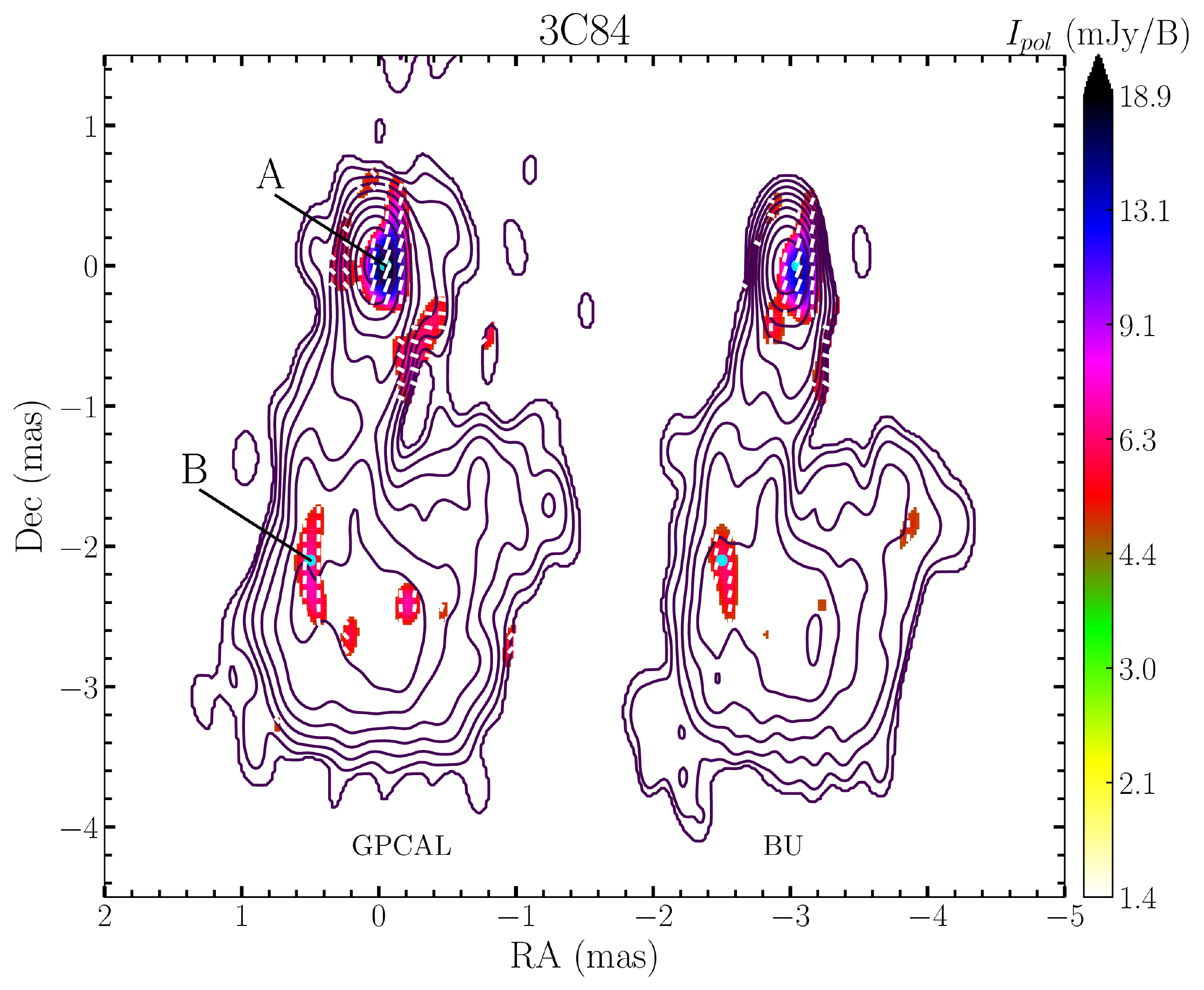}
\includegraphics[width = 0.43\textwidth, trim=0mm -5mm 0mm 0mm]{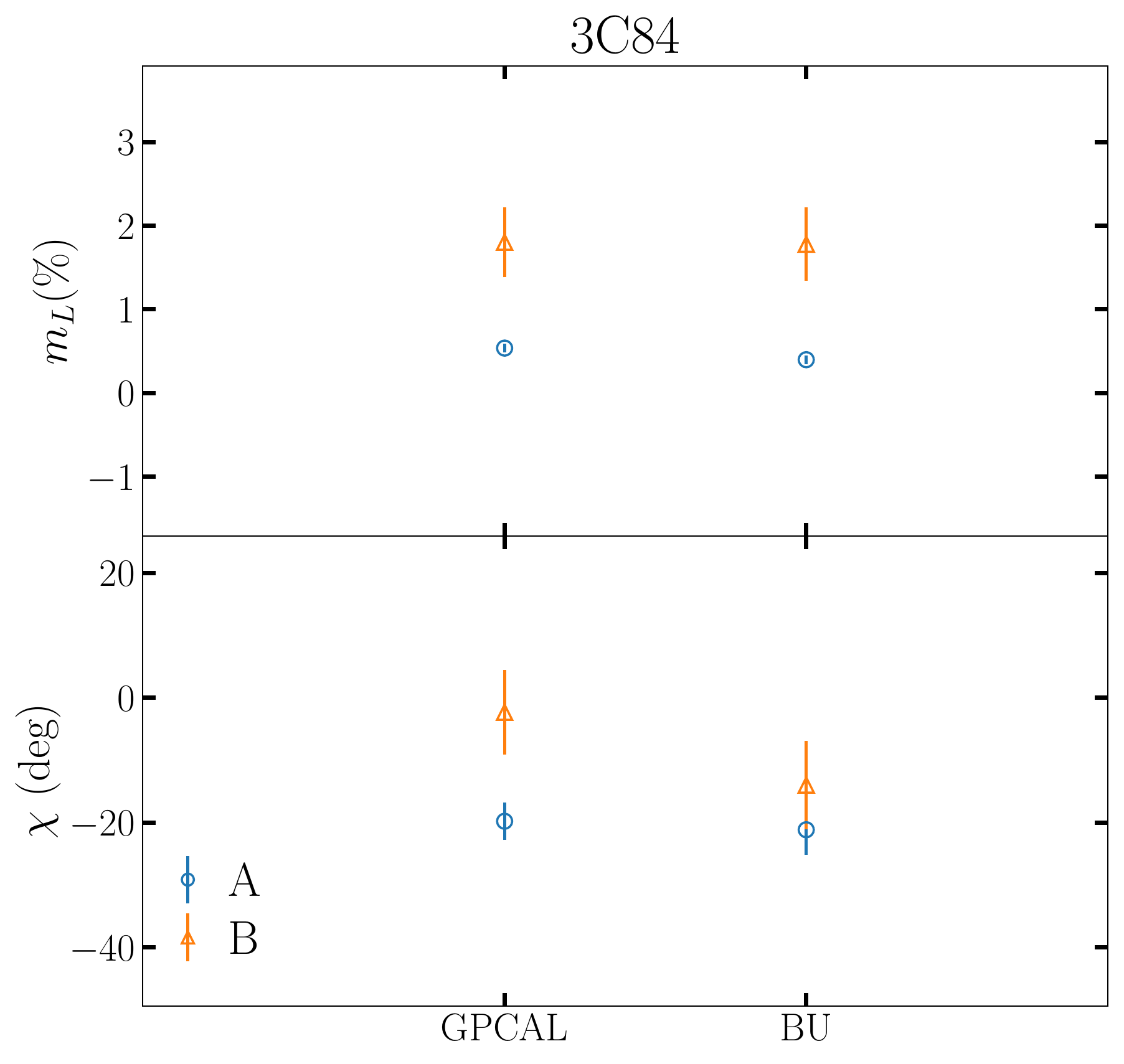}
\includegraphics[width = 0.55\textwidth]{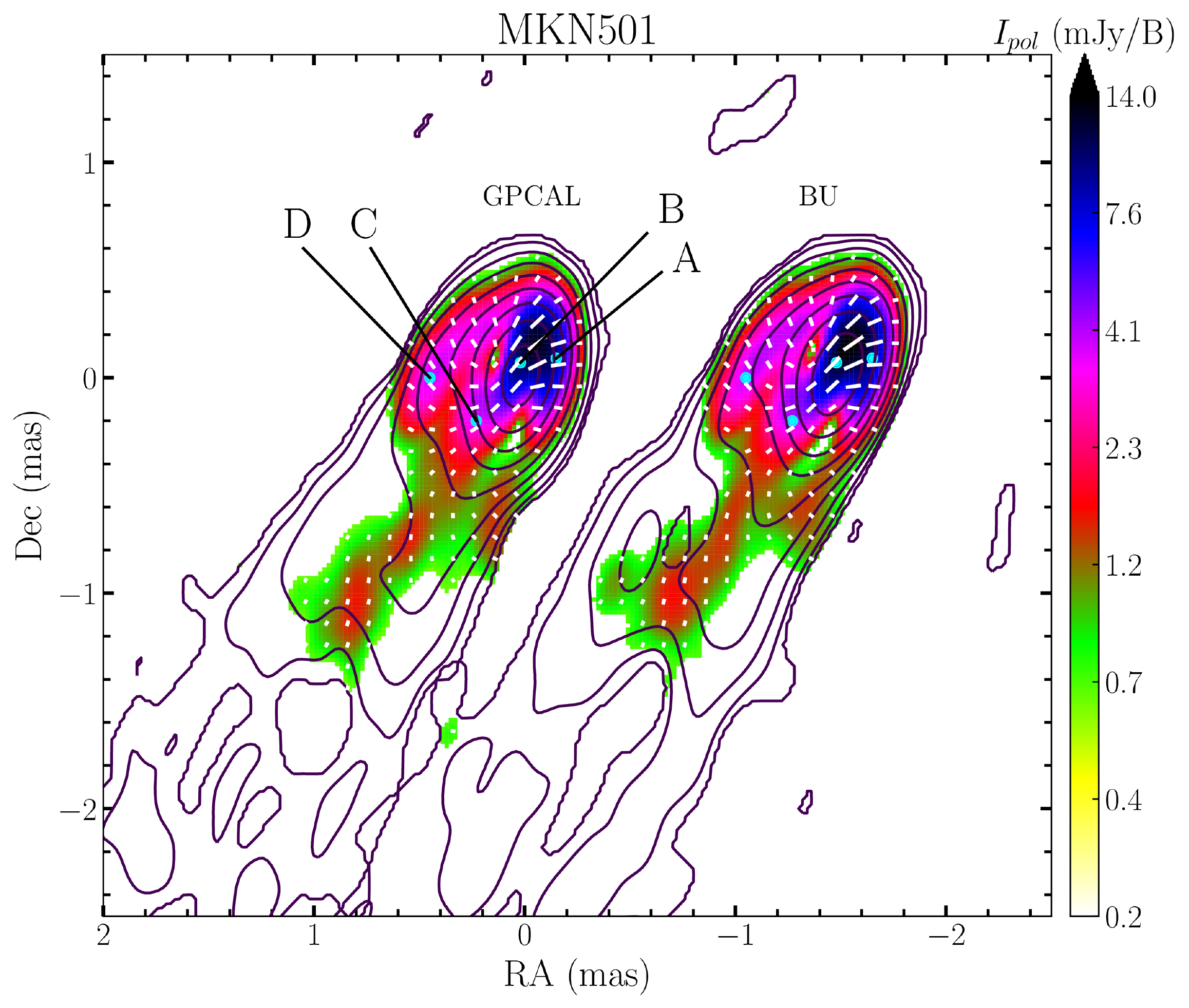}
\includegraphics[width = 0.43\textwidth, trim=0mm -15mm 0mm 0mm]{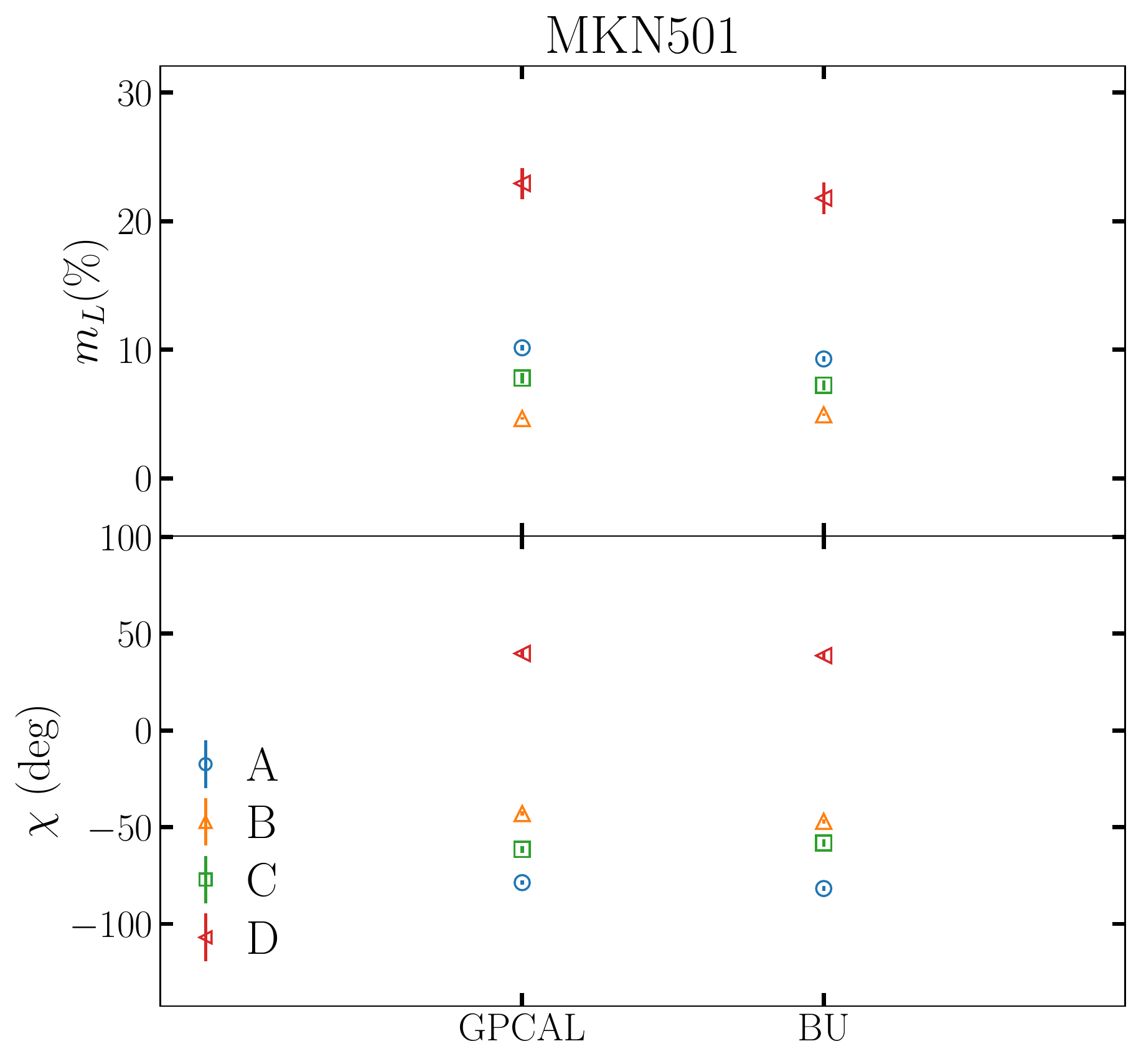}
\caption{Similar to Figure~\ref{fig:mojave_map} but for the BU 43 GHz data of 3C 84 (upper) and MKN 501 (lower). Contours start at 1.6 (0.24) and 2.3 (0.20) mJy per beam for the left and right maps of 3C 84 (MKN 501), respectively, which are four (three) times the off-source image rms-noise of each map, and increase by factors of two. The linear polarization maps are very similar to each other even for the weak polarization signals far from the cores. The fractional polarizations and EVPAs at several locations of the jets are also consistent between the maps. \label{fig:bu_map}}
\end{figure*}

Figure~\ref{fig:mojave_chisq} shows the reduced chi-squares of the fits for the different steps of calibration\footnote{The reduced chi-squares are slightly larger for IF 6--8 than the other IFs for the particular MOJAVE data. We found that the best-fit D-terms of several antennas become larger at higher frequencies; this trend could also be seen in the cross-power spectra for the cross-hand visibilities. If the D-terms are large, systematic errors such as the time-dependent D-terms could also be more severe (The D-terms can easily change by a few percents during the observations if the D-terms are several tens of percent). This can result in the poor fits for those IFs.}. The statistics are significantly improved between using the similarity assumption (iteration zero) and the first iteration of instrumental polarization self-calibration (iteration one). They are gradually improved with more iterations and become saturated after four or five iterations. This result demonstrates that the similarity assumption does not perfectly hold even for the calibrators having core-dominated linear polarization structures for this data.

After the D-term correction, we corrected the remaining RCP and LCP phase offset at the reference antenna for each IF by comparing the integrated EVPA of 0851+202 (OJ 287) with that in the MOJAVE database\footnote{\url{http://www.physics.purdue.edu/astro/MOJAVE/sourcepages/0851+202.shtml}}. We produce linear polarization maps of 1226+023 (3C 273) and 0430+052 (3C 120), which are known to have very complex total intensity and linear polarization structures \citep[e.g.,][]{Gomez2000, Asada2002}, and compare them with the MOJAVE maps in the left panels of Figure~\ref{fig:mojave_map}. We found that the distributions of linearly polarized intensity and EVPA are very consistent between the GPCAL and MOJAVE results for both sources. In the right panels of Figure~\ref{fig:mojave_map}, we also present the fractional polarizations and EVPAs at several locations of the jets indicated in the maps. Both quantities are in good agreement between the maps. The MOJAVE program has obtained D-terms with a high accuracy by combining the LPCAL results of many individual sources showing similar D-terms observed in the same run \citep[e.g.,][]{LH2005, Hovatta2012}. The fact that GPCAL could reproduce nearly identical linear polarization maps to the MOJAVE results demonstrates its capability of achieving a high degree of accuracy in D-term estimation. Normal VLBI programs usually do not observe such a number of sources. GPCAL will be especially useful in those cases where many calibrators suitable for LPCAL are not available.

It may not be straightforward for users to decide that they should request instrumental polarization self-calibration. In that case, one can check the reduced chi-square plot as shown in Figure~\ref{fig:mojave_chisq} and makes a decision based on whether the statistics improve with instrumental polarization self-calibration or not. Also, if there are calibrators having complex linear polarization structures, it may be questionable whether including those calibrators for instrumental polarization self-calibration would improve or degrade the D-term solutions. We perform a simple test to address this question using simulated data in Appendix~\ref{appendix:synthetic}. The result suggests that adding more calibrators can improve the D-term solutions even though they have complex polarization structures. However, this result is based on the simulated data having simple source structures and no antenna gain errors in the data. As one can naturally imagine, the D-term solutions may be degraded if calibrators having poor field rotation angle coverages, low SNRs, and antenna gain errors not well corrected are included for calibration. It is recommended for users to try different combinations of calibrators for the initial D-term estimation and instrumental polarization self-calibration, check the reduced chi-square values, the fitting residual plots (Figure~\ref{fig:mojave_resplot}), and the resulting linear polarization maps of the sources, and find the best combination.

\subsubsection{VLBA-BU-BLAZAR data at 43 GHz}
\label{sec:bu}

We evaluate the performance of GPCAL by using another VLBA data observed as a part of the VLBA-BU-BLAZAR (BU) monitoring program at 43 GHz\footnote{\url{https://www.bu.edu/blazars/VLBAproject.html}} \citep{Jorstad2017}. We analyzed the data observed on 2015 Jul 02 (Project code: BM413I) and obtained CLEAN images and self-calibrated data, similar to the MOJAVE data analysis. We ran GPCAL using six bright and compact calibrators, 0235+164, 0420-014, OJ 287, 1156+295, 1510-089, and 1749+096, for both the initial D-term estimation using the similarity assumption and additional instrumental polarization self-calibration with ten iterations. The reduced chi-squares of 1.7 -- 1.9 were obtained for different IFs. The EVPA calibration was done by referring to the integrated EVPA of OJ 287 in the BU database.

In Figure~\ref{fig:bu_map}, we compare the linear polarization maps of two sources, 3C 84 and MKN 501, also known for complex linear polarization structures \citep[e.g.,][]{Marscher2016, Nagai2017}, between the GPCAL and BU results. The distributions of linear polarization intensity and EVPA in the jets are consistent between the maps, even for the very weak polarization far from the cores. The consistency can also be seen in the right panels showing the fractional polarizations and EVPAs at several locations of the jets. This result demonstrates the capability of GPCAL for achieving a high D-term estimation accuracy.

\begin{figure*}[t!]
\centering
\includegraphics[width = 0.65\textwidth, trim=0mm -15mm 0mm 0mm]{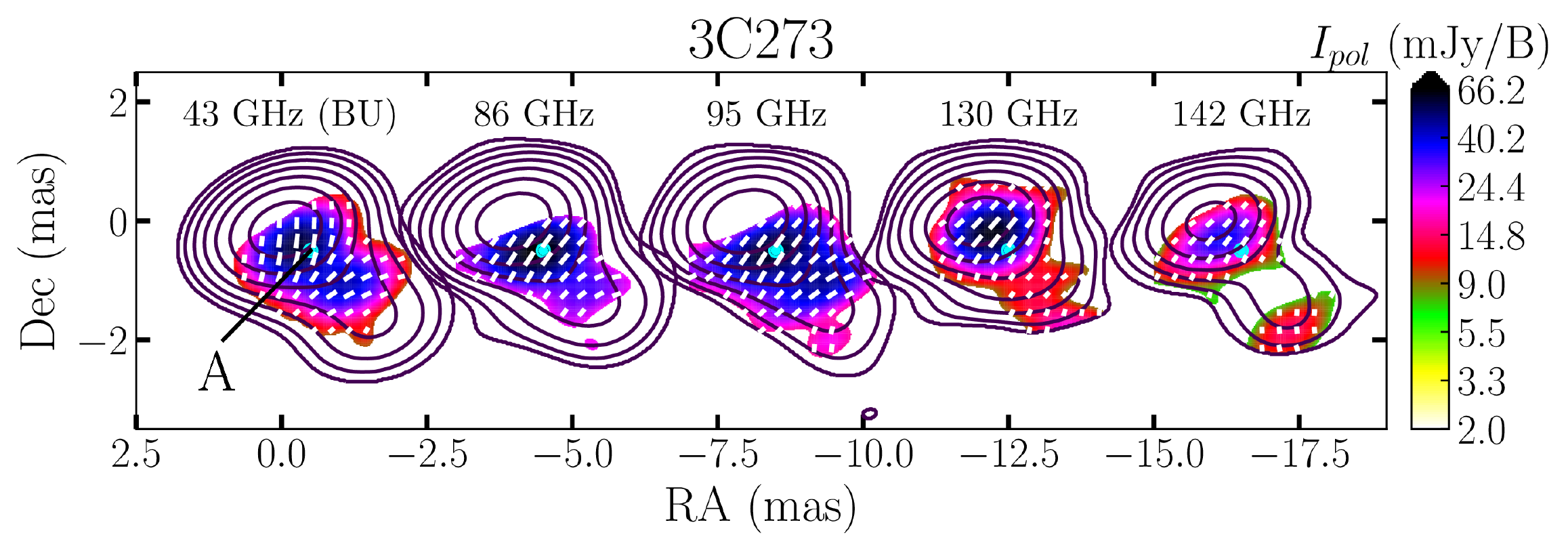}
\includegraphics[width = 0.33\textwidth]{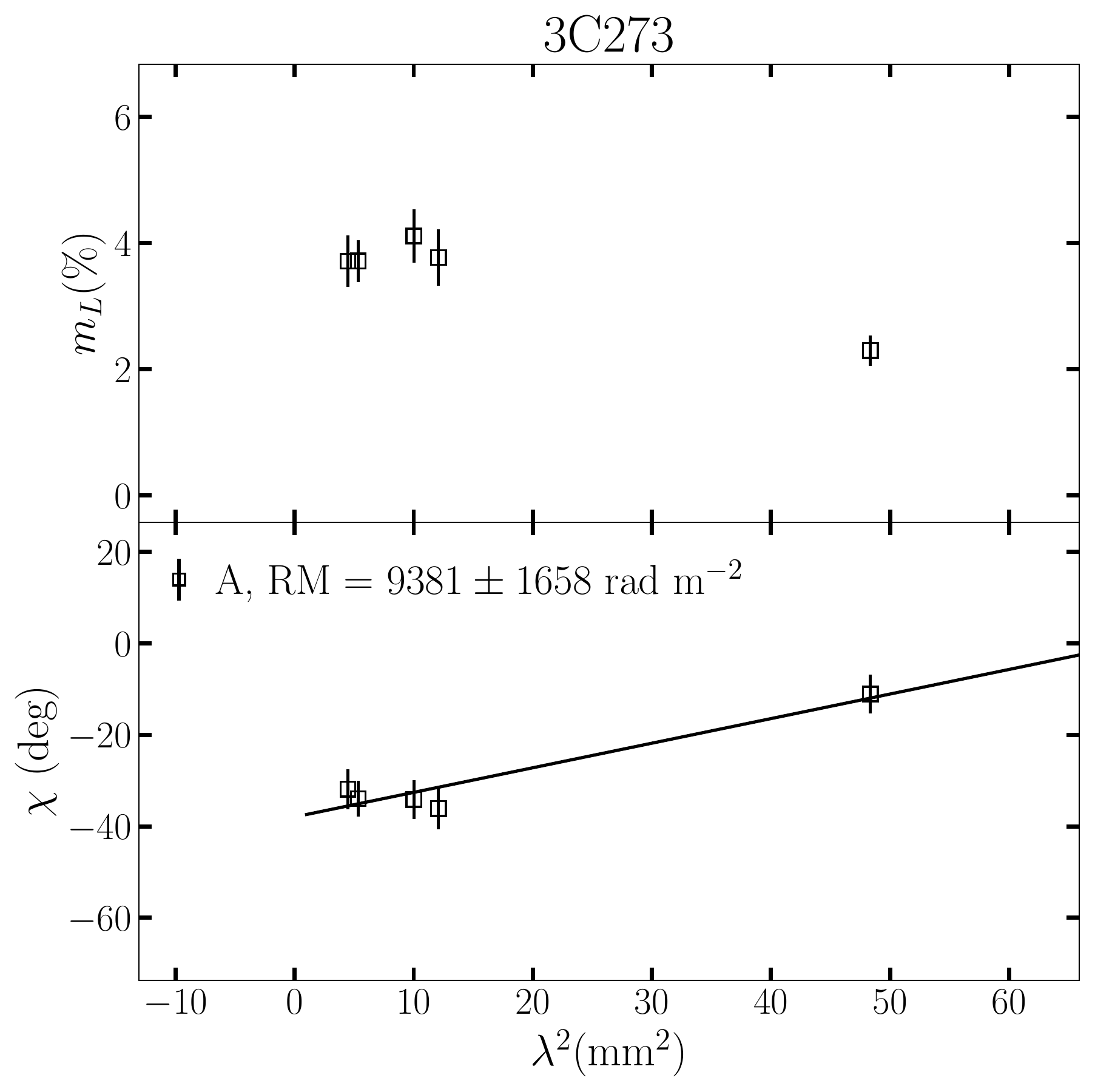}
\caption{\emph{Left:} linear polarization maps of 3C 273 observed on 2018 Feb 17 at 43 GHz taken from the BU database and on 2018 Feb 20--21 with the KVN at 86, 95, 130, and 142 GHz. All the maps are restored with the synthesized beam at 86 GHz for comparison. The fine scale polarization structures consisting of two polarized knots at distances of $\approx0.3$ and $\approx0.7$ mas from the core, observed in the high-resolution BU map, are blurred in this map due to the large convolving beam. \emph{Right:} fractional polarizations ($m_L$) and EVPAs ($\chi$) as functions of $\lambda^2$, where $\lambda$ is the observing wavelength, at $\approx0.7$ mas from the core marked as "A" in the left figure. The solid black line in the bottom panel shows the best-fit $\lambda^2$ law to the EVPAs. The polarization structures at all frequencies are generally consistent with each other, including the "shifts" of the linear polarization peaks from the total intensity peaks, as seen in the high-resolution BU image. The amounts of shift are different at 86--95 and 130--142 GHz, which indicates that the polarization structure of 3C 273 at higher frequencies could be different from that seen at 43 GHz presumably due to less depolarization. \label{fig:kvn_map}}
\end{figure*}

\subsubsection{KVN data at 86--142 GHz}

GPCAL is almost ideally suited for instrumental polarization calibration for the KVN because of its small (three) number of antennas. The limited number of baselines makes calibration quite sensitive to residual systematic errors in the data, and the D-terms from different calibrators often show large dispersion \citep[$\gtrsim1-3\%$,][]{Park2018}. In this case, using many calibrators in a single observing run with GPCAL would help to improve the D-term estimation accuracy. This is because the number of measurements at different antenna field rotation angles increases roughly by a factor of the number of calibrators, while the number of free parameters in the fitting increases by a factor of less than a few in the case of using the similarity assumption and does not increase at all when using instrumental polarization self-calibration.

We observed 11 AGN jets on 2018 Feb 20--21 with the KVN at 86, 95, 130, and 142 GHz and applied GPCAL to this data. One of the reasons for selecting this data is to verify the performance of the KVN polarimetry at very high frequencies up to 142 GHz, which is a unique advantage of the KVN, with assistance from GPCAL for an accurate D-term estimation. The observations at 86/130 GHz and 95/142 GHz were performed on the first (Feb 20) and second days (Feb 21), respectively. The data was taken at a recording rate of 8 Gbps and divided into a single IF for each frequency. More detailed descriptions of the data analysis and results will be presented in a forthcoming paper (Park et al. 2020, in preparation). We performed data reduction, imaging, self-calibration\footnote{Self-calibration is performed for visibility phases only because there are three antennas.}, and instrumental polarization calibration with GPCAL in a similar manner to the VLBA data analysis (Sections~\ref{sec:mojave} and ~\ref{sec:bu}). We used four to seven calibrators, which are bright and weakly-polarized or moderately polarized with relatively simple structures, for the initial D-term estimation using the similarity assumption. We included a few more calibrators having relatively complex polarization structures for instrumental polarization self-calibration with ten iterations. We obtained the reduced chi-squares of 1.3 -- 2.1 for different bands. The relatively bad goodness of fit for some bands may be understandable as we could not perform amplitude self-calibration because of the lack of antennas. Thus, there could be non-negligible gain errors affecting the fitting. 

We performed the EVPA calibration by comparing the integrated EVPAs of 3C 279 and OJ 287 with their EVPAs obtained by contemporaneous KVN single-dish observations, as described in \cite{Park2018}. We found that 3C 273, among our targets, would be a good source to test the performance of GPCAL. This flat-spectrum radio quasar shows an un-polarized core and moderately polarized knots at $\approx0.3$ and $\approx0.7$ mas from the core in the contemporaneous high-resolution BU observation on 2018 Feb 17 at 43 GHz\footnote{\url{https://www.bu.edu/blazars/VLBA_GLAST/3c273/3C273feb18_map.jpg}}. This structure would be difficult to obtain if there are significant D-term residuals in the data because the residuals tend to appear as artificial polarization signals in proportion to the total intensity emission \cite[][]{Leppanen1995}.

In Figure~\ref{fig:kvn_map}, we present the linear polarization maps of 3C 273 at four KVN frequencies. We also include the BU polarization map as a reference. All the maps are convolved with the synthesized beam at 86 GHz for a proper comparison. We ignored a possible core-shift between frequencies for image alignment because the expected core-shift between 43 and 130 GHz is $\lesssim0.05$ mas \citep{Lisakov2017}, which is much smaller than the convolving beam size. We found that the peak polarization positions are shifted from the cores at all four frequencies, although the large beam size of the KVN does not allow us to see the fine structures that were observed in the high-resolution BU map. Interestingly, the shifts at 86--95 GHz are larger than those at 130--142 GHz. 

One of the possible explanations for the different shifts is less depolarization of the jet at higher frequencies \citep[e.g.,][]{Sokoloff1998}. If the linear polarization intensity of the inner knot at $\approx0.3$ mas seen in the BU map becomes larger or the core polarization starts to be detected at higher frequencies due to less depolarization, then the positions of the peak polarization intensity could be shifted towards the core at higher frequencies. The higher degrees of linear polarization at $\approx0.7$ mas (marked as "A" in the map) at 86--142 GHz than at 43 GHz may support this interpretation, although it is difficult to investigate the exact origin of the different shifts with the low resolution maps only. Our results of the 3C 273 polarization using the KVN, showing a misalignment between the linearly polarized intensity peak and total intensity peak positions, demonstrate that GPCAL is capable of achieving a high D-term estimation accuracy and will be useful for future polarimetric studies using the KVN.

\section{Summary and Conclusion}
\label{sec:summary}

We have presented GPCAL, an automated pipeline for instrumental polarization calibration of VLBI data based on AIPS and Difmap. The general calibration procedure of the pipeline follows LPCAL in AIPS, which has been successful for a multitude of studies using various VLBI arrays for a long time. GPCAL provides several new functions that can enhance the D-term estimation accuracy. 

Firstly, it can fit the D-term model to multiple calibrators data simultaneously. This means that GPCAL properly considers the visibility weights of various sources, which should provide statistically more robust results compared to taking averages of the D-terms from individual sources estimated by LPCAL. Secondly, it allows using more accurate linear polarization models of calibrators for D-term estimation than the conventional way using the similarity assumption, which assumes that the linear polarization structures are proportional to the total intensity structures. This assumption may not hold in many cases, especially at high frequencies, and could be a source of significant uncertainties in the estimated D-terms. Thirdly, it includes the second-order terms in the model and can deal with the case of large D-terms and high source fractional polarization. Lastly, it provides many useful functions such as (i) changing the visibility weights of some stations for fitting, (ii) fixing the D-terms of some stations to be certain values when external constraints on those D-terms are available, (iii) estimating the D-terms of the stations comprising very short baselines and using them for fitting for the rest of the array, and (iv) providing plots showing the fitting results and statistics, which are useful for identifying some problematic scans or stations.

We have illustrated the capabilities of GPCAL by employing the simulated data and the real data sets observed with different VLBI arrays and at different frequencies. We produced the data simulated with \texttt{PolSimulate} in CASA, assuming simple source geometries consisting of several point sources for Stokes $I$, $Q$, and $U$, a VLBA-like array, and for uv-coverages of three sources. We assumed two cases for the source geometries; one with the locations of the total intensity models being coincident with those of the linearly polarized intensity models ($P \propto I$) and the other with the locations being significantly shifted from each other ($P \not\propto I$). We show that GPCAL can reproduce the ground-truth D-terms assumed in the simulation very well for both cases by using the data of three sources simultaneously. The latter case was difficult to model with the conventional way using the similarity assumption but could be successfully modeled thanks to the instrumental polarization self-calibration mode implemented in GPCAL.

We have applied GPCAL to the data of the monitoring programs of AGN jets with the VLBA at 15 GHz (the MOJAVE program) and 43 GHz (the VLBA-BU-BLAZAR program). We have shown that GPCAL can fit the model to the data of multiple calibrators simultaneously. The sources having complex linear polarization structures could also be used by performing instrumental polarization self-calibration. This result demonstrates that GPCAL will be very useful when there are no or few calibrators suitable for applying the similarity assumption in the data. We have compared the linear polarization maps of the sources showing complex polarization features obtained by GPCAL and taken from the monitoring program databases. The results are very consistent. These programs have achieved a high D-term estimation accuracy thanks to many good calibrators satisfying the similarity assumption well in their programs. The fact that GPCAL could reproduce nearly identical results to those programs demonstrates its capability of achieving a high D-term estimation accuracy. GPCAL will be especially useful for normal VLBI programs, for which it is difficult to have many good calibrators.

We have also applied GPCAL to the data of many AGN jets observed with the KVN at 86, 95, 130, and 142 GHz. Accurate D-term calibration for the KVN is understandably challenging because of the small number of antennas. GPCAL is well-suited to this type of data as well because one can increase the number of measurements by using many calibrators, while the number of free parameters is slightly increased or does not increase. We have shown that the linear polarization maps of 3C 273 at different frequencies obtained by GPCAL successfully recover its complex polarization structure seen in the contemporaneous high-resolution VLBA image at 43 GHz, namely the un-polarized core and the moderately polarized knots downstream of the core. The images even showed an indication of different amounts of depolarization at different frequencies, although a detailed interpretation is challenging because of the large synthesized beam of the KVN. This result demonstrates that GPCAL would be very useful for instrumental polarization calibration of VLBI arrays having not many antennas.

We note that the pipeline's current implementation does not take into account possible residual complex antenna gains and time-dependent D-terms for fitting. Also, GPCAL assumes that all calibrators are on an equal footing, while, in reality, some calibrators should be better than others, depending on field rotation angle coverages, SNRs, and so on, and one should put more weight on the good calibrators for fitting. These will be considered for future developments to enhance the polarization calibration accuracy further. We conclude with a remark on the importance of careful planning of observations. Although GPCAL provides useful functions to overcome the limitations of the existing calibraiton packages and to enhance the calibration accuracy, it is always important to have as many "good" calibrators as possible in the observations. This is especially important for the initial D-term estimation using the similarity assumption, which can affect the D-term estimates in the next steps and the final D-term estimates (Appendix~\ref{appendix:goodcal}).

\newpage

\acknowledgments

We thank the anonymous ApJ referee for detailed comments that significantly improved the manuscript. J.P. acknowledges financial support from the Korean National Research Foundation (NRF) via Global PhD Fellowship Grant 2014H1A2A1018695. J.P. is supported by an EACOA Fellowship awarded by the East Asia Core Observatories Association, which consists of the Academia Sinica Institute of Astronomy and Astrophysics,
the National Astronomical Observatory of Japan, Center for Astronomical Mega-Science, Chinese Academy of Sciences, and the Korea Astronomy and Space Science Institute. This work is supported by the Ministry of Science and Technology of Taiwan grant MOST 109-2112-M-001-025 (K.A). This research has made use of data from the MOJAVE database that is maintained by the MOJAVE team \citep{Lister2018}. This study makes use of 43 GHz VLBA data from the VLBA-BU Blazar Monitoring Program (VLBA-BU-BLAZAR; \url{http://www.bu.edu/blazars/VLBAproject.html}), funded by NASA through the Fermi Guest Investigator Program. The VLBA is an instrument of the National Radio Astronomy Observatory. The National Radio Astronomy Observatory is a facility of the National Science Foundation operated by Associated Universities, Inc. We are grateful to the staff of the KVN who helped to operate the array and to correlate the data. The KVN and a high-performance computing cluster are facilities operated by the KASI (Korea Astronomy and Space Science Institute). The KVN observations and correlations are supported through the high-speed network connections among the KVN sites provided by the KREONET (Korea Research Environment Open NETwork), which is managed and operated by the KISTI (Korea Institute of Science and Technology Information).

\facilities{VLBA(NRAO), KVN(KASI)}

\software{AIPS \citep{Greisen2003}, Difmap \citep{Shepherd1997}, ParselTongue \citep{Kettenis2006}, CASA \citep{CASA2007}, Numpy \citep{Numpy2011}, Scipy \citep{Scipy2020}, Pandas \citep{Pandas2010}, Astropy \citep{Astropy2013, Astropy2018}, Matplotlib \citep{Matplotlib2007}}

\appendix


\section{Importance of selecting good calibrators}
\label{appendix:goodcal}

We have presented that GPCAL can take into account complex linear polarization structures of calibrators by using instrumental polarization self-calibration (Section~\ref{sec:selfpol}). However, this procedure is based on the initial D-term estimation using the similarity assumption. Thus, the initial D-term estimates need to be reasonably accurate to achieve a high accuracy for the final D-term estimates. We performed a simple test to demonstrate the importance of selecting good calibrators for the initial D-term estimation. We obtained D-terms for the BU 43 GHz data by using two sources, 3C 279 and 1156+295, individually, in the same manner as in Section~\ref{sec:bu}. 3C 279 is known for having a high ($\approx10\%$) fractional polarization \citep[e.g.,][]{Park2018} and a complex jet structure at mm wavelengths \citep[e.g.,][]{Jorstad2017, Kim2020}, and thus is not a good calibrator for the initial D-term estimation. On the other hand, 1156+295 has a core-dominated linear polarization structure with a fractional polarization of a few \% \citep{Jorstad2017}, which can serve as a good calibrator for the initial D-term estimation.

\begin{figure}[t!]
\centering
\includegraphics[width = 0.325\textwidth]{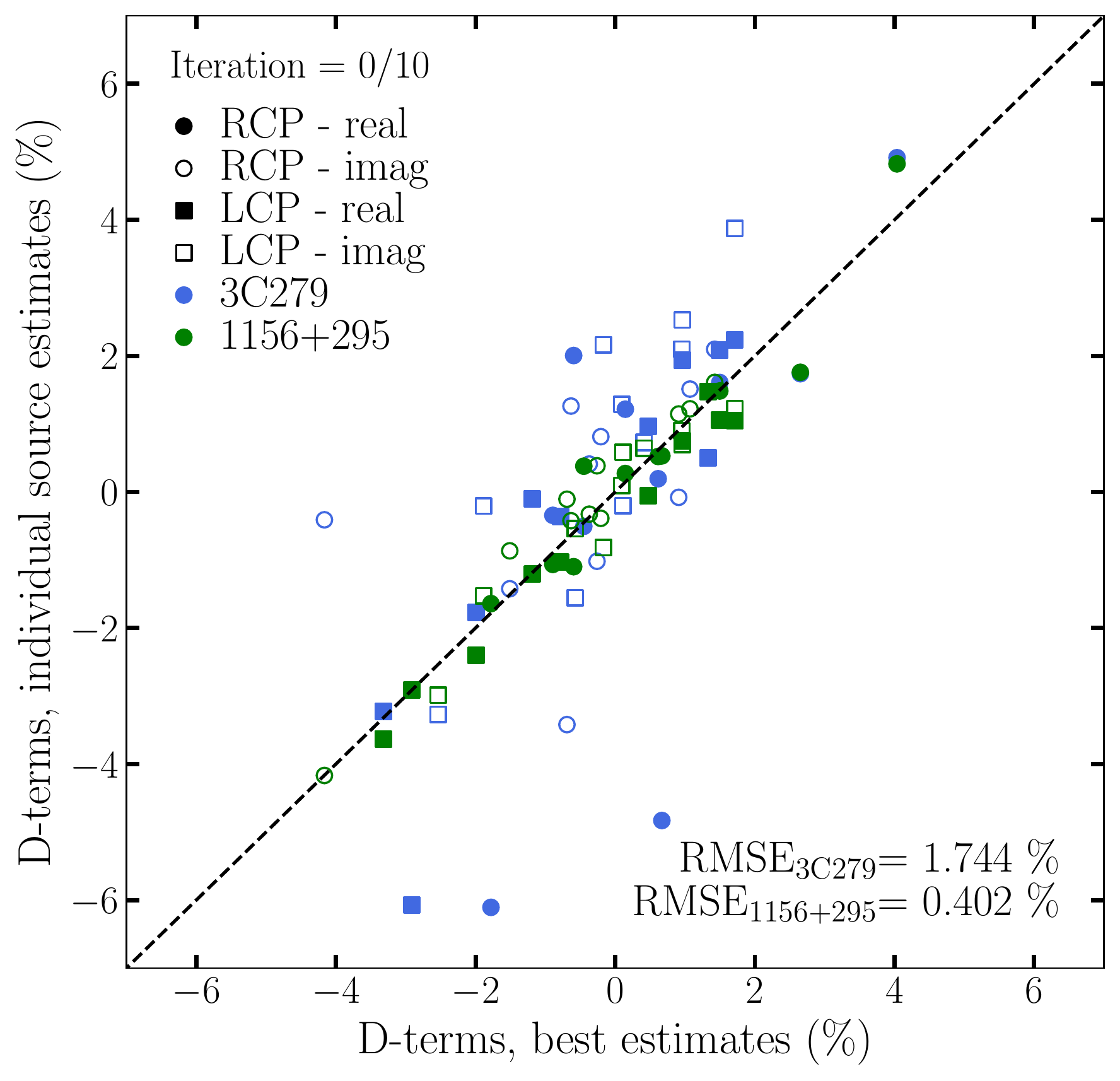}
\includegraphics[width = 0.325\textwidth]{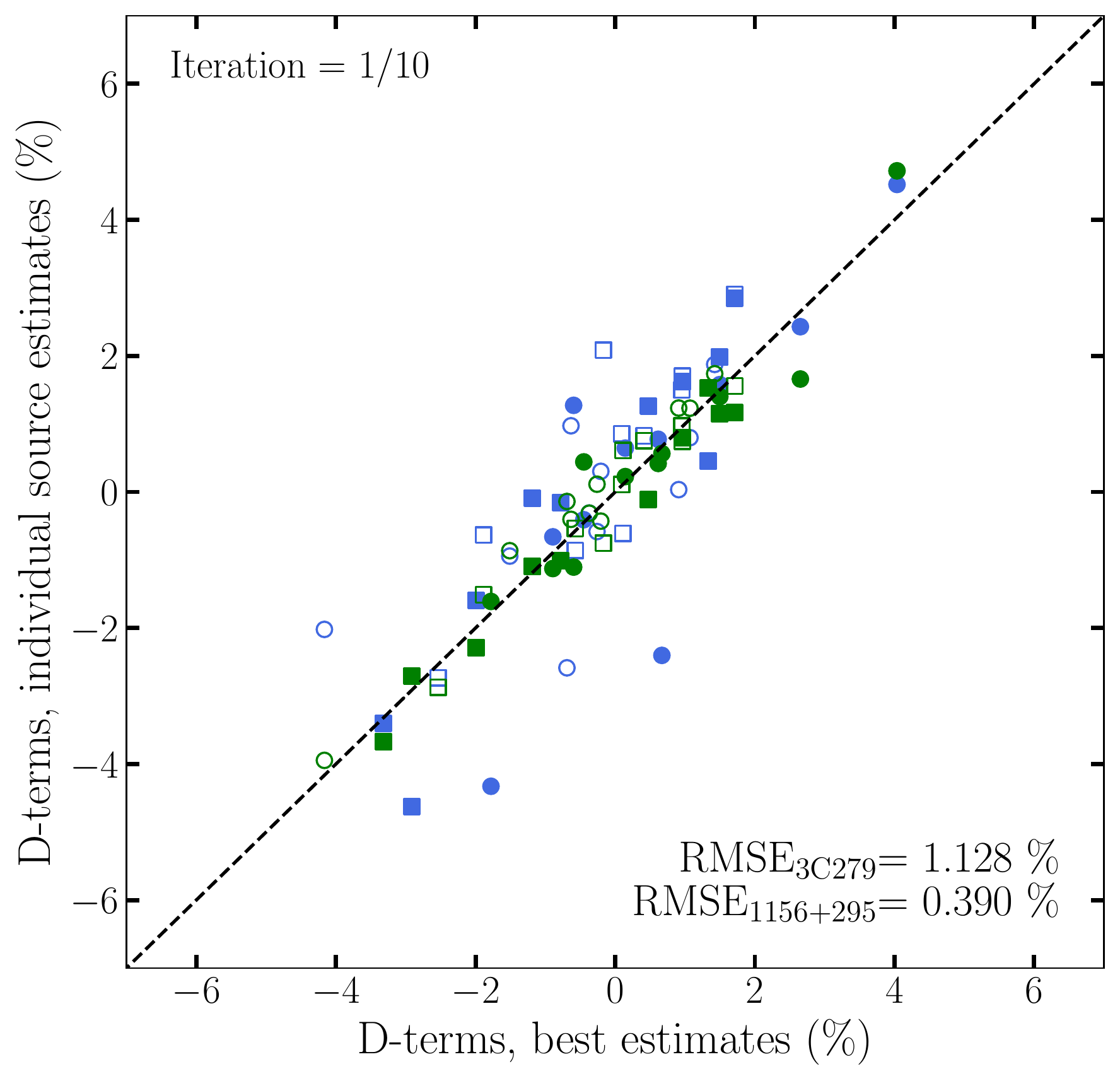}
\includegraphics[width = 0.325\textwidth]{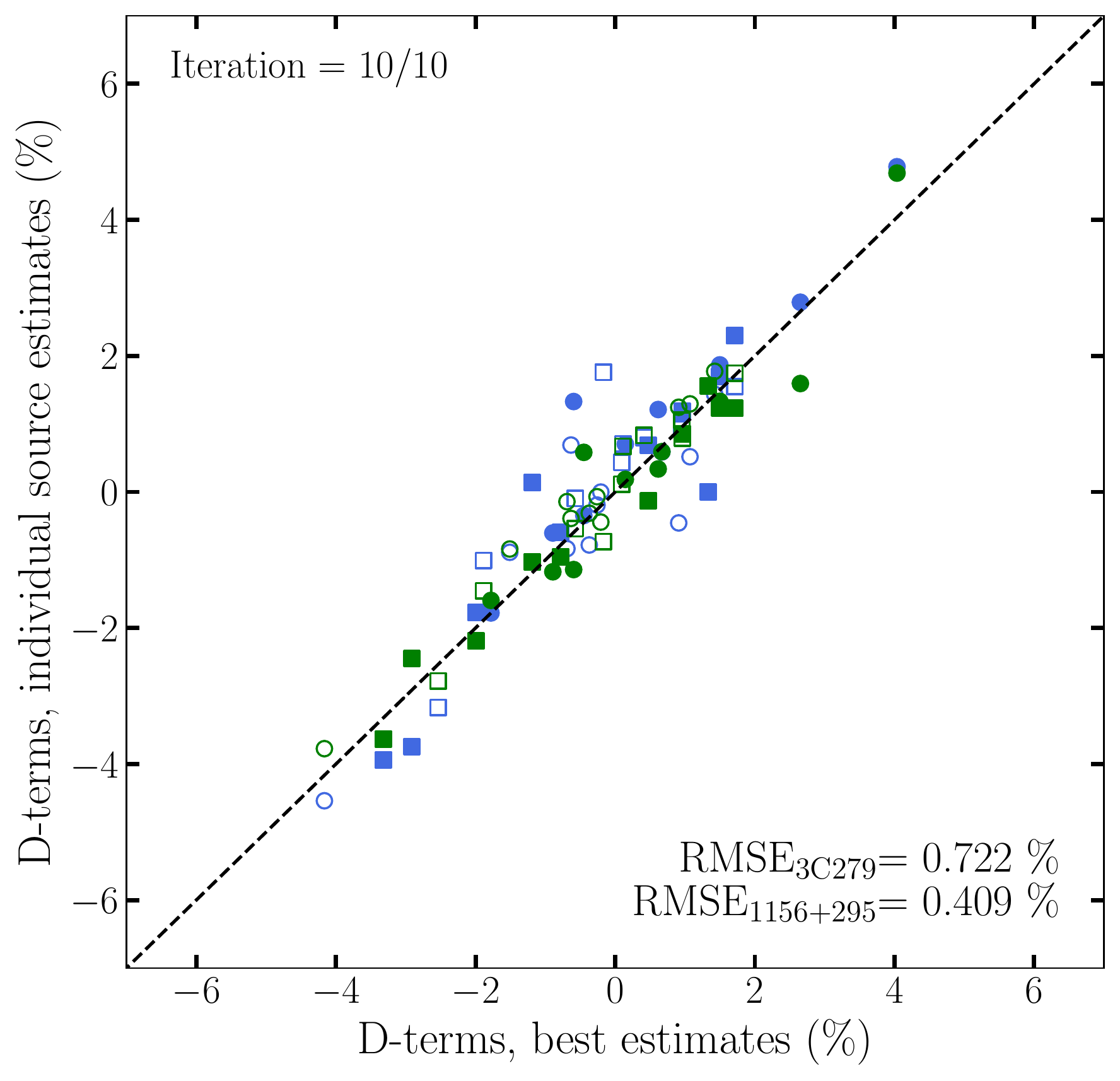}
\caption{Comparison of D-terms for the BU 43 GHz data obtained with GPCAL by using individual sources, 3C 279 (blue) and 1156+295 (green) on the y-axis, with the best D-term estimates obtained by using multiple sources simultaneously presented in Section~\ref{sec:bu} on the x-axis. The initial D-term estimates obtained by using the similarity assumption (left), the estimates after one (middle) and ten (right) iterations of instrumental polarization self-calibraiton for the individual source estimates are shown, while the same best D-term estimates obtained after ten iterations of instrumental polarization self-calibration are shown in all three panels. Filled and open symbols indicate the real and imaginary parts of D-terms, respectively. Circles and squares denote D-terms of RCP and LCP, respectively. The black dashed lines are one-to-one lines between the x and y-axes. \label{fig:test_3c279}}
\end{figure}

In Figure~\ref{fig:test_3c279}, we compare the D-terms obtained with those individual sources with our best D-term estimates using multiple sources simultaneously presented in Section~\ref{sec:bu}. The initial D-term estimates obtained with 3C 279 deviate a lot from the best estimates with a RMSE of $\approx1.7\%$, while those with 1156+295 show a good consistency with a RMSE of $\approx0.4\%$, as expected. The 3C 279 D-terms do improve with more iterations of instrumental polarization self-calibration, reaching a RMSE of $\approx0.7\%$ after ten iterations. The 1156+295 D-terms do not change much with instrumental polarization self-calibration, indicating that the similarity assumption does work well for this source. The final D-term estimates obtained with 3C 279 deviate more from the best estimates than do those with 1156+295, which demonstrates the importance of selecting good calibrators having either very low degrees of polarization or simple linear polarization structures for the initial D-term estimation. Similarly, it is not recommended to use any calibrators having poor antenna field rotation angle coverages for many stations or low SNRs for both the initial D-term estimation and instrumental polarization self-calibration procedures. They would naturally degrade the D-term estimates, affecting the estimates in the next steps, resulting in poor final D-term estimates.

\section{Validation of using calibrators having complex linear polarization structures}
\label{appendix:synthetic}

\begin{figure}[t!]
\centering
\includegraphics[width = 0.26\textwidth]{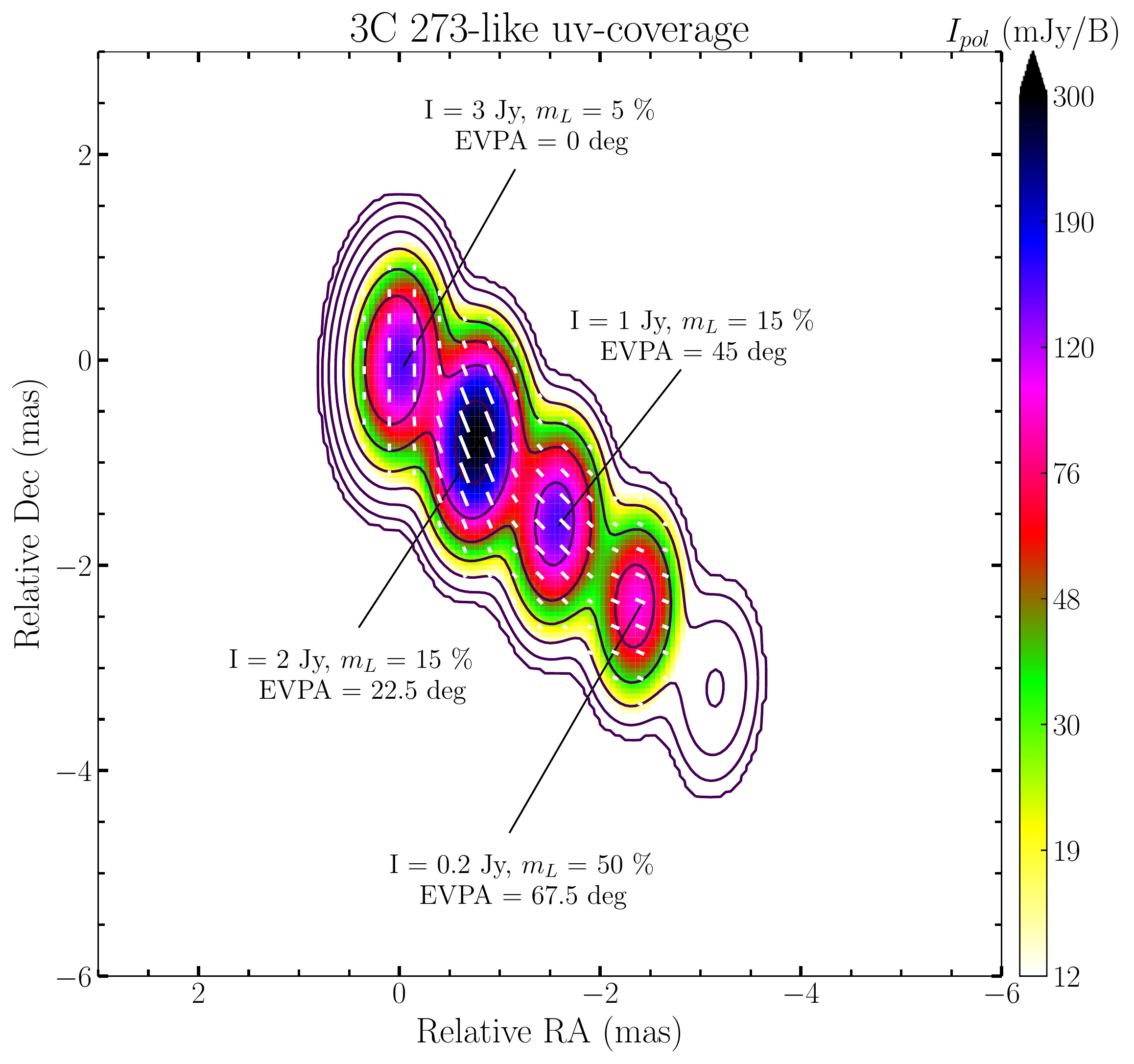}
\includegraphics[width = 0.24\textwidth]{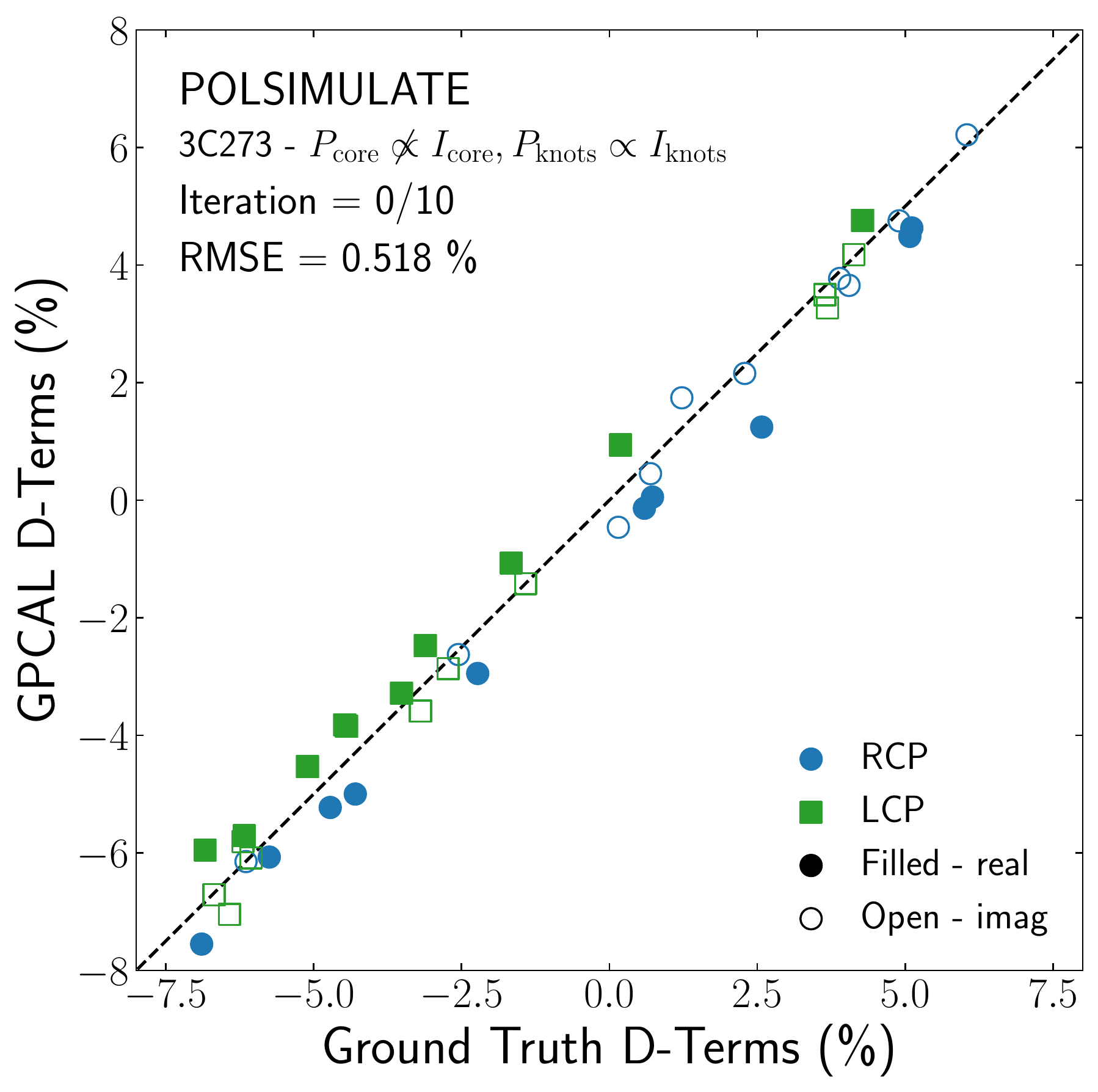}
\includegraphics[width = 0.24\textwidth]{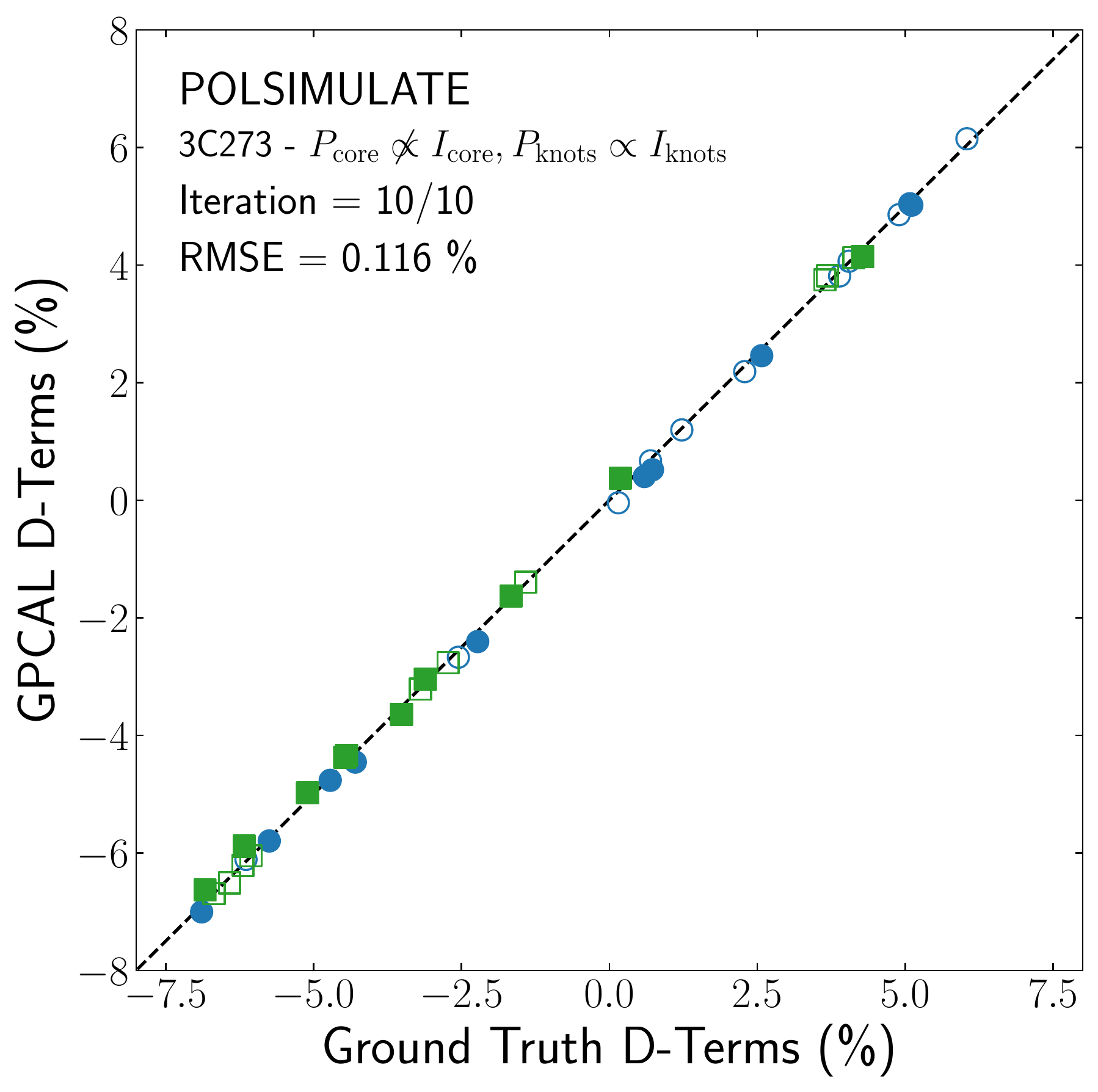}
\includegraphics[width = 0.24\textwidth]{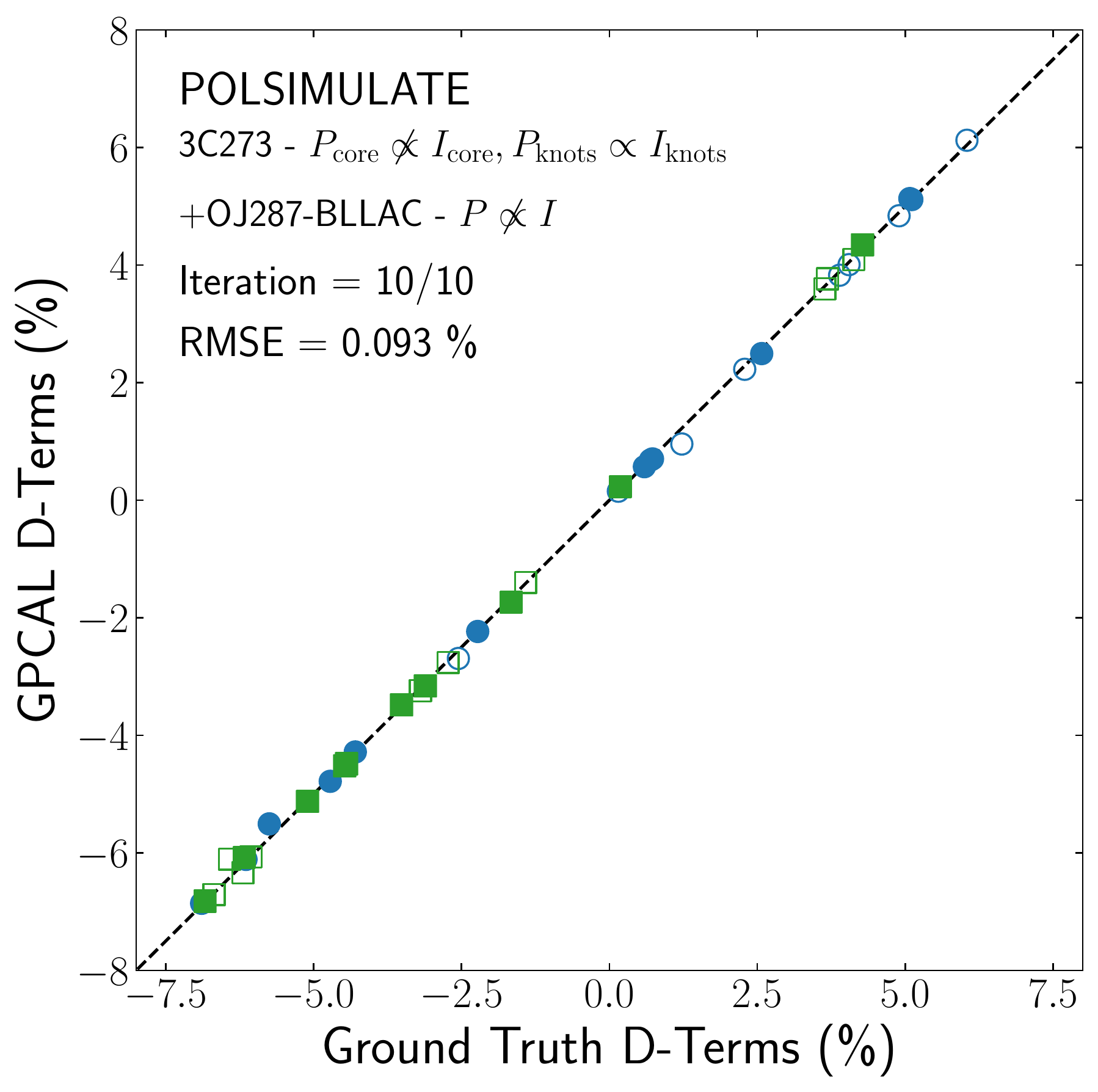}
\caption{Colormap on the left: same as Figure~\ref{fig:synthetic_example} but for the simulated data with a 3C 273-like uv-coverage, produced as explained in Appendix~\ref{appendix:synthetic}. Both the total intensity and linear polarization emission consist of several point sources. The total intensity and linear polarization models are slightly shifted from each other by $\approx0.071$ mas for the component at the origin of the map ("core"), while the models are coincident for the rest of the components ("knots"). Diagrams on the right: comparison between the ground-truth D-terms assumed in the simulated data and the reconstructed D-terms by GPCAL. The results for three cases are shown: the reconstructed D-terms obtained by the initial D-term estimation using the 3C 273 data (left), additional ten iterations of instrumental polarization self-calibration using the 3C 273 data (middle), and additional ten iterations of instrumental polarization self-calibration using all three data sets (right). The RMSE values are noted on the top left of each figure. The black dashed lines show one-to-one lines. \label{fig:synthetic_realistic}}
\end{figure}

In this appendix, we address the question whether including calibrators having complex linear polarization structures for instrumental polarization self-calibration can improve or degrade the D-term estimates. We consider a realistic case that there is a calibrator which satisfies the similarity assumption reasonably well and there are two other calibrators having complex linear polarization structures. We produce a simulated data with a 3C 273-like uv-coverage by using \texttt{PolSimulate}, as explained in Section~\ref{sec:simul}. We assume that the source's total intensity emission consists of five point sources and four of them are linearly polarized (the left panel of Figure~\ref{fig:synthetic_realistic}). The Stokes $Q$ and $U$ emission also consist of several point sources and their positions are the same as the total intensity model components' positions, except for the one at the origin of the map. For that component, there is a small shift in the positions between the total intensity and linear polarization models by $\approx0.071$ mas, which is less than 1/10 of the synthesized beam size. This kind of calibrator would normally be treated as a good calibrator that satisfy the similarity assumption well.

We ran GPCAL considering two cases. One is to use this data for the initial D-term estimation using the similarity assumption and perform additional ten iterations of instrumental polarization self-calibration using the same data. The other is to repeat the same procedure but including the simulated data with OJ 287 and BL Lac-like uv-coverages having significant shifts between the Stokes $I$ and linear polarization model components, which were used for the test of the $P \not \propto$ I case in Section~\ref{sec:simul}. We compare the reconstructed D-terms, obtained by (i) the initial D-term estimation using the 3C 273 data, (ii) additional ten iterations of instrumental polarization self-calibration using the 3C 273 data, and (iii) additional ten iterations of instrumental polarization self-calibration using all three data sets simultaneously, with the ground-truth D-terms in Figure~\ref{fig:synthetic_realistic}.

The initial D-term estimates have an RMSE of $\approx0.5\%$. This result demonstrates that even a very small positional shift ($\lesssim10\%$ of the synthesized beam size) between total intensity and linear polarization peaks can prevent an accurate D-term estimation. The RMSE values become smaller with the additional instrumental polarization self-calibration, reaching $\approx0.12\%$ after ten iterations when only the 3C 273 data is used. A smaller RMSE value of $\approx0.093\%$ is obtained when we include two more sources having the linear polarization structures that are significant different from the total intensity structures for instrumental polarization self-calibraiton. This test suggests that it is generally recommended to use as many calibrators as possible even though the calibrators have complex linear polarization structures. However, as we suggest in Appendix~\ref{appendix:goodcal}, this statement may be valid only when one uses calibrators that satisfy the similarity assumption reasonably well for the initial D-term estimation. Also, one should keep in mind that this result is based on the simulated data assuming simple source structures and no antenna gain errors. As pointed out in Section~\ref{sec:mojave}, if calibrators having poor antenna field rotation angle coverages, low SNRs, and antenna gain errors not well corrected are used, then they would degrade the D-term estimates. Users are recommended to try different combinations of calibrators for the initial D-term estimation and instrumental polarization self-calibration, check the results, and determine the lists of the best calibrators.

\bibliography{gpcal.bib}{}

\begin{thebibliography}{}
\expandafter\ifx\csname natexlab\endcsname\relax\def\natexlab#1{#1}\fi
\providecommand{\url}[1]{\href{#1}{#1}}
\providecommand{\dodoi}[1]{doi:~\href{http://doi.org/#1}{\nolinkurl{#1}}}
\providecommand{\doeprint}[1]{\href{http://ascl.net/#1}{\nolinkurl{http://ascl.net/#1}}}
\providecommand{\doarXiv}[1]{\href{https://arxiv.org/abs/#1}{\nolinkurl{https://arxiv.org/abs/#1}}}

\bibitem[{{Asada} {et~al.}(2002){Asada}, {Inoue}, {Uchida}, {Kameno},
  {Fujisawa}, {Iguchi}, \& {Mutoh}}]{Asada2002}
{Asada}, K., {Inoue}, M., {Uchida}, Y., {et~al.} 2002, \pasj, 54, L39,
  \dodoi{10.1093/pasj/54.3.L39}

\bibitem[{{Astropy Collaboration} {et~al.}(2013){Astropy Collaboration},
  {Robitaille}, {Tollerud}, {Greenfield}, {Droettboom}, {Bray}, {Aldcroft},
  {Davis}, {Ginsburg}, {Price-Whelan}, {Kerzendorf}, {Conley}, {Crighton},
  {Barbary}, {Muna}, {Ferguson}, {Grollier}, {Parikh}, {Nair}, {Unther},
  {Deil}, {Woillez}, {Conseil}, {Kramer}, {Turner}, {Singer}, {Fox}, {Weaver},
  {Zabalza}, {Edwards}, {Azalee Bostroem}, {Burke}, {Casey}, {Crawford},
  {Dencheva}, {Ely}, {Jenness}, {Labrie}, {Lim}, {Pierfederici}, {Pontzen},
  {Ptak}, {Refsdal}, {Servillat}, \& {Streicher}}]{Astropy2013}
{Astropy Collaboration}, {Robitaille}, T.~P., {Tollerud}, E.~J., {et~al.} 2013,
  \aap, 558, A33, \dodoi{10.1051/0004-6361/201322068}

\bibitem[{{Astropy Collaboration} {et~al.}(2018){Astropy Collaboration},
  {Price-Whelan}, {Sip{\H{o}}cz}, {G{\"u}nther}, {Lim}, {Crawford}, {Conseil},
  {Shupe}, {Craig}, {Dencheva}, {Ginsburg}, {Vand erPlas}, {Bradley},
  {P{\'e}rez-Su{\'a}rez}, {de Val-Borro}, {Aldcroft}, {Cruz}, {Robitaille},
  {Tollerud}, {Ardelean}, {Babej}, {Bach}, {Bachetti}, {Bakanov}, {Bamford},
  {Barentsen}, {Barmby}, {Baumbach}, {Berry}, {Biscani}, {Boquien}, {Bostroem},
  {Bouma}, {Brammer}, {Bray}, {Breytenbach}, {Buddelmeijer}, {Burke},
  {Calderone}, {Cano Rodr{\'\i}guez}, {Cara}, {Cardoso}, {Cheedella}, {Copin},
  {Corrales}, {Crichton}, {D'Avella}, {Deil}, {Depagne}, {Dietrich}, {Donath},
  {Droettboom}, {Earl}, {Erben}, {Fabbro}, {Ferreira}, {Finethy}, {Fox},
  {Garrison}, {Gibbons}, {Goldstein}, {Gommers}, {Greco}, {Greenfield},
  {Groener}, {Grollier}, {Hagen}, {Hirst}, {Homeier}, {Horton}, {Hosseinzadeh},
  {Hu}, {Hunkeler}, {Ivezi{\'c}}, {Jain}, {Jenness}, {Kanarek}, {Kendrew},
  {Kern}, {Kerzendorf}, {Khvalko}, {King}, {Kirkby}, {Kulkarni}, {Kumar},
  {Lee}, {Lenz}, {Littlefair}, {Ma}, {Macleod}, {Mastropietro}, {McCully},
  {Montagnac}, {Morris}, {Mueller}, {Mumford}, {Muna}, {Murphy}, {Nelson},
  {Nguyen}, {Ninan}, {N{\"o}the}, {Ogaz}, {Oh}, {Parejko}, {Parley}, {Pascual},
  {Patil}, {Patil}, {Plunkett}, {Prochaska}, {Rastogi}, {Reddy Janga},
  {Sabater}, {Sakurikar}, {Seifert}, {Sherbert}, {Sherwood-Taylor}, {Shih},
  {Sick}, {Silbiger}, {Singanamalla}, {Singer}, {Sladen}, {Sooley},
  {Sornarajah}, {Streicher}, {Teuben}, {Thomas}, {Tremblay}, {Turner},
  {Terr{\'o}n}, {van Kerkwijk}, {de la Vega}, {Watkins}, {Weaver}, {Whitmore},
  {Woillez}, {Zabalza}, \& {Astropy Contributors}}]{Astropy2018}
{Astropy Collaboration}, {Price-Whelan}, A.~M., {Sip{\H{o}}cz}, B.~M., {et~al.}
  2018, \aj, 156, 123, \dodoi{10.3847/1538-3881/aabc4f}

\bibitem[{{Attridge} {et~al.}(2005){Attridge}, {Wardle}, \&
  {Homan}}]{Attridge2005}
{Attridge}, J.~M., {Wardle}, J. F.~C., \& {Homan}, D.~C. 2005, \apjl, 633, L85,
  \dodoi{10.1086/498392}

\bibitem[{{Casadio} {et~al.}(2017){Casadio}, {Krichbaum}, {Marscher},
  {Jorstad}, {G{\'o}mez}, {Agudo}, {Bach}, {Kim}, {Hodgson}, \&
  {Zensus}}]{Casadio2017}
{Casadio}, C., {Krichbaum}, T., {Marscher}, A., {et~al.} 2017, Galaxies, 5, 67,
  \dodoi{10.3390/galaxies5040067}

\bibitem[{{Cotton}(1993)}]{Cotton1993}
{Cotton}, W.~D. 1993, \aj, 106, 1241, \dodoi{10.1086/116723}

\bibitem[{{Cotton}(1995{\natexlab{a}})}]{Cotton1995a}
{Cotton}, W.~D. 1995{\natexlab{a}}, in Astronomical Society of the Pacific
  Conference Series, Vol.~82, Very Long Baseline Interferometry and the VLBA,
  ed. J.~A. {Zensus}, P.~J. {Diamond}, \& P.~J. {Napier}, 189

\bibitem[{{Cotton}(1995{\natexlab{b}})}]{Cotton1995b}
{Cotton}, W.~D. 1995{\natexlab{b}}, in Astronomical Society of the Pacific
  Conference Series, Vol.~82, Very Long Baseline Interferometry and the VLBA,
  ed. J.~A. {Zensus}, P.~J. {Diamond}, \& P.~J. {Napier}, 289

\bibitem[{{Event Horizon Telescope Collaboration}
  {et~al.}(2019{\natexlab{a}}){Event Horizon Telescope Collaboration},
  {Akiyama}, {Alberdi}, {Alef}, {Asada}, {Azulay}, {Baczko}, {Ball},
  {Balokovi{\'c}}, {Barrett}, {Bintley}, {Blackburn}, {Boland}, {Bouman},
  {Bower}, {Bremer}, {Brinkerink}, {Brissenden}, {Britzen}, {Broderick},
  {Broguiere}, {Bronzwaer}, {Byun}, {Carlstrom}, {Chael}, {Chan}, {Chatterjee},
  {Chatterjee}, {Chen}, {Chen}, {Cho}, {Christian}, {Conway}, {Cordes}, {Crew},
  {Cui}, {Davelaar}, {De Laurentis}, {Deane}, {Dempsey}, {Desvignes}, {Dexter},
  {Doeleman}, {Eatough}, {Falcke}, {Fish}, {Fomalont}, {Fraga-Encinas},
  {Freeman}, {Friberg}, {Fromm}, {G{\'o}mez}, {Galison}, {Gammie},
  {Garc{\'\i}a}, {Gentaz}, {Georgiev}, {Goddi}, {Gold}, {Gu}, {Gurwell},
  {Hada}, {Hecht}, {Hesper}, {Ho}, {Ho}, {Honma}, {Huang}, {Huang}, {Hughes},
  {Ikeda}, {Inoue}, {Issaoun}, {James}, {Jannuzi}, {Janssen}, {Jeter}, {Jiang},
  {Johnson}, {Jorstad}, {Jung}, {Karami}, {Karuppusamy}, {Kawashima},
  {Keating}, {Kettenis}, {Kim}, {Kim}, {Kim}, {Kino}, {Koay}, {Koch}, {Koyama},
  {Kramer}, {Kramer}, {Krichbaum}, {Kuo}, {Lauer}, {Lee}, {Li}, {Li},
  {Lindqvist}, {Liu}, {Liuzzo}, {Lo}, {Lobanov}, {Loinard}, {Lonsdale}, {Lu},
  {MacDonald}, {Mao}, {Markoff}, {Marrone}, {Marscher}, {Mart{\'\i}-Vidal},
  {Matsushita}, {Matthews}, {Medeiros}, {Menten}, {Mizuno}, {Mizuno}, {Moran},
  {Moriyama}, {Moscibrodzka}, {M{\"u}ller}, {Nagai}, {Nagar}, {Nakamura},
  {Narayan}, {Narayanan}, {Natarajan}, {Neri}, {Ni}, {Noutsos}, {Okino},
  {Olivares}, {Ortiz-Le{\'o}n}, {Oyama}, {{\"O}zel}, {Palumbo}, {Patel}, {Pen},
  {Pesce}, {Pi{\'e}tu}, {Plambeck}, {PopStefanija}, {Porth}, {Prather},
  {Preciado-L{\'o}pez}, {Psaltis}, {Pu}, {Ramakrishnan}, {Rao}, {Rawlings},
  {Raymond}, {Rezzolla}, {Ripperda}, {Roelofs}, {Rogers}, {Ros}, {Rose},
  {Roshanineshat}, {Rottmann}, {Roy}, {Ruszczyk}, {Ryan}, {Rygl},
  {S{\'a}nchez}, {S{\'a}nchez-Arguelles}, {Sasada}, {Savolainen}, {Schloerb},
  {Schuster}, {Shao}, {Shen}, {Small}, {Sohn}, {SooHoo}, {Tazaki}, {Tiede},
  {Tilanus}, {Titus}, {Toma}, {Torne}, {Trent}, {Trippe}, {Tsuda}, {van
  Bemmel}, {van Langevelde}, {van Rossum}, {Wagner}, {Wardle}, {Weintroub},
  {Wex}, {Wharton}, {Wielgus}, {Wong}, {Wu}, {Young}, {Young}, {Younsi},
  {Yuan}, {Yuan}, {Zensus}, {Zhao}, {Zhao}, {Zhu}, {Algaba}, {Allardi},
  {Amestica}, {Anczarski}, {Bach}, {Baganoff}, {Beaudoin}, {Benson},
  {Berthold}, {Blanchard}, {Blundell}, {Bustamente}, {Cappallo},
  {Castillo-Dom{\'\i}nguez}, {Chang}, {Chang}, {Chang}, {Chen}, {Chilson},
  {Chuter}, {C{\'o}rdova Rosado}, {Coulson}, {Crawford}, {Crowley}, {David},
  {Derome}, {Dexter}, {Dornbusch}, {Dudevoir}, {Dzib}, {Eckart}, {Eckert},
  {Erickson}, {Everett}, {Faber}, {Farah}, {Fath}, {Folkers}, {Forbes},
  {Freund}, {G{\'o}mez-Ruiz}, {Gale}, {Gao}, {Geertsema}, {Graham}, {Greer},
  {Grosslein}, {Gueth}, {Haggard}, {Halverson}, {Han}, {Han}, {Hao},
  {Hasegawa}, {Henning}, {Hern{\'a}ndez-G{\'o}mez}, {Herrero-Illana},
  {Heyminck}, {Hirota}, {Hoge}, {Huang}, {Impellizzeri}, {Jiang}, {Kamble},
  {Keisler}, {Kimura}, {Kono}, {Kubo}, {Kuroda}, {Lacasse}, {Laing}, {Leitch},
  {Li}, {Lin}, {Liu}, {Liu}, {Lu}, {Marson}, {Martin-Cocher}, {Massingill},
  {Matulonis}, {McColl}, {McWhirter}, {Messias}, {Meyer-Zhao}, {Michalik},
  {Monta{\~n}a}, {Montgomerie}, {Mora-Klein}, {Muders}, {Nadolski}, {Navarro},
  {Neilsen}, {Nguyen}, {Nishioka}, {Norton}, {Nowak}, {Nystrom}, {Ogawa},
  {Oshiro}, {Oyama}, {Parsons}, {Paine}, {Pe{\~n}alver}, {Phillips}, {Poirier},
  {Pradel}, {Primiani}, {Raffin}, {Rahlin}, {Reiland}, {Risacher}, {Ruiz},
  {S{\'a}ez-Mada{\'\i}n}, {Sassella}, {Schellart}, {Shaw}, {Silva}, {Shiokawa},
  {Smith}, {Snow}, {Souccar}, {Sousa}, {Sridharan}, {Srinivasan}, {Stahm},
  {Stark}, {Story}, {Timmer}, {Vertatschitsch}, {Walther}, {Wei}, {Whitehorn},
  {Whitney}, {Woody}, {Wouterloot}, {Wright}, {Yamaguchi}, {Yu}, {Zeballos},
  {Zhang}, \& {Ziurys}}]{EHT2019a}
{Event Horizon Telescope Collaboration}, {Akiyama}, K., {Alberdi}, A., {et~al.}
  2019{\natexlab{a}}, \apjl, 875, L1, \dodoi{10.3847/2041-8213/ab0ec7}

\bibitem[{{Event Horizon Telescope Collaboration}
  {et~al.}(2019{\natexlab{b}}){Event Horizon Telescope Collaboration},
  {Akiyama}, {Alberdi}, {Alef}, {Asada}, {Azulay}, {Baczko}, {Ball},
  {Balokovi{\'c}}, {Barrett}, {Bintley}, {Blackburn}, {Boland}, {Bouman},
  {Bower}, {Bremer}, {Brinkerink}, {Brissenden}, {Britzen}, {Broderick},
  {Broguiere}, {Bronzwaer}, {Byun}, {Carlstrom}, {Chael}, {Chan}, {Chatterjee},
  {Chatterjee}, {Chen}, {Chen}, {Cho}, {Christian}, {Conway}, {Cordes}, {Crew},
  {Cui}, {Davelaar}, {De Laurentis}, {Deane}, {Dempsey}, {Desvignes}, {Dexter},
  {Doeleman}, {Eatough}, {Falcke}, {Fish}, {Fomalont}, {Fraga-Encinas},
  {Friberg}, {Fromm}, {G{\'o}mez}, {Galison}, {Gammie}, {Garc{\'\i}a},
  {Gentaz}, {Georgiev}, {Goddi}, {Gold}, {Gu}, {Gurwell}, {Hada}, {Hecht},
  {Hesper}, {Ho}, {Ho}, {Honma}, {Huang}, {Huang}, {Hughes}, {Ikeda}, {Inoue},
  {Issaoun}, {James}, {Jannuzi}, {Janssen}, {Jeter}, {Jiang}, {Johnson},
  {Jorstad}, {Jung}, {Karami}, {Karuppusamy}, {Kawashima}, {Keating},
  {Kettenis}, {Kim}, {Kim}, {Kim}, {Kino}, {Koay}, {Koch}, {Koyama}, {Kramer},
  {Kramer}, {Krichbaum}, {Kuo}, {Lauer}, {Lee}, {Li}, {Li}, {Lindqvist}, {Liu},
  {Liuzzo}, {Lo}, {Lobanov}, {Loinard}, {Lonsdale}, {Lu}, {MacDonald}, {Mao},
  {Markoff}, {Marrone}, {Marscher}, {Mart{\'\i}-Vidal}, {Matsushita},
  {Matthews}, {Medeiros}, {Menten}, {Mizuno}, {Mizuno}, {Moran}, {Moriyama},
  {Moscibrodzka}, {M{\"u}ller}, {Nagai}, {Nagar}, {Nakamura}, {Narayan},
  {Narayanan}, {Natarajan}, {Neri}, {Ni}, {Noutsos}, {Okino}, {Olivares},
  {Ortiz-Le{\'o}n}, {Oyama}, {{\"O}zel}, {Palumbo}, {Patel}, {Pen}, {Pesce},
  {Pi{\'e}tu}, {Plambeck}, {PopStefanija}, {Porth}, {Prather},
  {Preciado-L{\'o}pez}, {Psaltis}, {Pu}, {Ramakrishnan}, {Rao}, {Rawlings},
  {Raymond}, {Rezzolla}, {Ripperda}, {Roelofs}, {Rogers}, {Ros}, {Rose},
  {Roshanineshat}, {Rottmann}, {Roy}, {Ruszczyk}, {Ryan}, {Rygl},
  {S{\'a}nchez}, {S{\'a}nchez-Arguelles}, {Sasada}, {Savolainen}, {Schloerb},
  {Schuster}, {Shao}, {Shen}, {Small}, {Sohn}, {SooHoo}, {Tazaki}, {Tiede},
  {Tilanus}, {Titus}, {Toma}, {Torne}, {Trent}, {Trippe}, {Tsuda}, {van
  Bemmel}, {van Langevelde}, {van Rossum}, {Wagner}, {Wardle}, {Weintroub},
  {Wex}, {Wharton}, {Wielgus}, {Wong}, {Wu}, {Young}, {Young}, {Younsi},
  {Yuan}, {Yuan}, {Zensus}, {Zhao}, {Zhao}, {Zhu}, {Algaba}, {Allardi},
  {Amestica}, {Bach}, {Beaudoin}, {Benson}, {Berthold}, {Blanchard},
  {Blundell}, {Bustamente}, {Cappallo}, {Castillo-Dom{\'\i}nguez}, {Chang},
  {Chang}, {Chang}, {Chen}, {Chilson}, {Chuter}, {C{\'o}rdova Rosado},
  {Coulson}, {Crawford}, {Crowley}, {David}, {Derome}, {Dexter}, {Dornbusch},
  {Dudevoir}, {Dzib}, {Eckert}, {Erickson}, {Everett}, {Faber}, {Farah},
  {Fath}, {Folkers}, {Forbes}, {Freund}, {G{\'o}mez-Ruiz}, {Gale}, {Gao},
  {Geertsema}, {Graham}, {Greer}, {Grosslein}, {Gueth}, {Halverson}, {Han},
  {Han}, {Hao}, {Hasegawa}, {Henning}, {Hern{\'a}ndez-G{\'o}mez},
  {Herrero-Illana}, {Heyminck}, {Hirota}, {Hoge}, {Huang}, {Impellizzeri},
  {Jiang}, {Kamble}, {Keisler}, {Kimura}, {Kono}, {Kubo}, {Kuroda}, {Lacasse},
  {Laing}, {Leitch}, {Li}, {Lin}, {Liu}, {Liu}, {Lu}, {Marson},
  {Martin-Cocher}, {Massingill}, {Matulonis}, {McColl}, {McWhirter}, {Messias},
  {Meyer-Zhao}, {Michalik}, {Monta{\~n}a}, {Montgomerie}, {Mora-Klein},
  {Muders}, {Nadolski}, {Navarro}, {Nguyen}, {Nishioka}, {Norton}, {Nystrom},
  {Ogawa}, {Oshiro}, {Oyama}, {Padin}, {Parsons}, {Paine}, {Pe{\~n}alver},
  {Phillips}, {Poirier}, {Pradel}, {Primiani}, {Raffin}, {Rahlin}, {Reiland},
  {Risacher}, {Ruiz}, {S{\'a}ez-Mada{\'\i}n}, {Sassella}, {Schellart}, {Shaw},
  {Silva}, {Shiokawa}, {Smith}, {Snow}, {Souccar}, {Sousa}, {Sridharan},
  {Srinivasan}, {Stahm}, {Stark}, {Story}, {Timmer}, {Vertatschitsch},
  {Walther}, {Wei}, {Whitehorn}, {Whitney}, {Woody}, {Wouterloot}, {Wright},
  {Yamaguchi}, {Yu}, {Zeballos}, \& {Ziurys}}]{EHT2019b}
---. 2019{\natexlab{b}}, \apjl, 875, L2, \dodoi{10.3847/2041-8213/ab0c96}

\bibitem[{{Event Horizon Telescope Collaboration}
  {et~al.}(2019{\natexlab{c}}){Event Horizon Telescope Collaboration},
  {Akiyama}, {Alberdi}, {Alef}, {Asada}, {Azulay}, {Baczko}, {Ball},
  {Balokovi{\'c}}, {Barrett}, {Bintley}, {Blackburn}, {Boland}, {Bouman},
  {Bower}, {Bremer}, {Brinkerink}, {Brissenden}, {Britzen}, {Broderick},
  {Broguiere}, {Bronzwaer}, {Byun}, {Carlstrom}, {Chael}, {Chan}, {Chatterjee},
  {Chatterjee}, {Chen}, {Chen}, {Cho}, {Christian}, {Conway}, {Cordes}, {Crew},
  {Cui}, {Davelaar}, {De Laurentis}, {Deane}, {Dempsey}, {Desvignes}, {Dexter},
  {Doeleman}, {Eatough}, {Falcke}, {Fish}, {Fomalont}, {Fraga-Encinas},
  {Friberg}, {Fromm}, {G{\'o}mez}, {Galison}, {Gammie}, {Garc{\'\i}a},
  {Gentaz}, {Georgiev}, {Goddi}, {Gold}, {Gu}, {Gurwell}, {Hada}, {Hecht},
  {Hesper}, {Ho}, {Ho}, {Honma}, {Huang}, {Huang}, {Hughes}, {Ikeda}, {Inoue},
  {Issaoun}, {James}, {Jannuzi}, {Janssen}, {Jeter}, {Jiang}, {Johnson},
  {Jorstad}, {Jung}, {Karami}, {Karuppusamy}, {Kawashima}, {Keating},
  {Kettenis}, {Kim}, {Kim}, {Kim}, {Kino}, {Koay}, {Koch}, {Koyama}, {Kramer},
  {Kramer}, {Krichbaum}, {Kuo}, {Lauer}, {Lee}, {Li}, {Li}, {Lindqvist}, {Liu},
  {Liuzzo}, {Lo}, {Lobanov}, {Loinard}, {Lonsdale}, {Lu}, {MacDonald}, {Mao},
  {Markoff}, {Marrone}, {Marscher}, {Mart{\'\i}-Vidal}, {Matsushita},
  {Matthews}, {Medeiros}, {Menten}, {Mizuno}, {Mizuno}, {Moran}, {Moriyama},
  {Moscibrodzka}, {M{\"u}ller}, {Nagai}, {Nagar}, {Nakamura}, {Narayan},
  {Narayanan}, {Natarajan}, {Neri}, {Ni}, {Noutsos}, {Okino}, {Olivares},
  {Ortiz-Le{\'o}n}, {Oyama}, {{\"O}zel}, {Palumbo}, {Patel}, {Pen}, {Pesce},
  {Pi{\'e}tu}, {Plambeck}, {PopStefanija}, {Porth}, {Prather},
  {Preciado-L{\'o}pez}, {Psaltis}, {Pu}, {Ramakrishnan}, {Rao}, {Rawlings},
  {Raymond}, {Rezzolla}, {Ripperda}, {Roelofs}, {Rogers}, {Ros}, {Rose},
  {Roshanineshat}, {Rottmann}, {Roy}, {Ruszczyk}, {Ryan}, {Rygl},
  {S{\'a}nchez}, {S{\'a}nchez-Arguelles}, {Sasada}, {Savolainen}, {Schloerb},
  {Schuster}, {Shao}, {Shen}, {Small}, {Sohn}, {SooHoo}, {Tazaki}, {Tiede},
  {Tilanus}, {Titus}, {Toma}, {Torne}, {Trent}, {Trippe}, {Tsuda}, {van
  Bemmel}, {van Langevelde}, {van Rossum}, {Wagner}, {Wardle}, {Weintroub},
  {Wex}, {Wharton}, {Wielgus}, {Wong}, {Wu}, {Young}, {Young}, {Younsi},
  {Yuan}, {Yuan}, {Zensus}, {Zhao}, {Zhao}, {Zhu}, {Cappallo}, {Farah},
  {Folkers}, {Meyer-Zhao}, {Michalik}, {Nadolski}, {Nishioka}, {Pradel},
  {Primiani}, {Souccar}, {Vertatschitsch}, \& {Yamaguchi}}]{EHT2019c}
---. 2019{\natexlab{c}}, \apjl, 875, L3, \dodoi{10.3847/2041-8213/ab0c57}

\bibitem[{{Event Horizon Telescope Collaboration}
  {et~al.}(2019{\natexlab{d}}){Event Horizon Telescope Collaboration},
  {Akiyama}, {Alberdi}, {Alef}, {Asada}, {Azulay}, {Baczko}, {Ball},
  {Balokovi{\'c}}, {Barrett}, {Bintley}, {Blackburn}, {Boland}, {Bouman},
  {Bower}, {Bremer}, {Brinkerink}, {Brissenden}, {Britzen}, {Broderick},
  {Broguiere}, {Bronzwaer}, {Byun}, {Carlstrom}, {Chael}, {Chan}, {Chatterjee},
  {Chatterjee}, {Chen}, {Chen}, {Cho}, {Christian}, {Conway}, {Cordes}, {Crew},
  {Cui}, {Davelaar}, {De Laurentis}, {Deane}, {Dempsey}, {Desvignes}, {Dexter},
  {Doeleman}, {Eatough}, {Falcke}, {Fish}, {Fomalont}, {Fraga-Encinas},
  {Freeman}, {Friberg}, {Fromm}, {G{\'o}mez}, {Galison}, {Gammie},
  {Garc{\'\i}a}, {Gentaz}, {Georgiev}, {Goddi}, {Gold}, {Gu}, {Gurwell},
  {Hada}, {Hecht}, {Hesper}, {Ho}, {Ho}, {Honma}, {Huang}, {Huang}, {Hughes},
  {Ikeda}, {Inoue}, {Issaoun}, {James}, {Jannuzi}, {Janssen}, {Jeter}, {Jiang},
  {Johnson}, {Jorstad}, {Jung}, {Karami}, {Karuppusamy}, {Kawashima},
  {Keating}, {Kettenis}, {Kim}, {Kim}, {Kim}, {Kino}, {Koay}, {Koch}, {Koyama},
  {Kramer}, {Kramer}, {Krichbaum}, {Kuo}, {Lauer}, {Lee}, {Li}, {Li},
  {Lindqvist}, {Liu}, {Liuzzo}, {Lo}, {Lobanov}, {Loinard}, {Lonsdale}, {Lu},
  {MacDonald}, {Mao}, {Markoff}, {Marrone}, {Marscher}, {Mart{\'\i}-Vidal},
  {Matsushita}, {Matthews}, {Medeiros}, {Menten}, {Mizuno}, {Mizuno}, {Moran},
  {Moriyama}, {Moscibrodzka}, {M{\"u}ller}, {Nagai}, {Nagar}, {Nakamura},
  {Narayan}, {Narayanan}, {Natarajan}, {Neri}, {Ni}, {Noutsos}, {Okino},
  {Olivares}, {Oyama}, {{\"O}zel}, {Palumbo}, {Patel}, {Pen}, {Pesce},
  {Pi{\'e}tu}, {Plambeck}, {PopStefanija}, {Porth}, {Prather},
  {Preciado-L{\'o}pez}, {Psaltis}, {Pu}, {Ramakrishnan}, {Rao}, {Rawlings},
  {Raymond}, {Rezzolla}, {Ripperda}, {Roelofs}, {Rogers}, {Ros}, {Rose},
  {Roshanineshat}, {Rottmann}, {Roy}, {Ruszczyk}, {Ryan}, {Rygl},
  {S{\'a}nchez}, {S{\'a}nchez-Arguelles}, {Sasada}, {Savolainen}, {Schloerb},
  {Schuster}, {Shao}, {Shen}, {Small}, {Sohn}, {SooHoo}, {Tazaki}, {Tiede},
  {Tilanus}, {Titus}, {Toma}, {Torne}, {Trent}, {Trippe}, {Tsuda}, {van
  Bemmel}, {van Langevelde}, {van Rossum}, {Wagner}, {Wardle}, {Weintroub},
  {Wex}, {Wharton}, {Wielgus}, {Wong}, {Wu}, {Young}, {Young}, {Younsi},
  {Yuan}, {Yuan}, {Zensus}, {Zhao}, {Zhao}, {Zhu}, {Farah}, {Meyer-Zhao},
  {Michalik}, {Nadolski}, {Nishioka}, {Pradel}, {Primiani}, {Souccar},
  {Vertatschitsch}, \& {Yamaguchi}}]{EHT2019d}
---. 2019{\natexlab{d}}, \apjl, 875, L4, \dodoi{10.3847/2041-8213/ab0e85}

\bibitem[{{Event Horizon Telescope Collaboration}
  {et~al.}(2019{\natexlab{e}}){Event Horizon Telescope Collaboration},
  {Akiyama}, {Alberdi}, {Alef}, {Asada}, {Azulay}, {Baczko}, {Ball},
  {Balokovi{\'c}}, {Barrett}, {Bintley}, {Blackburn}, {Boland}, {Bouman},
  {Bower}, {Bremer}, {Brinkerink}, {Brissenden}, {Britzen}, {Broderick},
  {Broguiere}, {Bronzwaer}, {Byun}, {Carlstrom}, {Chael}, {Chan}, {Chatterjee},
  {Chatterjee}, {Chen}, {Chen}, {Cho}, {Christian}, {Conway}, {Cordes}, {Crew},
  {Cui}, {Davelaar}, {De Laurentis}, {Deane}, {Dempsey}, {Desvignes}, {Dexter},
  {Doeleman}, {Eatough}, {Falcke}, {Fish}, {Fomalont}, {Fraga-Encinas},
  {Friberg}, {Fromm}, {G{\'o}mez}, {Galison}, {Gammie}, {Garc{\'\i}a},
  {Gentaz}, {Georgiev}, {Goddi}, {Gold}, {Gu}, {Gurwell}, {Hada}, {Hecht},
  {Hesper}, {Ho}, {Ho}, {Honma}, {Huang}, {Huang}, {Hughes}, {Ikeda}, {Inoue},
  {Issaoun}, {James}, {Jannuzi}, {Janssen}, {Jeter}, {Jiang}, {Johnson},
  {Jorstad}, {Jung}, {Karami}, {Karuppusamy}, {Kawashima}, {Keating},
  {Kettenis}, {Kim}, {Kim}, {Kim}, {Kino}, {Koay}, {Koch}, {Koyama}, {Kramer},
  {Kramer}, {Krichbaum}, {Kuo}, {Lauer}, {Lee}, {Li}, {Li}, {Lindqvist}, {Liu},
  {Liuzzo}, {Lo}, {Lobanov}, {Loinard}, {Lonsdale}, {Lu}, {MacDonald}, {Mao},
  {Markoff}, {Marrone}, {Marscher}, {Mart{\'\i}-Vidal}, {Matsushita},
  {Matthews}, {Medeiros}, {Menten}, {Mizuno}, {Mizuno}, {Moran}, {Moriyama},
  {Moscibrodzka}, {Mul{\ensuremath{\ddot{}}}ler}, {Nagai}, {Nagar}, {Nakamura},
  {Narayan}, {Narayanan}, {Natarajan}, {Neri}, {Ni}, {Noutsos}, {Okino},
  {Olivares}, {Oyama}, {{\"O}zel}, {Palumbo}, {Patel}, {Pen}, {Pesce},
  {Pi{\'e}tu}, {Plambeck}, {PopStefanija}, {Porth}, {Prather},
  {Preciado-L{\'o}pez}, {Psaltis}, {Pu}, {Ramakrishnan}, {Rao}, {Rawlings},
  {Raymond}, {Rezzolla}, {Ripperda}, {Roelofs}, {Rogers}, {Ros}, {Rose},
  {Roshanineshat}, {Rottmann}, {Roy}, {Ruszczyk}, {Ryan}, {Rygl},
  {S{\'a}nchez}, {S{\'a}nchez-Arguelles}, {Sasada}, {Savolainen}, {Schloerb},
  {Schuster}, {Shao}, {Shen}, {Small}, {Sohn}, {SooHoo}, {Tazaki}, {Tiede},
  {Tilanus}, {Titus}, {Toma}, {Torne}, {Trent}, {Trippe}, {Tsuda}, {van
  Bemmel}, {van Langevelde}, {van Rossum}, {Wagner}, {Wardle}, {Weintroub},
  {Wex}, {Wharton}, {Wielgus}, {Wong}, {Wu}, {Young}, {Young}, {Younsi},
  {Yuan}, {Yuan}, {Zensus}, {Zhao}, {Zhao}, {Zhu}, {Anczarski}, {Baganoff},
  {Eckart}, {Farah}, {Haggard}, {Meyer-Zhao}, {Michalik}, {Nadolski},
  {Neilsen}, {Nishioka}, {Nowak}, {Pradel}, {Primiani}, {Souccar},
  {Vertatschitsch}, {Yamaguchi}, \& {Zhang}}]{EHT2019e}
---. 2019{\natexlab{e}}, \apjl, 875, L5, \dodoi{10.3847/2041-8213/ab0f43}

\bibitem[{{Event Horizon Telescope Collaboration}
  {et~al.}(2019{\natexlab{f}}){Event Horizon Telescope Collaboration},
  {Akiyama}, {Alberdi}, {Alef}, {Asada}, {Azulay}, {Baczko}, {Ball},
  {Balokovi{\'c}}, {Barrett}, {Bintley}, {Blackburn}, {Boland}, {Bouman},
  {Bower}, {Bremer}, {Brinkerink}, {Brissenden}, {Britzen}, {Broderick},
  {Broguiere}, {Bronzwaer}, {Byun}, {Carlstrom}, {Chael}, {Chan}, {Chatterjee},
  {Chatterjee}, {Chen}, {Chen}, {Cho}, {Christian}, {Conway}, {Cordes}, {Crew},
  {Cui}, {Davelaar}, {De Laurentis}, {Deane}, {Dempsey}, {Desvignes}, {Dexter},
  {Doeleman}, {Eatough}, {Falcke}, {Fish}, {Fomalont}, {Fraga-Encinas},
  {Friberg}, {Fromm}, {G{\'o}mez}, {Galison}, {Gammie}, {Garc{\'\i}a},
  {Gentaz}, {Georgiev}, {Goddi}, {Gold}, {Gu}, {Gurwell}, {Hada}, {Hecht},
  {Hesper}, {Ho}, {Ho}, {Honma}, {Huang}, {Huang}, {Hughes}, {Ikeda}, {Inoue},
  {Issaoun}, {James}, {Jannuzi}, {Janssen}, {Jeter}, {Jiang}, {Johnson},
  {Jorstad}, {Jung}, {Karami}, {Karuppusamy}, {Kawashima}, {Keating},
  {Kettenis}, {Kim}, {Kim}, {Kim}, {Kino}, {Koay}, {Koch}, {Koyama}, {Kramer},
  {Kramer}, {Krichbaum}, {Kuo}, {Lauer}, {Lee}, {Li}, {Li}, {Lindqvist}, {Liu},
  {Liuzzo}, {Lo}, {Lobanov}, {Loinard}, {Lonsdale}, {Lu}, {MacDonald}, {Mao},
  {Markoff}, {Marrone}, {Marscher}, {Mart{\'\i}-Vidal}, {Matsushita},
  {Matthews}, {Medeiros}, {Menten}, {Mizuno}, {Mizuno}, {Moran}, {Moriyama},
  {Moscibrodzka}, {M{\"u}ller}, {Nagai}, {Nagar}, {Nakamura}, {Narayan},
  {Narayanan}, {Natarajan}, {Neri}, {Ni}, {Noutsos}, {Okino}, {Olivares},
  {Oyama}, {{\"O}zel}, {Palumbo}, {Patel}, {Pen}, {Pesce}, {Pi{\'e}tu},
  {Plambeck}, {PopStefanija}, {Porth}, {Prather}, {Preciado-L{\'o}pez},
  {Psaltis}, {Pu}, {Ramakrishnan}, {Rao}, {Rawlings}, {Raymond}, {Rezzolla},
  {Ripperda}, {Roelofs}, {Rogers}, {Ros}, {Rose}, {Roshanineshat}, {Rottmann},
  {Roy}, {Ruszczyk}, {Ryan}, {Rygl}, {S{\'a}nchez}, {S{\'a}nchez-Arguelles},
  {Sasada}, {Savolainen}, {Schloerb}, {Schuster}, {Shao}, {Shen}, {Small},
  {Sohn}, {SooHoo}, {Tazaki}, {Tiede}, {Tilanus}, {Titus}, {Toma}, {Torne},
  {Trent}, {Trippe}, {Tsuda}, {van Bemmel}, {van Langevelde}, {van Rossum},
  {Wagner}, {Wardle}, {Weintroub}, {Wex}, {Wharton}, {Wielgus}, {Wong}, {Wu},
  {Young}, {Young}, {Younsi}, {Yuan}, {Yuan}, {Zensus}, {Zhao}, {Zhao}, {Zhu},
  {Farah}, {Meyer-Zhao}, {Michalik}, {Nadolski}, {Nishioka}, {Pradel},
  {Primiani}, {Souccar}, {Vertatschitsch}, \& {Yamaguchi}}]{EHT2019f}
---. 2019{\natexlab{f}}, \apjl, 875, L6, \dodoi{10.3847/2041-8213/ab1141}

\bibitem[{{Gabuzda} {et~al.}(2004){Gabuzda}, {Murray}, \&
  {Cronin}}]{Gabuzda2004}
{Gabuzda}, D.~C., {Murray}, {\'E}., \& {Cronin}, P. 2004, \mnras, 351, L89,
  \dodoi{10.1111/j.1365-2966.2004.08037.x}

\bibitem[{{G{\'o}mez} {et~al.}(2000){G{\'o}mez}, {Marscher}, {Alberdi},
  {Jorstad}, \& {Garc{\'\i}a-Mir{\'o}}}]{Gomez2000}
{G{\'o}mez}, J.-L., {Marscher}, A.~P., {Alberdi}, A., {Jorstad}, S.~G., \&
  {Garc{\'\i}a-Mir{\'o}}, C. 2000, Science, 289, 2317,
  \dodoi{10.1126/science.289.5488.2317}

\bibitem[{{G{\'o}mez} {et~al.}(2016){G{\'o}mez}, {Lobanov}, {Bruni}, {Kovalev},
  {Marscher}, {Jorstad}, {Mizuno}, {Bach}, {Sokolovsky}, {Anderson}, {Galindo},
  {Kardashev}, \& {Lisakov}}]{Gomez2016}
{G{\'o}mez}, J.~L., {Lobanov}, A.~P., {Bruni}, G., {et~al.} 2016, \apj, 817,
  96, \dodoi{10.3847/0004-637X/817/2/96}

\bibitem[{{Greisen}(2003)}]{Greisen2003}
{Greisen}, E.~W. 2003, Astrophysics and Space Science Library, Vol. 285, {AIPS,
  the VLA, and the VLBA}, ed. A.~{Heck}, 109, \dodoi{10.1007/0-306-48080-8_7}

\bibitem[{{Hada}(2017)}]{Hada2017}
{Hada}, K. 2017, Galaxies, 5, 2, \dodoi{10.3390/galaxies5010002}

\bibitem[{{Hada} {et~al.}(2016){Hada}, {Kino}, {Doi}, {Nagai}, {Honma},
  {Akiyama}, {Tazaki}, {Lico}, {Giroletti}, {Giovannini}, {Orienti}, \&
  {Hagiwara}}]{Hada2016}
{Hada}, K., {Kino}, M., {Doi}, A., {et~al.} 2016, \apj, 817, 131,
  \dodoi{10.3847/0004-637X/817/2/131}

\bibitem[{{H{\"o}gbom}(1974)}]{Hogbom1974}
{H{\"o}gbom}, J.~A. 1974, \aaps, 15, 417

\bibitem[{{Homan} \& {Lister}(2006)}]{HL2006}
{Homan}, D.~C., \& {Lister}, M.~L. 2006, \aj, 131, 1262, \dodoi{10.1086/500256}

\bibitem[{{Homan} \& {Wardle}(1999)}]{HW1999}
{Homan}, D.~C., \& {Wardle}, J.~F.~C. 1999, \aj, 118, 1942,
  \dodoi{10.1086/301108}

\bibitem[{{Hovatta} {et~al.}(2012){Hovatta}, {Lister}, {Aller}, {Aller},
  {Homan}, {Kovalev}, {Pushkarev}, \& {Savolainen}}]{Hovatta2012}
{Hovatta}, T., {Lister}, M.~L., {Aller}, M.~F., {et~al.} 2012, \aj, 144, 105,
  \dodoi{10.1088/0004-6256/144/4/105}

\bibitem[{{Hovatta} {et~al.}(2019){Hovatta}, {O'Sullivan}, {Mart{\'\i}-Vidal},
  {Savolainen}, \& {Tchekhovskoy}}]{Hovatta2019}
{Hovatta}, T., {O'Sullivan}, S., {Mart{\'\i}-Vidal}, I., {Savolainen}, T., \&
  {Tchekhovskoy}, A. 2019, \aap, 623, A111, \dodoi{10.1051/0004-6361/201832587}

\bibitem[{{Hunter}(2007)}]{Matplotlib2007}
{Hunter}, J.~D. 2007, Computing in Science Engineering, 9, 90

\bibitem[{{Issaoun} {et~al.}(2019){Issaoun}, {Johnson}, {Blackburn},
  {Brinkerink}, {Mo{\'s}cibrodzka}, {Chael}, {Goddi}, {Mart{\'\i}-Vidal},
  {Wagner}, {Doeleman}, {Falcke}, {Krichbaum}, {Akiyama}, {Bach}, {Bouman},
  {Bower}, {Broderick}, {Cho}, {Crew}, {Dexter}, {Fish}, {Gold}, {G{\'o}mez},
  {Hada}, {Hern{\'a}ndez-G{\'o}mez}, {Jan{\ss}en}, {Kino}, {Kramer}, {Loinard},
  {Lu}, {Markoff}, {Marrone}, {Matthews}, {Moran}, {M{\"u}ller}, {Roelofs},
  {Ros}, {Rottmann}, {Sanchez}, {Tilanus}, {de Vicente}, {Wielgus}, {Zensus},
  \& {Zhao}}]{Issaoun2019}
{Issaoun}, S., {Johnson}, M.~D., {Blackburn}, L., {et~al.} 2019, \apj, 871, 30,
  \dodoi{10.3847/1538-4357/aaf732}

\bibitem[{{Jorstad} {et~al.}(2005){Jorstad}, {Marscher}, {Lister}, {Stirling},
  {Cawthorne}, {Gear}, {G{\'o}mez}, {Stevens}, {Smith}, {Forster}, \&
  {Robson}}]{Jorstad2005}
{Jorstad}, S.~G., {Marscher}, A.~P., {Lister}, M.~L., {et~al.} 2005, \aj, 130,
  1418, \dodoi{10.1086/444593}

\bibitem[{{Jorstad} {et~al.}(2007){Jorstad}, {Marscher}, {Stevens}, {Smith},
  {Forster}, {Gear}, {Cawthorne}, {Lister}, {Stirling}, {G{\'o}mez}, {Greaves},
  \& {Robson}}]{Jorstad2007}
{Jorstad}, S.~G., {Marscher}, A.~P., {Stevens}, J.~A., {et~al.} 2007, \aj, 134,
  799, \dodoi{10.1086/519996}

\bibitem[{{Jorstad} {et~al.}(2017){Jorstad}, {Marscher}, {Morozova},
  {Troitsky}, {Agudo}, {Casadio}, {Foord}, {G{\'o}mez}, {MacDonald}, {Molina},
  {L{\"a}hteenm{\"a}ki}, {Tammi}, \& {Tornikoski}}]{Jorstad2017}
{Jorstad}, S.~G., {Marscher}, A.~P., {Morozova}, D.~A., {et~al.} 2017, \apj,
  846, 98, \dodoi{10.3847/1538-4357/aa8407}

\bibitem[{{Kettenis} {et~al.}(2006){Kettenis}, {van Langevelde}, {Reynolds}, \&
  {Cotton}}]{Kettenis2006}
{Kettenis}, M., {van Langevelde}, H.~J., {Reynolds}, C., \& {Cotton}, B. 2006,
  in Astronomical Society of the Pacific Conference Series, Vol. 351,
  Astronomical Data Analysis Software and Systems XV, ed. C.~{Gabriel},
  C.~{Arviset}, D.~{Ponz}, \& S.~{Enrique}, 497

\bibitem[{{Kim} {et~al.}(2020){Kim}, {Krichbaum}, {Broderick}, {Wielgus},
  {Blackburn}, {G{\'o}mez}, {Johnson}, {Bouman}, {Chael}, {Akiyama}, {Jorstad},
  {Marscher}, {Issaoun}, {Janssen}, {Chan}, {Savolainen}, {Pesce}, {{\"O}zel},
  {Alberdi}, {Alef}, {Asada}, {Azulay}, {Baczko}, {Ball}, {Balokovi{\'c}},
  {Barrett}, {Bintley}, {Boland }, {Bower}, {Bremer}, {Brinkerink},
  {Brissenden}, {Britzen}, {Broguiere}, {Bronzwaer}, {Byun}, {Carlstrom},
  {Chatterjee}, {Chatterjee}, {Chen}, {Chen}, {Cho}, {Christian}, {Conway},
  {Cordes}, {Crew}, {Cui}, {Davelaar}, {De Laurentis}, {Deane}, {Dempsey},
  {Desvignes}, {Dexter}, {Doeleman}, {Eatough}, {Falcke}, {Fish}, {Fomalont},
  {Fraga-Encinas}, {Friberg}, {Fromm}, {Galison}, {Gammie}, {Garc{\'\i}a},
  {Gentaz}, {Georgiev}, {Goddi}, {Gold}, {G{\'o}mez-Ruiz}, {Gu}, {Gurwell},
  {Hada}, {Hecht}, {Hesper}, {Ho}, {Ho}, {Honma}, {Huang}, {Huang}, {Hughes},
  {Ikeda}, {Inoue}, {James}, {Jannuzi}, {Jeter}, {Jiang}, {Jimenez-Rosales},
  {Jung}, {Karami}, {Karuppusamy}, {Kawashima}, {Keating}, {Kettenis}, {Kim},
  {Kim}, {Kino}, {Koay}, {Koch}, {Koyama}, {Kramer}, {Kramer}, {Kuo}, {Lauer},
  {Lee}, {Li}, {Li}, {Lindqvist}, {Lico}, {Liu}, {Liuzzo}, {Lo}, {Lobanov},
  {Loinard}, {Lonsdale}, {Lu}, {MacDonald}, {Mao}, {Markoff}, {Marrone},
  {Mart{\'\i}-Vidal}, {Matsushita}, {Matthews}, {Medeiros}, {Menten}, {Mizuno},
  {Mizuno}, {Moran}, {Moriyama}, {Moscibrodzka}, {Musoke}, {M{\"u}ller},
  {Nagai}, {Nagar}, {Nakamura}, {Narayan}, {Narayanan}, {Natarajan}, {Neri},
  {Ni}, {Noutsos}, {Okino}, {Olivares}, {Ortiz-Le{\'o}n}, {Oyama}, {Palumbo},
  {Park}, {Patel}, {Pen}, {Pi{\'e}tu}, {Plambeck}, {PopStefanija}, {Porth},
  {Prather}, {Preciado-L{\'o}pez}, {Psaltis}, {Pu}, {Ramakrishnan}, {Rao},
  {Rawlings}, {Raymond}, {Rezzolla}, {Ripperda}, {Roelofs}, {Rogers}, {Ros},
  {Rose}, {Roshanineshat}, {Rottmann}, {Roy}, {Ruszczyk}, {Ryan}, {Rygl},
  {S{\'a}nchez}, {S{\'a}nchez-Arguelles}, {Sasada}, {Schloerb}, {Schuster},
  {Shao}, {Shen}, {Small}, {Sohn}, {SooHoo}, {Tazaki}, {Tiede}, {Tilanus},
  {Titus}, {Toma}, {Torne}, {Trent}, {Traianou}, {Trippe}, {Tsuda}, {van
  Bemmel}, {van Langevelde}, {van Rossum}, {Wagner}, {Wardle}, {Ward-Thompson},
  {Weintroub}, {Wex}, {Wharton}, {Wong}, {Wu}, {Yoon}, {Young}, {Young},
  {Younsi}, {Yuan}, {Yuan}, {Zensus}, {Zhao}, {Zhao}, {Zhu}, {Algaba},
  {Allardi}, {Amestica}, {Anczarski}, {Bach}, {Baganoff}, {Beaudoin}, {Benson},
  {Berthold}, {Blanchard}, {Blundell}, {Bustamente}, {Cappallo},
  {Castillo-Dom{\'\i}nguez}, {Chang}, {Chang}, {Chang}, {Chen}, {Chilson},
  {Chuter}, {Rosado}, {Coulson}, {Crowley}, {Derome}, {Dexter}, {Dornbusch},
  {Dudevoir}, {Dzib}, {Eckart}, {Eckert}, {Erickson}, {Everett}, {Faber},
  {Farah}, {Fath}, {Folkers}, {Forbes}, {Freund}, {Gale}, {Gao}, {Geertsema},
  {Graham}, {Greer}, {Grosslein}, {Gueth}, {Haggard}, {Halverson}, {Han},
  {Han}, {Hao}, {Hasegawa}, {Henning}, {Hern{\'a}ndez-G{\'o}mez},
  {Herrero-Illana}, {Heyminck}, {Hirota}, {Hoge}, {Huang}, {Violette
  Impellizzeri}, {Jiang}, {John}, {Kamble}, {Keisler}, {Kimura}, {Kono},
  {Kubo}, {Kuroda}, {Lacasse}, {Laing}, {Leitch}, {Li}, {Lin}, {Liu}, {Liu},
  {Lu}, {Marson}, {Martin-Cocher}, {Massingill}, {Matulonis}, {McColl},
  {McWhirter}, {Messias}, {Meyer-Zhao}, {Michalik}, {Monta{\~n}a},
  {Montgomerie}, {Mora-Klein}, {Muders}, {Nadolski}, {Navarro}, {Neilsen},
  {Nguyen}, {Nishioka}, {Norton}, {Nowak}, {Nystrom}, {Ogawa}, {Oshiro},
  {Oyama}, {Parsons}, {Pe{\~n}alver}, {Phillips}, {Poirier}, {Pradel},
  {Primiani}, {Raffin}, {Rahlin}, {Reiland}, {Risacher}, {Ruiz},
  {S{\'a}ez-Mada{\'\i}n}, {Sassella}, {Schellart}, {Shaw}, {Silva}, {Shiokawa},
  {Smith}, {Snow}, {Souccar}, {Sousa}, {Sridharan}, {Srinivasan}, {Stahm},
  {Stark}, {Story}, {Timmer}, {Vertatschitsch}, {Walther}, {Wei}, {Whitehorn},
  {Whitney}, {Woody}, {Wouterloot}, {Wright}, {Yamaguchi}, {Yu}, {Zeballos},
  {Zhang}, {Ziurys}, \& {Event Horizon Telescope Collaboration}}]{Kim2020}
{Kim}, J.-Y., {Krichbaum}, T.~P., {Broderick}, A.~E., {et~al.} 2020, \aap, 640,
  A69, \dodoi{10.1051/0004-6361/202037493}

\bibitem[{{Kravchenko} {et~al.}(2017){Kravchenko}, {Kovalev}, \&
  {Sokolovsky}}]{Kravchenko2017}
{Kravchenko}, E.~V., {Kovalev}, Y.~Y., \& {Sokolovsky}, K.~V. 2017, \mnras,
  467, 83, \dodoi{10.1093/mnras/stx021}

\bibitem[{{Leppanen} {et~al.}(1995){Leppanen}, {Zensus}, \&
  {Diamond}}]{Leppanen1995}
{Leppanen}, K.~J., {Zensus}, J.~A., \& {Diamond}, P.~J. 1995, \aj, 110, 2479,
  \dodoi{10.1086/117706}

\bibitem[{{Lisakov} {et~al.}(2017){Lisakov}, {Kovalev}, {Savolainen},
  {Hovatta}, \& {Kutkin}}]{Lisakov2017}
{Lisakov}, M.~M., {Kovalev}, Y.~Y., {Savolainen}, T., {Hovatta}, T., \&
  {Kutkin}, A.~M. 2017, \mnras, 468, 4478, \dodoi{10.1093/mnras/stx710}

\bibitem[{{Lister} {et~al.}(2018){Lister}, {Aller}, {Aller}, {Hodge}, {Homan},
  {Kovalev}, {Pushkarev}, \& {Savolainen}}]{Lister2018}
{Lister}, M.~L., {Aller}, M.~F., {Aller}, H.~D., {et~al.} 2018, \apjs, 234, 12,
  \dodoi{10.3847/1538-4365/aa9c44}

\bibitem[{{Lister} \& {Homan}(2005)}]{LH2005}
{Lister}, M.~L., \& {Homan}, D.~C. 2005, \aj, 130, 1389, \dodoi{10.1086/432969}

\bibitem[{{Marscher}(2016)}]{Marscher2016}
{Marscher}, A. 2016, Galaxies, 4, 37, \dodoi{10.3390/galaxies4040037}

\bibitem[{{McMullin} {et~al.}(2007){McMullin}, {Waters}, {Schiebel}, {Young},
  \& {Golap}}]{CASA2007}
{McMullin}, J.~P., {Waters}, B., {Schiebel}, D., {Young}, W., \& {Golap}, K.
  2007, in Astronomical Society of the Pacific Conference Series, Vol. 376,
  Astronomical Data Analysis Software and Systems XVI, ed. R.~A. {Shaw},
  F.~{Hill}, \& D.~J. {Bell}, 127

\bibitem[{{Nagai} {et~al.}(2017){Nagai}, {Fujita}, {Nakamura}, {Orienti},
  {Kino}, {Asada}, \& {Giovannini}}]{Nagai2017}
{Nagai}, H., {Fujita}, Y., {Nakamura}, M., {et~al.} 2017, \apj, 849, 52,
  \dodoi{10.3847/1538-4357/aa8e43}

\bibitem[{{O'Sullivan} \& {Gabuzda}(2009)}]{OG2009}
{O'Sullivan}, S.~P., \& {Gabuzda}, D.~C. 2009, \mnras, 393, 429,
  \dodoi{10.1111/j.1365-2966.2008.14213.x}

\bibitem[{{Park} {et~al.}(2019){Park}, {Hada}, {Kino}, {Nakamura}, {Ro}, \&
  {Trippe}}]{Park2019}
{Park}, J., {Hada}, K., {Kino}, M., {et~al.} 2019, \apj, 871, 257,
  \dodoi{10.3847/1538-4357/aaf9a9}

\bibitem[{{Park} {et~al.}(2018){Park}, {Kam}, {Trippe}, {Kang}, {Byun}, {Kim},
  {Algaba}, {Lee}, {Zhao}, {Kino}, {Shin}, {Hada}, {Lee}, {Oh}, {Hodgson}, \&
  {Sohn}}]{Park2018}
{Park}, J., {Kam}, M., {Trippe}, S., {et~al.} 2018, \apj, 860, 112,
  \dodoi{10.3847/1538-4357/aac490}

\bibitem[{{Roberts} {et~al.}(1994){Roberts}, {Wardle}, \&
  {Brown}}]{Roberts1994}
{Roberts}, D.~H., {Wardle}, J.~F.~C., \& {Brown}, L.~F. 1994, \apj, 427, 718,
  \dodoi{10.1086/174180}

\bibitem[{{Schwab} \& {Cotton}(1983)}]{CS1983}
{Schwab}, F.~R., \& {Cotton}, W.~D. 1983, \aj, 88, 688, \dodoi{10.1086/113360}

\bibitem[{{Shepherd}(1997)}]{Shepherd1997}
{Shepherd}, M.~C. 1997, in Astronomical Society of the Pacific Conference
  Series, Vol. 125, Astronomical Data Analysis Software and Systems VI, ed.
  G.~{Hunt} \& H.~{Payne}, 77

\bibitem[{{Sokoloff} {et~al.}(1998){Sokoloff}, {Bykov}, {Shukurov},
  {Berkhuijsen}, {Beck}, \& {Poezd}}]{Sokoloff1998}
{Sokoloff}, D.~D., {Bykov}, A.~A., {Shukurov}, A., {et~al.} 1998, \mnras, 299,
  189, \dodoi{10.1046/j.1365-8711.1998.01782.x}

\bibitem[{{Trippe} {et~al.}(2010){Trippe}, {Neri}, {Krips}, {Castro-Carrizo},
  {Bremer}, {Pi{\'e}tu}, \& {Fontana}}]{Trippe2010}
{Trippe}, S., {Neri}, R., {Krips}, M., {et~al.} 2010, \aap, 515, A40,
  \dodoi{10.1051/0004-6361/200913871}

\bibitem[{{van der Walt} {et~al.}(2011){van der Walt}, {Colbert}, \&
  {Varoquaux}}]{Numpy2011}
{van der Walt}, S., {Colbert}, S.~C., \& {Varoquaux}, G. 2011, Computing in
  Science Engineering, 13, 22

\bibitem[{{Virtanen} {et~al.}(2020){Virtanen}, {Gommers}, {Oliphant},
  {Haberland}, {Reddy}, {Cournapeau}, {Burovski}, {Peterson}, {Weckesser},
  {Bright}, {van der Walt}, {Brett}, {Wilson}, {Jarrod Millman}, {Mayorov},
  {Nelson}, {Jones}, {Kern}, {Larson}, {Carey}, {Polat}, {Feng}, {Moore}, {Vand
  erPlas}, {Laxalde}, {Perktold}, {Cimrman}, {Henriksen}, {Quintero}, {Harris},
  {Archibald}, {Ribeiro}, {Pedregosa}, {van Mulbregt}, \&
  {Contributors}}]{Scipy2020}
{Virtanen}, P., {Gommers}, R., {Oliphant}, T.~E., {et~al.} 2020, Nature
  Methods, 17, 261, \dodoi{https://doi.org/10.1038/s41592-019-0686-2}

\bibitem[{{Wardle} {et~al.}(1998){Wardle}, {Homan}, {Ojha}, \&
  {Roberts}}]{Wardle1998}
{Wardle}, J.~F.~C., {Homan}, D.~C., {Ojha}, R., \& {Roberts}, D.~H. 1998, \nat,
  395, 457, \dodoi{10.1038/26675}

\bibitem[{{W}es {M}c{K}inney(2010)}]{Pandas2010}
{W}es {M}c{K}inney. 2010, in {P}roceedings of the 9th {P}ython in {S}cience
  {C}onference, ed. {S}t\'efan van~der {W}alt \& {J}arrod {M}illman, 56 -- 61,
  \dodoi{10.25080/Majora-92bf1922-00a}

\bibitem[{{Zavala} \& {Taylor}(2005)}]{ZT2005}
{Zavala}, R.~T., \& {Taylor}, G.~B. 2005, \apjl, 626, L73,
  \dodoi{10.1086/431901}

\end{thebibliography}
\bibliographystyle{aasjournal}


\end{document}